\documentclass{jfm}
\usepackage{graphicx}
\usepackage{epstopdf, epsfig}
\usepackage{amsmath,amssymb} 
\usepackage{lscape}
\usepackage{pgfplots}
\pgfplotsset{compat=1.5}
\usepackage{floatrow}
\usepackage{enumitem,lipsum}
\usepackage{arydshln}
\usepackage{float}
\usepackage{setspace}

\definecolor{pink}{HTML}{16ca04}

\def\IB#1{\boldsymbol{#1}} 
\def\bten#1{\mathsfbi{#1}} 
\def\sl#1{\mathfrak{s}_{x}({\tilde \psi }_{#1})} 

\pgfplotsset{every tick label/.append style={font=\small}}

\newcommand{\grad}[2]{\frac{\partial #1}{\partial #2}}
\newcommand{\lap}[2]{\frac{\partial^{2} #1}{\partial #2^{2}}}

\shorttitle{Non-Newtonian effects on the slip and mobility of an active particle}
\shortauthor{A. Choudhary, T. Renganathan and S. Pushpavanam}

\title{Non-Newtonian effects on the slip and mobility of a self-propelling active particle}

\author{Akash Choudhary\aff{1}, T. Renganathan\aff{1}
	\and S. Pushpavanam\aff{1}
	\corresp{\email{spush@iitm.ac.in}}}

\affiliation{\aff{1}Department of Chemical Engineering, Indian Institute of Technology, Chennai,
	
	TN 600036, India}

\begin{document}
	
	\maketitle
	
	
		\begin{abstract}
	Janus particles propel themselves by generating concentration gradients along their active surface. 
	This induces a flow near the surface, known as the diffusio-osmotic slip, which propels the particle even in the absence of externally applied concentration gradients.
	In this work, we study the influence of viscoelasticity and shear-thinning (described by the second-order-fluid and Carreau model, respectively) on the diffusio-osmotic slip on an active surface. Using matched asymptotic expansions, we provide an analytical expression for the modification of slip induced by the non-Newtonian behavior. 
	The results reveal that the modification in slip velocity, arising from polymer elasticity, is proportional to the second tangential derivative of the concentration field. 
	Using the reciprocal theorem, we estimate the influence of this modification on the mobility of a Janus sphere. 
	The current study also has direct implications on the understanding of the transport of complex fluids in diffusio-osmotic pumps.
	

\end{abstract}

\section {Introduction}
Synthetic active particles are micron and submicron sized colloidal particles which can propel themselves along predictable trajectories. 
The self-propulsion arises from the generation of local concentration gradient at the surface, which is induced by variation in the surface activity such as adsorption or reaction \citep{anderson1989,golestanian2005propulsion,golestanian2007designing,julicher2009generic}. 
These active particles play an important role in biomedical research; they offer possible applications in drug-delivery micromachines and controlled studies of microbial infections through bio-sensing \citep{gao2014synthetic,su2019janus}.
Furthermore, a suspension of active particles represents a non-equilibrium system which exhibits characteristics such as enhanced fluid mixing in complex fluids \citep{gomez2016dynamics} and inertia-less turbulence, which are also observed in microbial suspensions and granular matter \citep{patteson2016active}.
Therefore, understanding such novel self-propelled systems provides insights which extend to a wide range of physical phenomena, making it intriguing from a scientific standpoint. 

The notion of self-propulsion at the micro-scale was introduced via exploitation of electrochemical-mechanical transduction mechanism i.e. conversion of electrochemical energy (stored in electrolytes) to power mechanical motion. \cite{paxton2004catalytic,paxton2006chemical} and \cite{ozin2005synthetic} fabricated bimetallic micro rods by coating the two halves with platinum and silver. 
When immersed in an aqueous solution of hydrogen peroxide, the oxidation process generates excess electrons and protons (H\textsuperscript{+}) at the platinum end.
 The electrons are transported to the other end through conduction, where they are consumed to reduce hydrogen peroxide.
  The asymmetric distribution of protons drives the autonomous motion of the micro-rod, which was later called `self-electrophoresis'.


\cite{golestanian2005propulsion,howse2007self} studied the second mechanism of self-propulsion: conversion of chemical energy (arising from molecular interactions such as van der Waals) to power the mechanical motion in a non-electrolytic medium.
\cite{golestanian2007designing} proposed the first continuum framework to understand the underlying mechanism which drives the self-propulsion of Janus spheres. 
On the basis of existing experimental studies \citep{golestanian2005propulsion,ozin2005synthetic}, three major assumptions were made: (i) the interaction layer is asymptotically thin in comparison to the particle radius ($ \epsilon  \ll 1 $, where $ \epsilon $ is the ratio of the thickness of interaction layer to particle radius); (ii) the active (catalytic) surface adsorbs (or desorbs) the solute molecules at a fixed rate and diffusion time scale is much shorter than the reactive time scale; (iii) advective effects are negligible.
Borrowing insights from the seminal works of \cite{derjaguin1947kinetic}, Anderson, Prieve and co-workers \citep{anderson1982motion,anderson1989}, they showed that the interaction between the particle and the solute molecules creates a pressure gradient inside the thin interaction layer. 
This pressure gradient is balanced by the viscous stresses which is the driving mechanism for self-propulsion. 
An external variation in solute concentration around the colloid can trigger the pressure gradient which generates a surface slip. 
Consequently, the freely suspended particle moves in the direction of chemical gradient, known as diffusio-phoresis: a macro-scale motion manifested through an asymmetry at the micro-scale. 
A Janus sphere (possessing a chemically active cap coated on an inert particle) can create and sustain this pressure gradient as its asymmetrically active surface facilitates a tangential concentration gradient and is therefore also called a self-diffusiophoretic particle. 


The continuum description assumes that the solute molecules do not occupy volume and therefore loses its validity for nanometer sized Janus spheres. 
For such cases, \cite{cordova2008osmotic, brady2011particle} provided a colloidal perspective to diffusiophoresis: solute molecule and particle interact both energetically (through hard-sphere and van der Waals forces) and hydrodynamically with each other. 
Later \cite{sharifi2013} demonstrated that the results derived by \cite{brady2011particle} (for diffusiophoretic velocity) can be obtained through a continuum description, in the limit of asymptotically small solute molecules. 
They relaxed the assumption of thin interaction layer and incorporated the effects of an irreversible reaction (characterized by Damköhler number) at the active surface which was found to dampen the propulsion.

\cite{michelin2014phoretic} explored the effects of finite advection on self-propulsion. They found that the particles exhibiting attractive interaction with solute molecules show a maxima in swimming speed at a finite Péclet number ($ Pe \sim O(1)$, here $ Pe $ is the ratio of diffusive to advective time scales); whereas for repulsive interactions, the swimming speed always reduces with increase in advection. 
In agreement with \cite{sharifi2013}, they also reported a monotonic decrease in swimming velocity with an increase in Damköhler number.
Utilizing matched asymptotic expansions, they showed that the diffusive flux at the surface of the Janus sphere is equal to that at the outer edge of the interaction layer, provided $ \epsilon Pe \ll 1 $. 


\setlength{\dashlinedash}{2pt}

\begin{table}
	\renewcommand{\arraystretch}{0.8}
	\small
	\begin{center}
		\def~{\hphantom{0}}
		\begin{tabular}{c c l} \\ \hline \\
			Investigation                                                     & Regime        &  Description						 \\[10pt] \hline \\
			
			\begin{tabular}[t]{c} Golestanian \textit{et al.}\\ (2005, 2007) \end{tabular}        & \begin{tabular}[t]{c} $ \epsilon \rightarrow 0 $\\ $ Da \rightarrow 0 ; \, Pe \rightarrow 0 $\end{tabular}     & \begin{tabular}[t]{l} Built the first continuum-level frame-\\ work using the foundation laid by \\ \cite{anderson1982motion}. \end{tabular} \\ \\
			
			\begin{tabular}[t]{c} \cite{brady2011particle} \\ \cite{sharifi2013} \end{tabular}        & \begin{tabular}[t]{c}  $ \epsilon \sim O(1) $ \\ $ Da \ll 1 ; \, Pe \rightarrow 0 $ \end{tabular}     & \begin{tabular}[t]{l}  Bridged continuum and colloidal \\ perspectives. \end{tabular} \\ \\
			
			\begin{tabular}[t]{c} \cite{michelin2014phoretic} \end{tabular}         & \begin{tabular}[t]{c} $ \epsilon \ll 1 $\\$ Da > O(1); \, Pe > O(1) $ \end{tabular}     & \begin{tabular}[t]{l} Studied advective and reactive effects.\\ Showed well-posedness of flux bound-\\ary condition at the macro-scale. \end{tabular} \\ \\
			%
			
			\hdashline \\
			
			\begin{tabular}[t]{c}  \cite{zhu2012self} \\ \cite{corato2015locomotion}  \end{tabular}        & \begin{tabular}[t]{c} \; $ \epsilon \rightarrow 0 $; two modes \end{tabular}     & \begin{tabular}[t]{l} Swimming kinematics and hydrodyna- \\mics of swimmers in complex media.\\ \cite{zhu2012self} $ \rightarrow De \sim O(1) $ \\ \cite{corato2015locomotion} $ \rightarrow De \ll 1 $ \end{tabular} \\ \\
			
			\begin{tabular}[t]{c} \cite{datt2015squirmingST,datt2017activeComplex} \\ \cite{pietrzyk2019flow} \end{tabular}       & \begin{tabular}[t]{c} $ \epsilon \rightarrow 0 $\\ $ \chi \ll 1; \, De \ll 1 $ \end{tabular}     & \begin{tabular}[t]{l} Extended \cite{golestanian2007designing} \& \\showed the importance of higher swim-\\ ming modes and asymmetry of surface\\ activity, assuming the swimming gait\\ to be Newtonian. \end{tabular} \\ \\
			
			\begin{tabular}[t]{c}  Natale \textit{et al.} (2017) \end{tabular}       & \begin{tabular}[t]{c}  $ \epsilon \rightarrow 0 $ \\ $ Da \sim O(1); Pe \sim O(1) $ \\ $ \, De \ll 1 $  \end{tabular}     & \begin{tabular}[t]{l} Extended \cite{datt2017activeComplex} to \\ include advective and reactive effects. \\ FEM simulations revealed sharp grad-\\ient in stresses due to viscoelasticity.  \end{tabular} \\ \\
			
			
			\begin{tabular}[t]{c} {}\\ \\ This work \end{tabular}         & \begin{tabular}[t]{c} {}\\$ \epsilon \ll 1 $\\ $ Da \rightarrow 0 $ \\ $ \, De \ll 1; \, \chi \ll 1  $ \end{tabular}     & \begin{tabular}[t]{l}  Extends \cite{anderson1982motion} to inc-\\lude the effects of complex rheology.\\ Uses \cite{golestanian2007designing} frame-\\ work to study non-Newtonian effects\\ on the swimming gait and mobility\\ of Janus sphere for $ Pe \rightarrow 0 \, \& \, \epsilon Pe \ll 1 $. \end{tabular}\\ \hline

		\end{tabular}
		\caption{A summary of continuum-level studies in the past towards the modeling of a non-Brownian active particle; first section of the table describes the studies in Newtonian medium in various regimes; second section summarizes the studies which explored the effects of Non-Newtonian behavior. Here, $ \epsilon $ is the interaction layer thickness, normalized by the particle radius; $ Da $ is the Damköhler number which represents the ratio of time scale associated with diffusion to that of reaction; $ Pe $ is the Péclet number which is the ratio of time scale associated with diffusion to that of advection; $ De $ is the Deborah number which governs the ratio of time scale of polymer relaxation to that of advection; $ \chi $ represents the deviation of infinite shear rate viscosity from that of zero shear rate (the difference is normalized with zero shear rate viscosity).}
		\label{table1}
	\end{center}
\end{table}

\normalsize

All the aforementioned studies assume the surrounding medium to be Newtonian. Since a majority of potential applications of the synthesized active particles lie in drug-delivery and other areas of biological research \citep{patteson2016active,su2019janus}, understanding the influence of complex rheology on self-propulsion is essential. 
There has been limited progress towards this, both experimentally and theoretically.
Table \ref{table1} summarizes the theoretical studies which analyzed self-propulsion in various regimes.
The experimental and theoretical analysis of \cite{gomez2016dynamics, aragones2018diffusion} demonstrated that the translational swimming of a Janus particle is coupled to the rotational motion in a viscoelastic medium. 
Recently, \cite{saad2019diffusiophoresis} studied the effect of polymer entanglements on active motion and showed that the particle can escape the entangled confinements at time-scales significantly shorter than the polymer relaxation time.

There have been a few theoretical studies in the past which have used the continuum-level framework and studied the effect of bulk non-Newtonian stresses on active propulsion.
\cite{zhu2012self} employed the Giesekus model to study squirming in a viscoelastic fluid, considering the first two modes of swimming. 
They reported that swimming speed (and the swimming power required) is always lower than that in Newtonian fluids, with a minimum at a moderate value of Weissenberg number. Using a second-order fluid model, \cite{corato2015locomotion} also studied locomotion of a two-mode squirmers. They reported that `pullers' are slowed, `pushers' are hastened and `neutral' squirmers are unaffected in viscoelastic fluids.
Later \cite{datt2015squirmingST} showed that it is essential to consider higher modes while modeling squirmers in a non-Newtonian medium as the constitutive equation is non-linear. Using a Carreau-Yasuda model, they demonstrated that the swimming may be faster or slower in shear-thinning fluids (compared to Newtonian fluids) depending upon the rate of actuation i.e. strength of the swimming modes. 
Their subsequent study \citep{datt2017activeComplex} extended the analysis of \cite{golestanian2007designing} to understand the self-propulsion of Janus sphere in complex media.
Through an approach based on reciprocal theorem \citep{elfring2016effect}, they showed the effect of bulk non-Newtonian stresses on the swimming velocity, assuming a prescribed slip velocity. 
They found: (i) swimming is always slower in weakly shear-thinning fluids; (ii) for weakly viscoelastic fluid, a Janus sphere swims faster if surface coverage of activity is more than half ($ > \pi/2 $) and vice-versa.
However, they have acknowledged that the assumption of Newtonian slip in a non-Newtonian medium may not hold true. A change in slip velocity may significantly alter self-propulsion and therefore entails further investigation.
A recent study by \cite{SOF_natale2017} extended the work of \cite{datt2017activeComplex} to investigate the effects of advection and reaction (i.e. finite Péclet and Damköhler numbers) on self-diffusiophoresis through complex media. 
Their FEM simulations demonstrated the presence of an extensional flow across the point of discontinuity in the surface activity. This behavior was triggered due to large gradients in viscoelastic stresses.
The origin of these viscoelasticity-triggered extensional flows is currently unknown. Given the recent interest in self-propulsion through complex medium with several open questions in the literature \citep{datt2017activeComplex,SOF_natale2017,pietrzyk2019flow}, it is natural to ask: how does the slip change in complex fluids and what is its relationship to the concentration field?


The importance of considering non-Newtonian effects in the thin interaction region around a diffusio-phoretic particle can be realized by considering the progress in the field of electrophoresis.
\cite{khair2012coupling} demonstrated that the modification in slip due to shear-thinning can alter the mobility of an electrophoretic particle and the flow field around it. 
Very recently, through the use of various continuum-level rheological models (Oldroyd-B, Giesekus, FENE-P, and FENE-CR), \cite{li2020electrophoresis} showed that the electrophoretic particle (similar to a neutral squirmer in Newtonian fluid) behaves like a puller-type squirmer at low Weissenberg numbers. Such behavior arises primarily due to the elastic effects in the thin electrical double layer. 
These studies show that the high shear rate inside the thin layer generates significant polymer extension and deformation which needs to be accounted in the prediction of phoretic motion.
Motivated by these recent developments in the context of electrophoresis \citep{khair2012coupling,zhao2013electrokinetics,li2020electrophoresis}, we aim to investigate the modification due to complex rheology in both slip and mobility of a self-propelling active particle.




In this work, we study the influence of viscoelasticity (second-order-fluid model) and shear-thinning (Carreau-fluid model) on the slip and mobility of an axisymmetric Janus particle. Using matched asymptotic expansions, we provide an analytical expression for the modification of the diffusio-osmotic slip due to complex rheology.
Employing the reciprocal theorem, we evaluate the modification in the swimming velocity. 
Our results are applicable to a general diffusio-osmotic flow of complex fluid and hence can be used to model the transport of biological fluids through narrow channels and confinements, where applying pressure drop is undesirable \citep{Pump_Geo_michelin2015SoftMatter,Pump_Geo_lisicki2016phoretic,Pump_Patch_michelin2019nature}.

\section{Active particle in a second-order-fluid}

\begin{figure}
	\centering
	{{\includegraphics[width=0.85\linewidth]{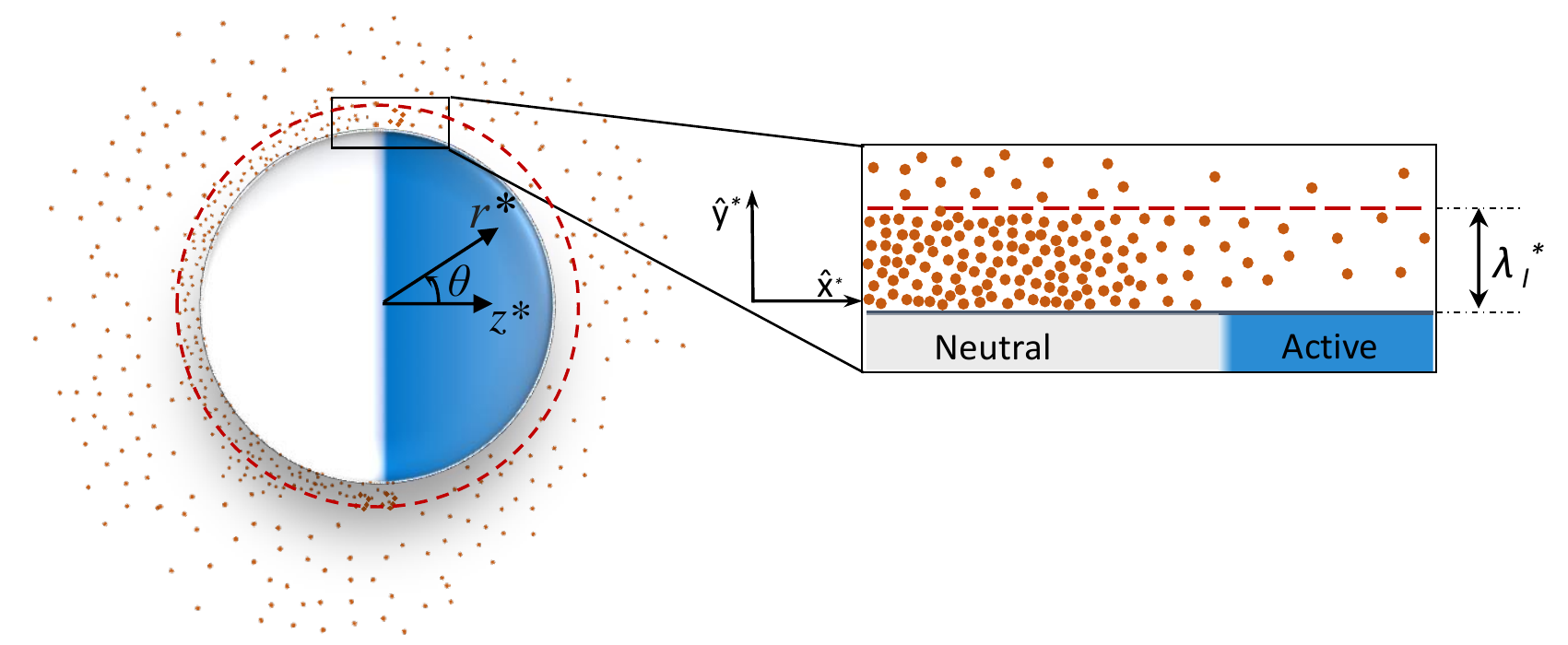} }}
	\caption{Schematic of an axisymmetric Janus particle suspended in a non-Newtonian medium with uniform external solute concentration. Zoomed-in view shows the solute in the thin interaction layer above a surface with varying activity. $ \lambda_{I}^{*} $ is the interaction layer thickness.}
	\label{fig:schematic}
\end{figure}

We consider an active particle surrounded by a non-Newtonian fluid medium consisting of a solute at uniform concentration $ C_{\infty}^{*} $. The solute, treated as a continuum, interacts with the active particle of radius $ a^{*} $. The short-range interaction potential is governed by
\begin{equation}
\mathcal{P}^{*}(r,\theta)=k_{B}^{*} T^{*} \psi(r,\theta) ,
\end{equation}
which acts over the length scale corresponding to the interaction layer thickness ($ \lambda_{I}^{*}) $, where $ \lambda_{I}^{*} \ll a^{*} $. 
Here $ {*} $ denotes the dimensional variables.
	$ \psi $ is interaction potential energy between the solute molecules and particle surface, scaled with the thermal energy ($ k_{B}^{*} T^{*} $).
This interaction generates a pressure field in the thin interaction layer, which decays to zero far away from the surface.
The particle surface is partially active; a fixed-flux adsorption ($ \mathcal{A}^{*} $) varies as a step function in the tangential direction.
Since the interaction layer ($ \lambda_{I}^{*} $) is very thin compared to the particle radius ($ a^{*} $), we model this as a the diffusio-osmotic flow over a flat surface (see fig.\ref{fig:schematic}).
This approximation neglects the curvature effects and introduces an error of $ \mathcal{O}(\lambda_{I}^{*}/ a^{*}) $ and has been employed earlier in studies involving particle electrophoresis \citep{anderson1989,o1983solution}.

We assume the rheology inside and outside the interaction layer to follow the second-order fluid (SOF) model \citep{bird1987dynamics}. This assumption is valid for slow and weakly non-Newtonian flows, i.e. polymers with low molecular weight ($ \sim 10^{4} $) and radius of gyration smaller than the interaction layer thickness which is typically $ 1-10 $ nm \citep{sharifi2013}. 
The qualitative insights obtained through such continuum models can also be extended to the cases of larger polymers as their segments would experience strong shear, which can result in modification of the dynamics in the interaction layer \citep{li2020electrophoresis}.
We also assume the solute molecules, inside and outside the interaction layer, to follow Fickian diffusion with constant diffusivity.
It has been reported that the presence of polymers significantly affects the diffusive mass transport in stagnant polymeric medium \citep{maldonado2016breakdown,makuch2020diffusion}. To our knowledge, there has been no study conducted on the mass transport in sheared complex flows.
As a first step, we focus on gaining qualitative insights and assume that the solute molecules and polymers are in the dilute concentration regime and the associated effects, such as volume exclusion and variable diffusivity, do not affect the system at leading order.

Following \cite{anderson1982motion,michelin2014phoretic}, we represent the system using the following non-dimensional equations
\begin{subequations}\label{NDmom}
	\begin{gather}
	\bnabla \bcdot \IB{u}=0, \quad -\bnabla p + \nabla^{2}\IB{u}+De \, \bnabla \bcdot  \mathsfbi{S}  -\frac{1}{\epsilon^{2}}  (c+C_{\infty})\bnabla \psi=\IB{0}, \label{NDmom:NS}\\
	Pe(\IB{u}\bcdot \bnabla c)=\bnabla \bcdot [ \bnabla c + \left( c+C_{\infty}\right) \bnabla \psi  ]  . \label{NDmom:conc}
	\end{gather}
\end{subequations}
The characteristic scales are
\begin{equation}
C_{ch}=\frac{|\mathcal{A^{*}}|a^{*}}{D^{*}}, \; U_{ch}=\frac{k^{*}_{B}T^{*}\lambda_{I}^{* \, 2} C_{ch}}{\mu^{*} a^{*}},\; p_{ch}=\frac{\mu^{*} U_{ch}}{a^{*}},\; y_{ch}=x_{ch}=a^{*}.
\label{char_scales}
\end{equation}
Here, $ D^{*} $ is the solute diffusion coefficient, $ |\mathcal{A}^{*}| $ is the maximum magnitude of activity, $ c $ denotes the disturbance to the uniform concentration $ C_{\infty} $ (defined as: $ c=C-C_{\infty} $). 
In eq. (\ref{NDmom:NS}), the polymeric stress $ \mathsfbi{S}=\mathsfbi{A} \bcdot \mathsfbi{A} + \delta \mathsfbi{B} $, where $ \mathsfbi{A} $ is the rate of strain tensor ($ \bnabla \IB{u} + (\bnabla \IB{u})^{T} $) and $ \mathsfbi{B} $ is the steady-state Rivlin-Ericksen tensor (covariant derivative of rate of strain tensor $ \bten{A} $) \citep{bird1987dynamics}.
\begin{equation}
 \bten{B} = \IB{U} \bcdot \bnabla \bten{A} + \bten{A}\bcdot \bnabla \IB{U}^{T} + \bnabla \IB{U} \bcdot \bten{A}.
\label{RE_tensor}
\end{equation}
Here $ T $ denotes the transpose.
The dimensionless quantities are defined as
\begin{equation}\label{definitions}
Pe=\frac{U_{ch}a^{*}}{D^{*}}, \;
De=\frac{\Psi_{1}^{*}+\Psi_{2}^{*}}{\mu^{*}}\left( \frac{U_{ch}}{a^{*}}\right), \; \delta=\frac{-\Psi_{1}^{*}}{2\left(\Psi_{1}^{*}+\Psi_{2}^{*} \right) } \mbox{ and\ } \epsilon=\frac{\lambda_{I}^{*}}{a^{*}},
\end{equation}
where $ Pe $ (Peclet number) is the ratio of diffusive to advective time scales;
$ De $ (Deborah number) is defined as the ratio of viscoelastic time scale to that based on shear arsing due to macro-scale movement of the Janus sphere; $ \delta $ is a viscometric parameter which compares first and second normal stress coefficients ($ \Psi_{1}^{*} $ and $ \Psi_{2}^{*} $, respectively); $ \epsilon $ is the dimensionless thickness of the interaction layer.

The boundary conditions are
\begin{subequations}\label{NDBC}
	\begin{gather}
	\left( \left.\frac{\partial c}{\partial y}\right\vert_{y=0} +\left.(c+C_{\infty}) \frac{\partial \psi}{\partial y}\right\vert_{y=0} \right) = \frac{\mathcal{A^{*}}(x)}{|\mathcal{A^{*}}|}=\mathcal{K}(x)\;
	\mbox{ and\ } \! \; \IB{u}= \IB{0} \quad \mbox{ at\ } y=0 ; \label{NDBC:surface}\\
	p \rightarrow 0 ,\; \psi \rightarrow 0  \mbox{ and\ } \! \; c \rightarrow 0 \qquad \mbox{ as\ } y\rightarrow \infty.  \label{NDBC:inf}
	\end{gather}
\end{subequations}
In the next sections, we investigate the influence of non-Newtonian effects on the diffusio-osmotic slip on a partially active surface using matched asymptotic expansions (MAE).

\subsection{Evaluation of the diffusio-osmotic slip}

For an asymptotically thin interaction region ($ \epsilon \ll 1 $), we expand the field variables ($ \IB{u},c, p, \psi $) as
\begin{equation}
f(x,y)=f^{(0)}(x,y)+\epsilon f^{(1)}(x,y)+\cdots.
\label{epsExpand}
\end{equation}
Following \cite{michelin2014phoretic}, we divide the domain into an `inner' ($ 0 \ll y \ll 1 $) and `outer' region ($  y \gg 1 $). 
We use MAE to replace the solution of the inner region with coarse-grained boundary conditions for the outer region.

\subsubsection{Outer region}
Neglecting the rapidly decaying interaction potential in the outer region i.e. $ \psi=0 $ \citep{sharifi2013,michelin2014phoretic}, the leading order equations read
\begin{subequations}\label{outerGE}
	\begin{gather}
	\bnabla  \bcdot \IB{u}^{(0)} = 0,  \quad	 -  \bnabla p^{(0)} + \nabla^{2} \IB{u}^{(0)} + De \, \bnabla \bcdot  \mathsfbi{S}^{(0)} = \IB{0}, \label{outerGE:flow}\\
	Pe(\IB{u}^{(0)}\bcdot \bnabla c^{(0)})= \nabla^{2}c^{(0)},  \label{outerGE:convdiff}
	\end{gather}
\end{subequations}
subject to the following boundary conditions
\begin{subequations}
	\begin{gather}
		\IB{u}^{(0)} \rightarrow \IB{0} , \; p^{(0)} \rightarrow 0 \; \mbox{ and\ } 
		c^{(0)} \rightarrow 0 \quad \mbox{ as\ } y \rightarrow \infty,\\
		\IB{n}\bcdot \bnabla c^{(0)} = \mathcal{K}(x) \quad \mbox{ at\ } y \rightarrow 0. \label{CONCouterBC}
	\end{gather}  
\label{outerBC}
\end{subequations}
The concentration boundary condition (\ref{CONCouterBC}) has been shown to be well-posed, provided $ \epsilon Pe\ll1 $ \cite[p. 580]{michelin2014phoretic}.
	To obtain the macro-scale boundary condition for velocity at the \textit{surface} (i.e. $ y \rightarrow 0 $), the outer solution must be matched with the inner solution.

\subsubsection{Inner region}
We first rescale the variables to derive the equations in the inner layer. The variables in the inner layer are defined as $ \hat{f} $.
In the thin interaction layer limit ($ \epsilon \ll 1 $), using boundary layer principles, we define the characteristic scales for $ y^{*}- $ direction, vertical velocity ($ v^{*} $) and pressure ($ p^{*} $) as $ \lambda_{I}^{*} $, $ \epsilon \, U_{ch} $ and $ p_{ch} \, \epsilon^{-2} $, respectively. 
Using (\ref{epsExpand}) and (\ref{NDmom}), we obtain the leading order equations in the inner region as
\begin{subequations}\label{innerGE}
	\begin{gather}
	\frac{\p \hat{u}^{(0)}}{\p \hat{x}} + \frac{\p \hat{v}^{(0)}}{\p \hat{y}} =0 , \label{innerGE:CE}\\
	- \frac{\p \hat{p}^{(0)}}{\p \hat{x}} +  \frac{\p^2 \hat{u}^{(0)}}{\p \hat{y}^2} +De \left( \frac{\p \hat{S}_{xx}^{(0)}}{\p \hat{x}}  +  \frac{\p \hat{S}_{yx}^{(0)}}{\p \hat{y}}   \right)  =  (\hat{c}^{(0)} + C_{\infty})\frac{\p \hat{\psi}^{(0)}}{\p \hat{x}}  ,  \label{innerGE:NSx}\\
	- \frac{\p \hat{p}^{(0)}}{\p \hat{y}} + De \left( \frac{\p \hat{S}_{yy}^{(0)}}{\p \hat{y}} \right) =  (\hat{c}^{(0)}+C_{\infty}) \frac{\p \hat{\psi}^{(0)}}{\p \hat{y}}    ,  \label{innerGE:NSy}\\
	\epsilon^{2}Pe  \left(  \hat{u}^{(0)} \frac{\p \hat{c}^{(0)}}{\p \hat{x}} + \hat{v}^{(0)} \frac{\p \hat{c}^{(0)}}{\p \hat{y}}  \right) =  \frac{\p}{\p \hat{y}} \left(  \frac{\p \hat{c}^{(0)}}{\p \hat{y}}  +  (\hat{c}^{(0)}+C_{\infty}) \frac{\p \hat{\psi}^{(0)}}{\p \hat{y}}  \right)    \label{innerGE:conc},
	\end{gather}
\end{subequations}
subject to the following surface boundary conditions at the leading order
\begin{equation}\label{innerBC}
\left.	\IB{u}^{(0)} \right\vert_{\hat{y}=0} = \IB{0} \quad  \mbox{and\ }
\left( \left.\frac{\partial \hat{c}^{(0)}}{\partial \hat{y}}\right\vert_{\hat{y}=0} +\left.(\hat{c}^{(0)}+C_{\infty}) \frac{\partial \psi}{\partial \hat{y}}\right\vert_{\hat{y}=0} \right) =0.
\end{equation}
For low to moderate advective effects (i.e. $ Pe \ll \epsilon^{-2} $), we can neglect the LHS of (\ref{innerGE:conc}). This decouples the solute concentration field from the hydrodynamics.
The pressure scaling in the inner region suggests a decay condition as $ \hat{y} \rightarrow \infty $ \citep[p.579]{michelin2014phoretic}.
The components of the polymeric stress tensor ($ \mathsfbi{S} $) are rescaled as: 
\begin{subequations}
	\begin{gather}\label{innerSscale}
	S_{xx} = \frac{\hat{S}_{xx}}{\epsilon^{2}} =\frac{1}{ \epsilon^{2}}\left( \frac{\partial \hat{u}}{\partial \hat{y}} \right)^{2},\quad  	S_{yy} =\frac{\hat{S}_{yy}}{\epsilon^{2}} =\frac{1}{\epsilon^{2}} \left( 1 +2 \delta \right)  \left( \frac{\partial \hat{u}}{\partial \hat{y}} \right)^{2}, \\
	S_{xy}=S_{yx} = \frac{\hat{S}_{xy}}{\epsilon} =\frac{\hat{S}_{yx}}{\epsilon} =\frac{2 \delta}{ \epsilon}\left( \frac{\partial \hat{u}}{\partial \hat{y}} \frac{\partial \hat{u}}{\partial \hat{x}} + \frac{\hat{v} \frac{\partial^2 \hat{u}}{\partial \hat{y}^2}  +  \hat{u} \frac{\partial^2 \hat{u}}{\partial \hat{y} \partial \hat{x}} }{2}  \right).
	\end{gather}
\end{subequations}

We now perform a perturbation expansion in Deborah number ($ De $) i.e. accounting for weakly non-linear viscoelastic effects such that $ \epsilon \ll De \ll 1 $. The field variables for each term in (\ref{epsExpand}) can be further expanded as
\begin{equation}
f^{(i)}(x,y)=f^{(i)}_{0}(x,y)+De \, f^{(i)}_{1}(x,y)+\cdots.
\label{DeExpand}
\end{equation}
Here $ f $ represents the velocity and pressure field. The concentration field is not expanded in $ De $ as it is decoupled from velocity field in the inner region. 

\vspace{2mm}
$\mathcal{O}$($ De^{0} $) \textit{solution}:
Since our objective is to obtain leading order change in the diffusio-osmotic slip, for convenience, we temporarily drop the superscript $ (0) $ from all the variables. 
The solution to leading order (i.e. $ O $($ De^{0} $)) governing equations is obtained as
\begin{subequations}\label{innersol}
	\begin{gather}
	\hat{c}(\hat{x},\hat{y})=\mathcal{I}(\hat{x}) e^{-\hat{\psi}(\hat{x},\hat{y})} - C_{\infty}  , \qquad
	\hat{p}_{0}(\hat{x},\hat{y})=\mathcal{I}(\hat{x}) \left(  e^{-\hat{\psi}(\hat{x},\hat{y})} -1  \right)   , \mbox{and\ }\label{innerConcSol}\\
	\hat{u}_{0}(\hat{x},\hat{y})=-\mathcal{I}'(\hat{x}) \int_{0}^{\hat{y}} \int_{t}^{\infty} \left(  e^{-\hat{\psi}(\hat{x},s)} -1  \right) {\rm{d}}s \, {\rm{d}}t + \mathcal{J}_{0}(\hat{x})\,\hat{y}. \label{innerVelSol}
	\end{gather}
\end{subequations}
Here, $ \mathcal{I}(\hat{x}) $ and $ \mathcal{J}_{0}(\hat{x}) $ are to be determined through matching, $ \mathcal{I}'(\hat{x}) $ represents $ {\rm{d}} \mathcal{I}/{\rm{d}} {\hat{x}} $.
The leading order velocity field ($ \hat{u}_{0} $) is represented in a form different from that reported by \cite{anderson1982motion,michelin2014phoretic}, as it helps in the evaluation of higher order velocity field. It can be seen in fig. \ref{fig:innerVel}(a) that this expression is equivalent to that provided in the literature.

\vspace{2mm}
$\mathcal{O}$($ De^{1} $) \textit{solution}:
For ease of calculation, we assume the interaction potential ($ \hat{\psi} $) to be independent of the tangential direction ($x$). 
The flow field at $ \mathcal{O}(De) $ is governed by
\begin{subequations}\label{innerDeGE}
	\begin{gather}
	\frac{\partial \hat{u}_{1}}{\partial \hat{x}} + \frac{\partial \hat{v}_{1}}{\partial \hat{y}} =0 , \label{innerDeGE:CE}\\
	- \frac{\partial \hat{p}_{1}}{\partial \hat{x}} +  \frac{\partial^2 \hat{u}_{1}}{\partial \hat{y}^2} + \frac{\partial \hat{S}_{xx \, 0}}{\partial \hat{x}}  +  \frac{\partial \hat{S}_{xy \, 0}}{\partial \hat{y}}   =  0  ,  \label{innerDeGE:NSx}\\
	- \frac{\partial \hat{p}_{1}}{\partial \hat{y}} +  \frac{\partial \hat{S}_{yy \, 0}}{\partial \hat{y}} =  0. \label{innerDeGE:NSy}
	\end{gather}
\end{subequations}
Using the pressure decay condition ($ \hat{p} \rightarrow 0  $ as $ \hat{y} \rightarrow \infty $), the solution to (\ref{innerDeGE:NSy}) yields $ \hat{p}_{1}=\hat{S}_{yy \, 0} - \mathcal{J}_{0}(\hat{x})^{2}(1+2\delta) $.
Simplifying (\ref{innerDeGE:NSx}), we obtain
\begin{equation}
\frac{\partial^{2} \hat{u}_{1}}{\partial \hat{y}^{2}} = - \delta  \left\lbrace   -\frac{\partial \hat{u}_{0}}{\partial \hat{y}} \frac{\partial^{2} \hat{u}_{0}}{\partial \hat{y} \partial \hat{x}} +  \frac{\partial \hat{u}_{0}}{\partial \hat{x}} \frac{\partial^{2} \hat{u}_{0}}{\partial \hat{y}^{2}}  +  \hat{v}_{0} \frac{\partial^{3} \hat{u}_{0}}{\partial \hat{y}^{3}}  +  \hat{u}_{0} \frac{\partial^{3} \hat{u}_{0}}{\partial \hat{x}  \partial \hat{y}^{2}}         \right\rbrace   +  \frac{\partial \hat{p}_{1}}{\partial \hat x}   .
\label{Deu1eq}
\end{equation}
Here, $ \hat{v}_{0} $ is found by substituting (\ref{innerVelSol}) in the continuity equation and integrating it over $ \hat{y} $ direction.
Substituting (\ref{innerVelSol}) and $ \hat{v}_{0} $ into (\ref{Deu1eq}), integrating twice over $ \hat{y} $ direction and using the no-slip condition, we obtain 
	\footnote{Details of the derivation can be found in the supplementary material.}
{\small{
\begin{align} \label{innerDeSol}
\hat{u}_{1}= \mathcal{J}_{1}(\hat{x}) \hat{y}
& -  \delta \mathcal{I}'(\hat{x})\mathcal{I}''(\hat{x}) \int_{0}^{\hat{y}}  dp \int_{p}^{\infty}  \left\lbrace   
\left(  \int_{r}^{\infty} \mathcal{F}(s) \, {\rm{d}}s  \right)^{2} 
+  2 \mathcal{F}(r) \left(  \int_{0}^{r}s\mathcal{F}(s) {\rm{d}}s   + r  \int_{r}^{\infty} \mathcal{F}(s) {\rm{d}}s  \right) {\rm{d}}r  
\right.  \nonumber \\
&  \qquad \qquad \qquad \qquad \qquad  \qquad \left.  + \, \hat{\psi}'(r) e^{-\hat{\psi}(r)} \left( \int_{0}^{r} \left( r - \frac{s}{2} \right)s\mathcal{F}(s) {\rm{d}}s \,
+ \, \frac{r^{2}}{2} \int_{r}^{\infty} \mathcal{F}(s) {\rm{d}}s  \right)	
\right. \nonumber \\
&  \qquad \qquad \qquad \qquad \qquad \qquad \left.  - \left( \frac{\mathcal{J}_{0}(\hat{x})}{\mathcal{I}'(\hat{x})} +  \frac{\mathcal{J}_{0}'(\hat{x})}{\mathcal{I}''(\hat{x})} \right) \left( r \mathcal{F}(r) + \int_{r}^{\infty} \mathcal{F}(s) ds  \right)
\right.  \nonumber \\
&\qquad \qquad \qquad \qquad \qquad \qquad \left.  +   \frac{\mathcal{J}_{0}(\hat{x})}{\mathcal{I}''(\hat{x})} \mathcal{F}'(r) \frac{r^{2}}{2}     +  \frac{\mathcal{J}_{0}(\hat{x}) \mathcal{J}_{(0)}'(\hat{x})}{\mathcal{I}'(\hat{x}) \mathcal{I}''(\hat{x})} 
- \frac{2\mathcal{J}_{0}(\hat{x}) \mathcal{J}_{0}'(\hat{x})(1+2\delta) }{\delta \mathcal{I}'(\hat{x}) \mathcal{I}''(\hat{x})} 
\right\rbrace dr . 		
\end{align}
}}
Here, $ \mathcal{F}(\xi)=-1+e^{-\hat{\psi}(\xi)} $, $ \mathcal{J}_{0},\, \mathcal{J}_{1} $, and $ \mathcal{I} $ are to be determined through matching.

%

\subsubsection{Matching}
For any field variable $ f $, the matching condition at $\mathcal{O}$($ \epsilon^{0} $) is
\begin{equation}
\lim\limits_{y \rightarrow 0} (f^{(0)}_{0}+De f^{(0)}_{1}+\cdots)=\lim\limits_{\hat{y} \rightarrow \infty}(\hat{f}^{(0)}_{0}+De \hat{f}^{(0)}_{1}+\cdots).
\label{Matching}
\end{equation}
The matching condition for the concentration field ($ \hat{c} $) yields	$\mathcal{I}(\hat{x}) = \lim\limits_{y \rightarrow 0} c^{(0)}(x,y) + C_{\infty}$.
We substitute (\ref{innerVelSol}) and (\ref{innerDeSol}) in the above matching condition and obtain: $	\mathcal{J}_{0} = \mathcal{J}_{1}=0$ at the leading order.
At the present order of approximation, the total slip velocity at the outer edge of the interaction layer is
\begin{equation}
\left. u^{(0)} \right\vert_{y=0} = \left. c^{(0)}_{x}\right\vert_{y=0} \left(  M_{0} + De \, \delta \, M_{1} \left.c^{(0)}_{xx}\right\vert_{y=0}  \right),
\label{FINALDeSol_alt}
\end{equation}
where $ M_{0} $ and $ M_{1} $ are the mobility coefficients representing the Newtonian and non-Newtonian contribution, respectively:
\begin{align}\label{FINALDeSol}
& M_{0}
 = - \int_{0}^{\infty} \int_{t}^{\infty} \mathcal{F}(s) \rm{d}s \, \rm{d}t 
 \; \;
\mbox{ and\ }  \; \;
M_{1}
= -  {\displaystyle\int_{0}^{\infty} }
\int_{p}^{\infty} \mathcal{G}(r) {\rm{d}}r \, {\rm{d}}p  ,\\
&\mbox{ where\ } \mathcal{G}(r)= 
\left\lbrace   
\left(  \int_{r}^{\infty} \mathcal{F}(s) \, {\rm{d}}s  \right)^{2} 
+  2\,\mathcal{F}(r) \left( 
\int_{0}^{r}  s \mathcal{F}(s) {\rm{d}} s   +   r  \int_{r}^{\infty}  \mathcal{F}(s) {\rm{d}} s
\right) \right.\nonumber \\
& \qquad \qquad \qquad \quad  \; \left.  	+  \hat{\psi}'(r) \, e^{-\hat{\psi}(r)}  \left(
\int_{0}^{r} \left( r - \frac{s}{2}\right) s\, \mathcal{F}(s) {\rm{d}}s + \frac{r^{2}}{2} \int_{r}^{\infty} \mathcal{F}(s) {\rm{d}}s
\right) 
\right\rbrace.
\end{align}
Modification to the diffusio-osmotic slip velocity (arising from the non-Newtonian effects) is found to be proportional to the first and second tangential derivative  of the bulk-scale concentration at the particle \textit{surface}. 
The dependency of viscoelastic effects on the second derivative can be intuitively understood by realizing that the polymers stretch only when there exists a spatial variation (or gradient) in the flow. In the current physics, as the flow itself is generated by the chemical gradient, an appearance of second derivative at the leading order is intuitive.
The proportionality to $ De \, \delta $  and the definitions described in (\ref{definitions}) suggests that the effect of viscoelasticity is solely due to first normal stress difference. 
Furthermore, the dimensional form of the slip velocity reveals that the first order slip has a dependency on the characteristic length scale ($ a^{*} $), whereas the Newtonian slip is independent of it.


\subsubsection{Velocity profile in the inner region}

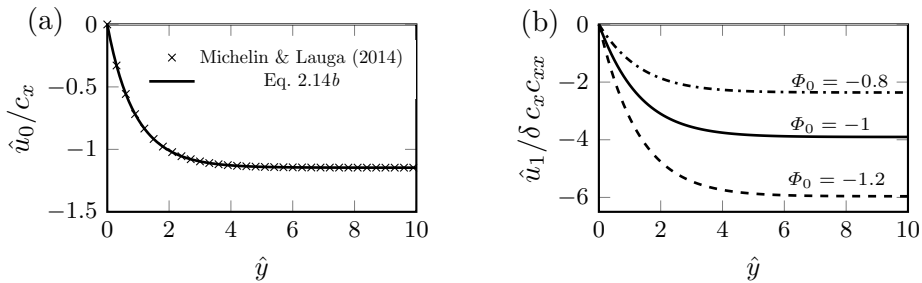
\begin{figure}
	\centering
	\begin{tikzpicture}

	\begin{axis}[clip=false,
	height=.30\textwidth,width=.42\textwidth, extra y ticks= 0, xshift=0cm,yshift=0cm,	ylabel shift= -3 pt,
	extra y tick labels = ,
	extra y tick style  = { grid = major }, 
	xlabel=$ \hat{y} $, ylabel= \large $ {\hat{u}_{0}}/{c_{x}} $ \normalsize,%
	, xmin=0, xmax=10, ymin=-1.5, ymax=0 , thick
	,	legend style={draw=none,at={(1,0.60)},anchor=south east},	legend style={nodes={scale=0.7, transform shape}}
	]
	\addplot[only marks,thin,mark =x, black] table[x=x, y=y] {Newt_Michelin.txt};
	\addplot[line width=1pt, black] table[x=x, y=y] {Newt_Akash.txt};
	\legend{Michelin \& Lauga (2014),Eq. \ref{innerVelSol}}
	\node[scale=1.1] at (axis cs: -2, 0.02) {(a)}; 
	\end{axis}

	\begin{axis}[clip=false,
	height=.30\textwidth,width=.42\textwidth, extra y ticks= 0,	xshift=6.5cm,ylabel shift= -3 pt,
	extra y tick labels = ,
	extra y tick style  = { grid = major }, 
	draw=none,
	xlabel=$ \hat{y} $, ylabel= \large $ {\hat{u}_{1}}/{\delta \,c_{x} c_{xx}} $ \normalsize,%
	, xmin=0, xmax=10, ymin=-6.5, ymax=0 , thick
	]
	\addplot[line width=1pt, black] table[x=x, y=y1] {Visco.txt};
	\node at (78,450) {{\scriptsize $ \Phi_{0}=-0.8 $}};
	\addplot[line width=1pt, black,dashed] table[x=x, y=y2] {Visco.txt};
	\node at (75,300) {{\scriptsize $ \Phi_{0}=-1 $}};
	\addplot[line width=1pt, black, dash dot] table[x=x, y=y3] {Visco.txt};
	\node at (77,110) {{\scriptsize $ \Phi_{0}=-1.2 $}};
	\node[scale=1.1] at (axis cs: -1.9, 0.02) {(b)}; 
	\end{axis}


	\end{tikzpicture}
	\caption{(a) Velocity field of Newtonian fluid inside the interaction layer (for $ \Phi_{0}=-1 $). (b) Variation of the modification to the velocity field. The results are for an exponentially decaying solute-surface interaction ($ \hat{\psi}(\xi)=\Phi_{0} e^{-\xi} $), where a negative $ \Phi_{0} $ represents an attractive interaction. Here $ c_{x} $ and $ c_{xx} $ are the concentration gradients at the outer edge of the inner region i.e. $ {  \left.c_{x}\right\vert_{y=0} } $ and $ {  \left.c_{xx}\right\vert_{y=0} } $.}
	\label{fig:innerVel}
\end{figure}

Fig. \ref{fig:innerVel}(a) shows the Newtonian component of the velocity field (\ref{innersol}b) inside the thin interaction layer for an exponentially decaying solute-surface attraction \citep{anderson1982motion,sharifi2013}. 
The velocity monotonically grows away from the surface and approaches an asymptotic value, which upon multiplication with the tangential concentration gradient provides the Newtonian slip velocity.
The viscoelastic component of the velocity field ($ \hat{u}^{(0)}_{1} $), shown in Fig. \ref{fig:innerVel}(b), also grows monotonically away from the surface and attains an asymptotic value.
The magnitude of this asymptotic value increases with the magnitude of interaction ($ \Phi_{0} $).

\begin{figure}
	\centering
	\begin{tikzpicture}
	
	\begin{axis}[clip=false,
	height=.30\textwidth,width=.40\textwidth, extra y ticks= 0, xshift=-6.8cm,yshift=-4cm,	ylabel shift= -3 pt,
	extra y tick labels = ,
	extra y tick style  = { grid = major }, 
	xlabel=$ \hat{y} $, ylabel= ,%
	, xmin=0, xmax=10, ymin= -2.8, ymax= +1.2 , thick
	,	legend style={draw=none,at={(1,0.25)},anchor=south east},	legend style={nodes={scale=0.9, transform shape}}
	]
	\addplot[line width=0.9pt, black] table[x=x, y=newt] {Repulsive_mobility.txt};
	\addplot[line width=1pt, black, dashed] table[x=x, y=SOF] {Repulsive_mobility.txt};
	\legend{$ {\hat{u}_{0}}/{c_{x}} $, $ {\hat{u}_{1}}/{\delta \,c_{x} c_{xx}} $}
	\node[scale=1.1] at (axis cs: -2, 1) {(a)}; 
	\end{axis}
	\begin{axis}[clip=false,
	height=.30\textwidth,width=.40\textwidth, extra y ticks= 0, xshift=0cm,yshift=-4cm,	ylabel shift= -3 pt,
	extra y tick labels = ,
	extra y tick style  = { grid = major }, 
	xlabel=$ \hat{y} $, ylabel =  ,%
	, xmin=0, xmax=50, ymin=-0.7, ymax=0 , thick
	,	legend style={draw=none,at={(1,0.35)},anchor=south east},	legend style={nodes={scale=0.9, transform shape}}
	]
	\addplot[line width=0.9pt, black] table[x=x, y=newt] {vdW_mobility2.txt};
	\addplot[line width=1pt, black,dashed] table[x=x, y=SOF] {vdW_mobility2.txt};
	\legend{$ {\hat{u}_{0}}/{c_{x}} $, $ {\hat{u}_{1}}/{\delta \,c_{x} c_{xx}} $}
	\node[scale=1.1] at (axis cs: -10, 0) {(b)}; 
	\end{axis}
	
	\end{tikzpicture}
	\caption{ Velocity field inside the interaction layer for: (a) repulsive exponential interaction ($ \Phi_{0}=+1 $); (b) attractive long-range van der Waals interaction ($ \Phi_{0}=-1 $). Here $ c_{x} $ and $ c_{xx} $ are the concentration gradients at the outer edge of the inner region i.e. $ {  \left.c_{x}\right\vert_{y=0} } $ and $ {  \left.c_{xx}\right\vert_{y=0} } $.}
	\label{fig:innerVel2}
\end{figure}
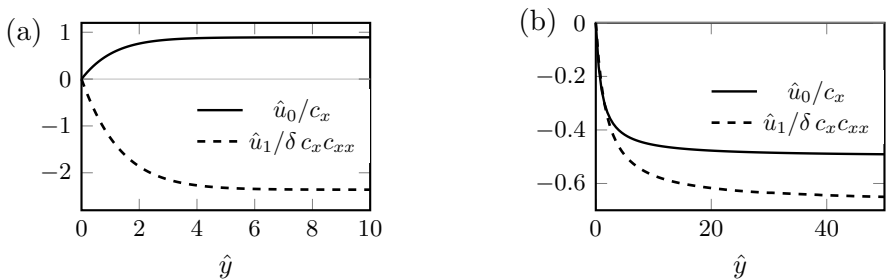

The interactions between solute molecules and the particle surface can be either attractive or repulsive; if the solute is more attracted to the surface than the solvent, the interaction coefficient ($ \Phi_{0} $) is negative and vice-versa. 
To show that the current analysis is valid for other forms of potential, we evaluate the velocity field inside the interaction layer for (i) an exponentially repulsive and (ii) a long-range van der Waals type interaction (arising primarily from the dipolar forces $ \sim 1/r^{6} $). 
(i) For repulsive interactions, the thin `layer'  is characterized by an absence of solute concentration and the direction of motion (and slip) is opposite to that observed for an attractive interaction \citep{michelin2014phoretic}. 
Fig \ref{fig:innerVel2} (a) shows that while the Newtonian slip is reversed (c.f. fig.\ref{fig:innerVel}a), the contribution from viscoelasticity is in the same direction as the case of attraction potential (c.f. fig.\ref{fig:innerVel}b). 
(ii) To incorporate long-range attractive van der Waals interactions, we follow \citet[p.~112]{anderson1982motion} and assume the dipolar interactions between the solute molecules to be pairwise additive. The potential is defined as
\begin{equation}\label{vdW}
	\hat{\psi}(\xi)=\Phi_{0} \left(- \frac{1}{(1+\xi)^{9}} + \frac{1}{(1+\xi)^{3}} \right).
\end{equation}
The velocity fields (Newtonian and viscoelastic) remain qualitatively similar to that obtained from the attractive exponential interactions. As (\ref{vdW}) decays slower than the exponential interaction, the asymptotic value (i.e. slip velocity) is obtained at larger distances from the surface.

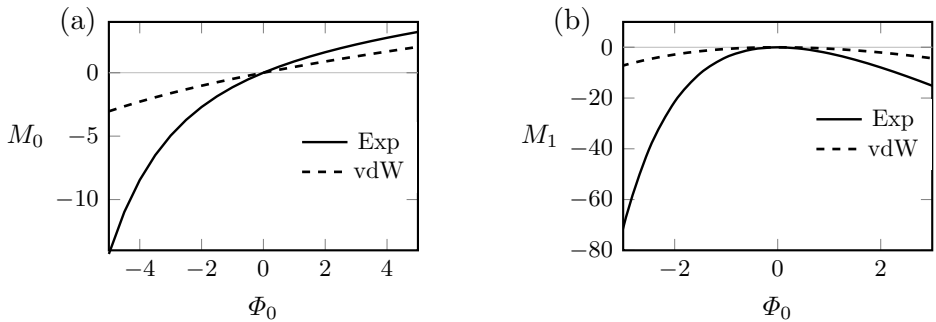
\begin{figure}
	\centering
	\begin{tikzpicture}
	
	\begin{axis}[clip=false,
	height=.34\textwidth,width=.42\textwidth, extra y ticks= 0, xshift=-6.8cm,yshift=-4cm,	ylabel shift= -3 pt,
		ylabel style={rotate=-90},
	extra y tick labels = ,
	extra y tick style  = { grid = major }, 
	xlabel=$ \Phi_{0} $, ylabel= $ M_{0} $,%
	, xmin=-5, xmax=5, ymin= -14, ymax= 4 , thick
	,	legend style={draw=none,at={(1,0.25)},anchor=south east},	legend style={nodes={scale=0.9, transform shape}}
	]
	\addplot[line width=0.9pt, black] table[x=x, y=M0exp] {VE_Mob.txt};
	\addplot[line width=1pt, black, dashed] table[x=x, y=M0vdw] {VE_Mob.txt};
	\legend{ Exp ,  vdW }
	\node[scale=1.1] at (axis cs: -6, 4) {(a)}; 
	\node[scale=1.1] at (axis cs: 10, 4) {(b)}; 
	\end{axis}
	\begin{axis}[
	height=.34\textwidth,width=.42\textwidth, extra y ticks= 0, xshift=0cm,yshift=-4cm,	ylabel shift= -3 pt,
		ylabel style={rotate=-90},
	extra y tick labels = ,
	extra y tick style  = { grid = major }, 
	xlabel=$ \Phi_{0} $, ylabel = $ M_{1} $ ,%
	, xmin=-3, xmax=3, ymin=-80, ymax=10 , thick
	,	legend style={draw=none,at={(1,0.35)},anchor=south east},	legend style={nodes={scale=0.9, transform shape}}
	]
	\addplot[line width=0.9pt, black,smooth] table[x=x, y=M1exp] {VE_Mob.txt};
	\addplot[line width=1pt, black,dashed,smooth] table[x=x, y=M1vdw] {VE_Mob.txt};
	\legend{ Exp ,  vdW }
	\node[scale=1.1] at (axis cs: -10, 0) {(b)}; 
	\end{axis}
	
	\end{tikzpicture}
	\caption{ Variation of mobility coefficient (a) $ M_{0} $ and (b) $ M_{1} $ with respect to $ \Phi_{0} $, for exponential and van der Waals interactions.}
	\label{fig:Mobility}
\end{figure}

The above results demonstrate that the velocity fields inside the inner region approach an asymptotic value, the slip velocity. The Newtonian component of this slip is equal to the product of concentration gradient ($ \left.c_{x}\right\vert_{y=0} $) and mobility coefficient $ M_{0} $. Similarly, the viscoelastic component of velocity field involves $ M_{1} $ (see eq.\ref{FINALDeSol_alt}). 
The magnitude of mobility coefficients depends on the nature of interaction (exponential or van der Waals) and magnitude of attraction or repulsion ($ \Phi_{0} $). Fig. \ref{fig:Mobility} quantifies the effect of the type and magnitude of interaction on $ M_{0} $ and $ M_{1} $.
Fig. \ref{fig:Mobility}(a) shows that the sign of Newtonian mobility coefficient is opposite for repulsive interactions ($ \Phi_{0} > 0 $) as compared to attractive ones ($ \Phi_{0} < 0 $), for both the type of interactions i.e. short-range exponential or long-range van der Waals solute-surface interaction.
Fig. \ref{fig:Mobility}(b) shows that the sign of $ M_{1} $ is always negative and does not depend on the nature of interaction.
Furthermore, for repulsive interactions, the magnitude of mobility coefficients ($ M_{0} $ and $ M_{1} $) is reduced because the adsorption coefficient in such cases is generally lower than that of attractive interactions \citep[p.69]{anderson1989}.

The above results, valid for moderate advective effects ($  \epsilon Pe \ll 1 $), are also applicable to diffusio-osmotic flows of complex fluids in micro-channels, arising from externally imposed concentration gradients or generated due to active `patches' \citep{Pump_Geo_michelin2015SoftMatter,Pump_Patch_michelin2019nature}.

Several experimental studies \citep{ebbens2011direct,baraban2012transport,ke2010motion} have shown that the particles are propelled with catalytic surface oriented at the rear-end, suggesting an attractive interaction between the solute molecules and active particle. 
Thus, we focus primarily on the exponentially attractive interaction because of its simplicity and ease of computation. We follow \cite{sabass2012dynamics,sharifi2013} and assume $ | \Phi_{0} |=1 $ in the results which follow.

\subsection{Diffusio-osmotic slip on an active particle}
We now extend the results of previous section (eq.\ref{FINALDeSol_alt}) to an axisymmetric Janus particle. Here, direction normal to the surface is $ r $; the tangential direction is polar angle $ \theta $, varying from $ 0 $ to $ \pi $. 
The surface activity $\mathcal{K}$ follows a step function \citep{golestanian2007designing,michelin2014phoretic,SOF_natale2017}
\begin{equation}
\mathcal{K} (\theta)=\left\{
\begin{array}{ll}
1 \qquad \theta < \theta_{c}\\
0 \qquad \theta > \theta_{c},
\end{array}
\right.
\label{stepK}
\end{equation}
where $ \theta_{c} $ is the angle at which the activity undergoes a step change; it represents the surface coverage of activity. 
To evaluate the slip velocity, we require solution to the concentration field ($ c $) in the outer region. In the absence of advection ($ Pe\rightarrow 0 $), the solution to (\ref{outerGE:convdiff}) is sought in terms of an expansion in Legendre polynomials. Following \cite{golestanian2007designing}, we obtain the solution to bulk-scale concentration field as
\begin{equation}
c(r,\theta)=\displaystyle\sum_{n=0}^{\infty} {\frac{-\mathcal{K}_{n}}{(n+1)}}\, \frac{P_{n}(\cos \theta)}{r^{n+1}},
\label{concGold}
\end{equation}
where $ P_{n} $ is the n\textsuperscript{th} order Legendre polynomial. Here $ \mathcal{K}_{n} $ are the spectral coefficients of the activity distribution:
\begin{equation}\label{spectralK}
	 \mathcal{K}(\theta)=\sum_{n=0}^{\infty} \mathcal{K}_{n} P_{n}(cos\theta) . 
\end{equation}
These coefficients are found by taking an inner product of (\ref{spectralK}) with the Legendre polynomials \citep{michelin2014phoretic}, and are obtained as
\begin{equation}
\mathcal{K}_{0}=\frac{(1-\cos \theta_{c})}{2} \; \mbox{ and\ } \;  \mathcal{K}_{n}=\frac{-1}{2} (P_{n+1}(\cos \theta_{c}) - P_{n-1}(\cos \theta_{c})) \; \mbox{for\ } n \geq 1.
\label{modesCoeff}
\end{equation}
Using the expression (\ref{concGold}) for disturbance concentration, we obtain the total tangential slip velocity ($ M_{0} c_{\theta}(1,\theta) + De \delta M_{1} c_{\theta}(1,\theta) c_{\theta \theta}(1,\theta) $)
\begin{equation}\label{uslip}
	\left.\IB{u}\right\vert_{r=1} = M_{0} \sum_{n=1}^{\infty} \frac{-\mathcal{K}_{n}}{n+1} \frac{\partial P_{n}}{\partial \theta} \IB{e}_{\theta}
	+ De \, \delta M_{1} \left(\sum_{n=1}^{\infty} \frac{-\mathcal{K}_{n}}{(n+1)} \frac{\partial P_{n}}{\partial \theta} \right) 
	 \left(\sum_{n=1}^{\infty} \frac{-\mathcal{K}_{n}}{(n+1)} \frac{\partial^{2} P_{n}}{\partial \theta^{2}}\right)
	 \IB{e}_{\theta}.
\end{equation}

\begin{figure}
	\centering
	
	{{\includegraphics[scale=0.36]{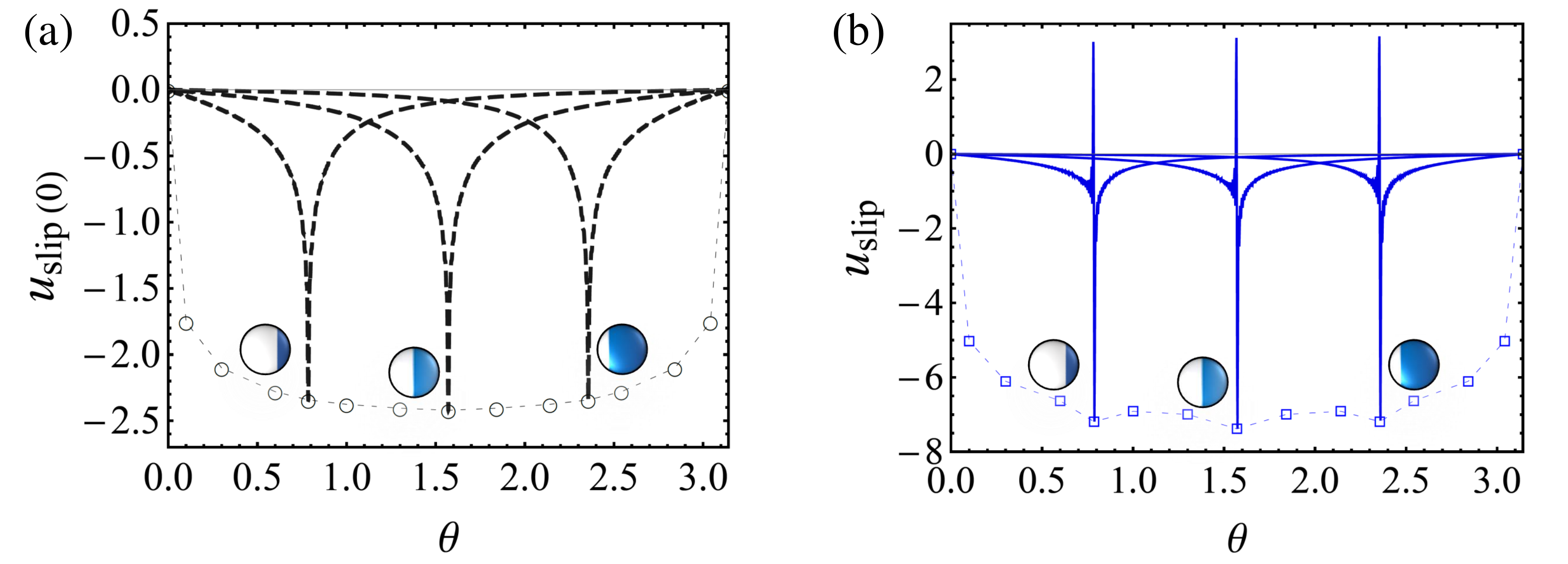} }}%
	$ \; \; \; $
	\begin{tikzpicture}

	\begin{axis}[clip=false,
	height=.24\textwidth,width=.38\textwidth, extra y ticks= 0, xshift=-7.2cm,yshift=0cm,	
	xtick={
		0, 0.5233, 1.0467, 1.5708, 2.093, 2.617, 3.14
	},
	xticklabels={
		0, $\frac{\pi}{6}$, $\frac{\pi}{3}$, $\frac{\pi}{2}$,	$\frac{2\pi}{3}$, $\frac{5\pi}{6}$, $\pi$
	},
	ylabel shift= -8 pt,
	xlabel=$ \theta $, 
	ylabel= {\large $ c_{\theta \theta}$ \normalsize},%
	, xmin=0, xmax=3.14, ymin=-200, ymax=200 , thick
	,	legend style={draw=none,at={(1,0.61)},anchor=south east},	legend style={nodes={scale=1, transform shape}}
	]
	\addplot[line width=0.25pt, blue] table[x=x, y=cxx] {Conc_700_0.002.txt};
	\node[scale=1.1] at (axis cs: -1.05, 180) {(d)}; 
	\end{axis}
	
	
	\begin{axis}[clip=false,
	height=0.24\textwidth,width=.38\textwidth, extra y ticks= 0, xshift=-13.5cm,yshift=-0cm,	ylabel shift= -8pt,
	xtick={
		0, 0.5233, 1.0467, 1.5708, 2.093, 2.617, 3.14
	},
	xticklabels={
		0, $\frac{\pi}{6}$, $\frac{\pi}{3}$, $\frac{\pi}{2}$,	$\frac{2\pi}{3}$, $\frac{5\pi}{6}$, $\pi$
	},
	extra y tick labels = ,
	extra y tick style  = { grid = major }, 
	ylabel= {\large $c,\, c_{\theta}$ \normalsize},%
	xlabel=$ \theta $
	, xmin=0, xmax=3.14, ymin=-1.2, ymax=2.5 , thick
	,	legend style={draw=none,at={(1,0.56)},anchor=south east},	legend style={nodes={scale=0.8, transform shape}}
	]
	\addplot[line width=1pt, black,dash dot] table[x=x, y=c] {Conc_700_0.002.txt};
	\addplot[line width=1pt, red, dashed] table[x=x, y=cx] {Conc_700_0.002.txt};
	\addplot[line width=0.1pt, black] coordinates { (0,0) (3.14,0) }; 
	\addplot[line width=0.1pt, black] coordinates { (1.5708,-1.2) (1.5708,2.5) }; 
	\legend{$ c $,$ c_{\theta} $}
	\node[scale=1.1] at (axis cs: -0.85, 2.3) {(c)}; 
	\end{axis}
	
	\end{tikzpicture}
	\caption{(a) The dashed line represents the Newtonian slip velocity along the polar angle for three different surface coverages ($ \theta_{c}=\pi/4, \pi/2, 3\pi/4 $). The empty circular markers represent the maximum magnitude of the slip velocity for a range of surface coverage: $ \theta_{c} \in (0,\pi) $. (b) The solid line represents the total slip velocity (for three different surface coverages) along the polar angle for $ \hat{\psi}(\xi)=- e^{-\xi} $, $ De=0.01 $, and $ \delta=-0.5 $. The empty square markers represent the maximum magnitude of the slip velocity for a range of surface coverage: $ \theta_{c} \in (0,\pi) $.. (c) Concentration profile and its gradient along the polar angle (on particle surface) for $ \theta_{c}=\pi/2 $. (d) Shows the profile for $ c_{\theta \theta} $. First 700 modes were used to describe the concentration field and its gradients.} 
	\label{fig:slip}
\end{figure}
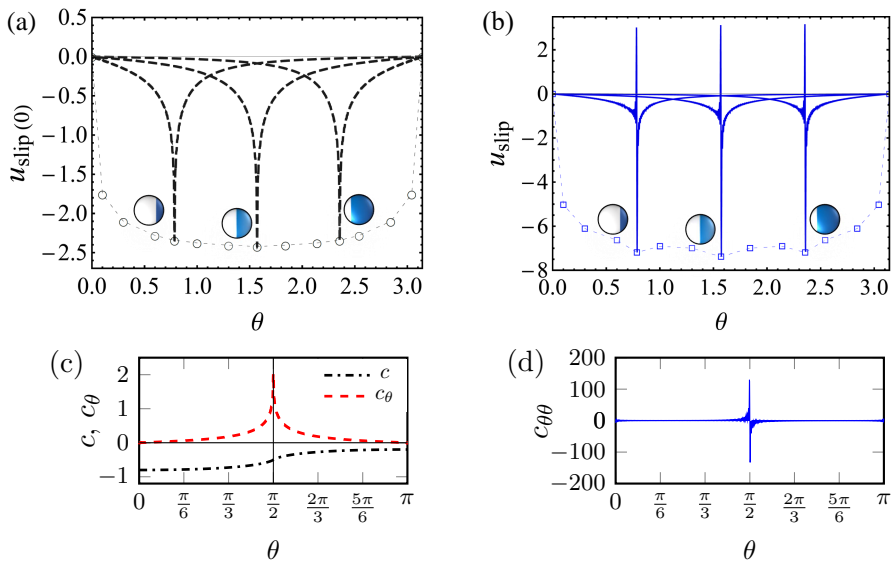

\vspace{2 mm}

Fig. \ref{fig:slip}(a) shows the Newtonian slip velocity varying across the polar angle for three different surface coverages ($ \theta_{c}=\pi/4, \, \pi/2, \, 3\pi/4 $). 
The negative value depicts that the diffusio-osmotic flow is towards the active region, and is highest in magnitude at the point of discontinuity of activity ($ \theta=\theta_{c} $) \footnote{These results are validated with those reported by \cite{michelin2014phoretic} and are described in the supplementary material.}.
For a second-order fluid, figure \ref{fig:slip}(b) shows a localized reversal of slip velocity across $ \theta_{c} $ for all three surface coverages i.e. the slip velocity exhibits sharp non-linear gradients across the point of discontinuity of activity. \cite{SOF_natale2017} also reported such behavior in viscoelastic stresses at $ \theta_{c} $ in their FEM simulations.
They reported a viscoelasticity triggered extensional flow, across the point of discontinuity in surface activity.

To analyze this slip reversal or the extensional flow across the point of transition in activity, we examine the behavior of concentration field around the particle, as the slip (\ref{FINALDeSol_alt}) depends on the first and second polar gradient of the concentration field.
In fig. \ref{fig:slip}(c,d), $ c $, $ c_{\theta} $ and $ c_{\theta \theta}$ profiles are shown for $ \theta_{c}=\pi/2 $.  
A step change in surface activity causes the second tangential gradient to undergo a sharp reversal at $ \theta=\theta_{c} $, which is also reflected in the slip velocity (in fig. \ref{fig:slip}b).
The profile of $ c_{\theta \theta} $, despite accounting for first 700 modes, exhibits oscillations near the point of discontinuity of surface activity.
This seemingly divergent behavior is because of the step function representation of surface activity (\ref{stepK}), which is widely employed in the literature \citep{golestanian2007designing,michelin2014phoretic,SOF_natale2017}.
This oscillatory behavior indicates a violation of scaling in the inner region: in section \S 2.1.2 the length scale in the tangential direction was assumed to be $ a^{*} $, which is not the true characteristic of the system as variations in the tangential direction are rapid for a step discontinuity in surface activity. 
Since such step discontinuities are unlikely to be realized in experiments, we now consider a smooth variation in the activity and demonstrate that the oscillatory behavior of $ c_{\theta \theta} $ for step activity is a mathematical artefact. 


\subsection{Diffusio-osmotic slip on an active particle: Sigmoidal function approximation}


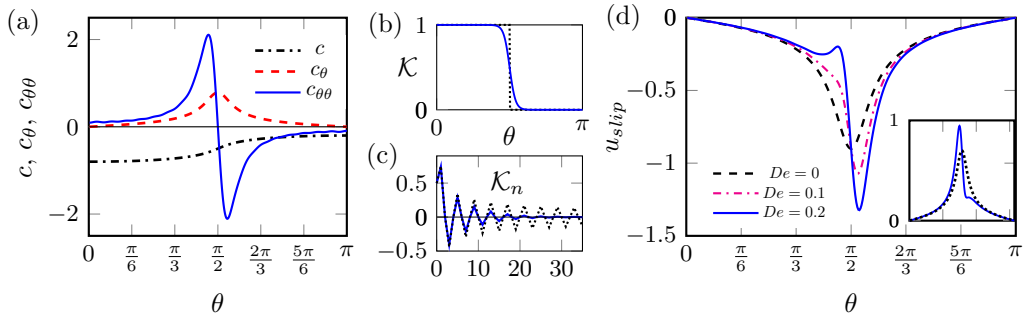
\begin{figure}
	\centering
	\begin{tikzpicture}
	
	\begin{axis}[clip=false,
	height=0.33\textwidth,width=.37\textwidth, extra y ticks= 0, xshift=-7.5cm,yshift=0cm,	ylabel shift= -3 pt,
	extra y tick labels = ,
	extra y tick style  = { grid = major }, 
	xtick={
		0, 0.5233, 1.0467, 1.5708, 2.093, 2.617, 3.14
	},
	xticklabels={
		0, $\frac{\pi}{6}$, $\frac{\pi}{3}$, $\frac{\pi}{2}$,	$\frac{2\pi}{3}$, $\frac{5\pi}{6}$, $\pi$
	},
	xlabel= $ \theta $, ylabel= { $c,\, c_{\theta}, \, c_{\theta \theta}$ },%
	, xmin=0, xmax=3.14, ymin=-2.5, ymax=2.5 , thick
	,	legend style={draw=none,at={(1,0.58)},anchor=south east},	legend style={nodes={scale=0.8, transform shape}}
	]
	\addplot[line width=1pt, black,dash dot] table[x=x, y=c] {Smooth_conc_suppl.txt};
	\addplot[line width=1pt, red,dashed] table[x=x, y=cx] {Smooth_conc_suppl.txt};
	\addplot[line width=0.8pt, blue] table[x=x, y=cxx] {Smooth_conc_suppl.txt};
	\addplot[line width=0.1pt, black] coordinates { (0,0) (3.14,0) }; 
	\node[scale=1] at (axis cs: -0.8, 2.4) {(a)}; 
	\legend{$ c $,$ c_{\theta} $, $ c_{\theta \theta} $}
	\end{axis}
	\begin{axis}[clip=false,
	height=.20\textwidth,width=.26\textwidth, extra y ticks= 0, xshift=-2.9cm,yshift=1.66cm,	
	xlabel shift= -10 pt,
	ylabel shift= -8 pt,
	yticklabel style = {font=\footnotesize},
	xticklabel style = {font=\footnotesize},
	ylabel style={rotate=-90},
	xlabel= $ \theta $, 
	ylabel= {$ \mathcal{K} $},%
	ytick={0, 1},
	yticklabels={0, 1},
	xtick={0, 3.14},
	xticklabels={0, $ \pi $},
	, xmin=0, xmax=3.14, ymin=0, ymax=1 ,
	,	legend style={draw=none,at={(1,0.61)},anchor=south east},	legend style={nodes={scale=0.5, transform shape}}
	]
	\node[scale=1] at (axis cs: -1.1, 1) {(b)}; 
	\addplot[line width=0.8pt, black,densely dotted] table[x=x, y=step] {stepVSsmooth.txt};
	\addplot[line width=0.7pt, blue] table[x=x, y=gam16] {stepVSsmooth.txt};
	\end{axis}
	\begin{axis}[clip=false,
	height=.21\textwidth,width=.26\textwidth, extra y ticks= 0, xshift=-2.9cm,yshift=-0.2cm,	
	ylabel style={rotate=-90},
	yticklabel style = {font=\footnotesize},
	xticklabel style = {font=\footnotesize},
	, xmin=0, xmax=35, ymin=-0.5, ymax=0.9 ,
	,	legend style={draw=none,at={(1,0.61)},anchor=south east},	legend style={nodes={scale=0.5, transform shape}}
	]
	\node[scale=1] at (axis cs: -13, 0.9) {(c)}; 
	\addplot[line width=0.8pt, blue] table[x=x, y=new] {Smooth_K_suppl.txt};
	\addplot[line width=0.8pt, black,densely dotted] table[x=x, y=old] {Smooth_K_suppl.txt};
	\addplot[line width=0.1pt, black] coordinates { (0,0) (35,0) }; 
	\node at (160,100) {{ $ \mathcal{K}_{n} $}};
	\end{axis}
	\begin{axis}[clip=false,
	height=.33\textwidth,width=.44\textwidth, extra y ticks= 0, xshift=0.4cm,yshift=0cm,	
	ylabel shift= -7 pt,
	xtick={
		0, 0.5233, 1.0467, 1.5708, 2.093, 2.617, 3.14
	},
	xticklabels={
		0, $\frac{\pi}{6}$, $\frac{\pi}{3}$, $\frac{\pi}{2}$,	$\frac{2\pi}{3}$, $\frac{5\pi}{6}$, $\pi$
	},
	xlabel=$ \theta $ \normalsize, ylabel= { $ u_{slip} $ \normalsize},%
	, xmin=0, xmax=3.14, ymin=-1.5, ymax=0 , thick
	,	legend style={draw=none,at={(0.46,0.03)},anchor=south east},	legend style={nodes={scale=0.6, transform shape}}
	]
	\node[scale=1] at (axis cs: -0.6,0) {(d)}; 
	\addplot[line width=0.9pt, black, dashed] table[x=x, y=michelin] {Smooth_slip_suppl.txt};
	\addplot[line width=0.8pt, magenta,dash dot] table[x=x, y=aku01] {Smooth_slip_suppl.txt};
	\addplot[line width=0.8pt, blue] table[x=x, y=aku02] {Smooth_slip_suppl.txt};
	\legend{$ De=0 $, $ De=0.1 $,$ De=0.2 $} 
	\end{axis}
	\begin{axis}[clip=false, 
	height=.215\textwidth,width=.22\textwidth, extra y ticks= 0, xshift=3.35cm,yshift=0.2cm,	
	ylabel shift= -3 pt,
	yticklabel style = {font=\scriptsize},
	ytick={
		0, 1
	},
	yticklabels={
		0, 1
	},
	xtick={
	},
	xticklabels={
	},
	xlabel= , ylabel= ,%
	, xmin=0, xmax=3.14, ymin=0, ymax=1 , thick
	,	legend style={draw=none,at={(0.477,0.03)},anchor=south east},	legend style={nodes={scale=0.48, transform shape}}
	]
	\addplot[line width=0.7pt, blue] table[x=x, y=aku02]
	{Smooth_slip_suppl_VE_repulsive.txt};\label{De02}
	\addplot[line width=1pt, black, densely dotted] table[x=x, y=michelin]
	{Smooth_slip_suppl_VE_repulsive.txt};\label{De00}
	\end{axis}
	\end{tikzpicture}
	\caption{(a) Profile of concentration and its first and second gradient along the polar angle on particle surface for $ \theta_{c}=\pi/2 $. (b) Variation of surface activity as a step (dotted line) and smooth function (solid line). (c) Convergence of spectral modes: solid (blue) line shows fast convergence for the sigmoidal activity; dotted (black) line depicts slow convergence for step activity. (d) Different curves represent the total slip velocity along the polar angle for $ \hat{\psi}(\xi)=-1 e^{-\xi} $, $ \zeta=16 $, $ \delta=-0.5 $ and different Deborah numbers. The inset shows the slip velocities for Newtonian and second-order fluid  for the case of repulsive interactions; $ \Phi_{0}=+1 $, $ De = 0 $ (\ref{De00}), and $ De = 0.2 $ (\ref{De02}).} 
	\label{fig:smooth}
\end{figure}

The step change in surface activity is approximated as a logistic function: 
\begin{equation} \label{sigmoid}
\mathcal{K}(\theta)={1}/\left( {1+ {\rm{exp}}\left[{-\zeta  \left(\theta_{c}- \theta  \right)}\right]}\right).
\end{equation}
Here $ \zeta $ (transition parameter) determines the sharpness of the transition of non-dimensional activity from 1 to 0. 
We have chosen $ \zeta=16 $ such that the transition length is large enough to maintain the consistency of the analysis with respect to the scaling used in \S 2.1.2  (i.e. $ l_{transition}^{*} \sim a^{*} $)\footnote{For a particle of  $ a^{*}=5 \mu $m, $ \zeta=16 $ corresponds to a circumferential length of $ \sim 2 \, \mu $m, which is much larger than the typical interaction layer thickness ($\sim 10 $nm).}.
The expression for concentration and slip velocity is identical to (\ref{concGold}) and (\ref{uslip}), respectively. The spectral modes can be found by substituting (\ref{sigmoid}) into (\ref{spectralK}) and taking the inner product with respect to $ P_{n} $ on both sides
\begin{equation} \label{sigmoid_modes}
\mathcal{K}_{n}=\frac{2n+1}{2} \int_{0}^{\pi} \mathcal{K}(\theta) P_{n}(cos \theta) {\rm{d}}\theta.
\end{equation}
For activity represented by (\ref{sigmoid}), the concentration field and its tangential gradients (at the surface) are depicted in fig.\ref{fig:smooth}(a). 
The variation of concentration field and its gradients along $ \theta $ is qualitatively similar to that reported for the step change; the magnitudes, however, are significantly reduced. 
Fig. \ref{fig:smooth}(b) $ \& $ (c) show the variation of activity and its faster convergence in comparison to the step function. 
Fig.\ref{fig:smooth}(d) compares the slip velocity for a second-order fluid with that of a Newtonian fluid.

We observe a reduction and an increase in slip velocity (relative to the Newtonian slip) across $ \theta_{c} $. 
The magnitude of modification in the slip is significantly reduced (as compared to that of step change in activity) which is a consequence of smooth transition of activity.
The above observations are for attractive interactions between solute and particle surface; for repulsive interactions (as shown in the inset of fig. \ref{fig:smooth}d), the effect of viscoelasticity is qualitatively reversed.

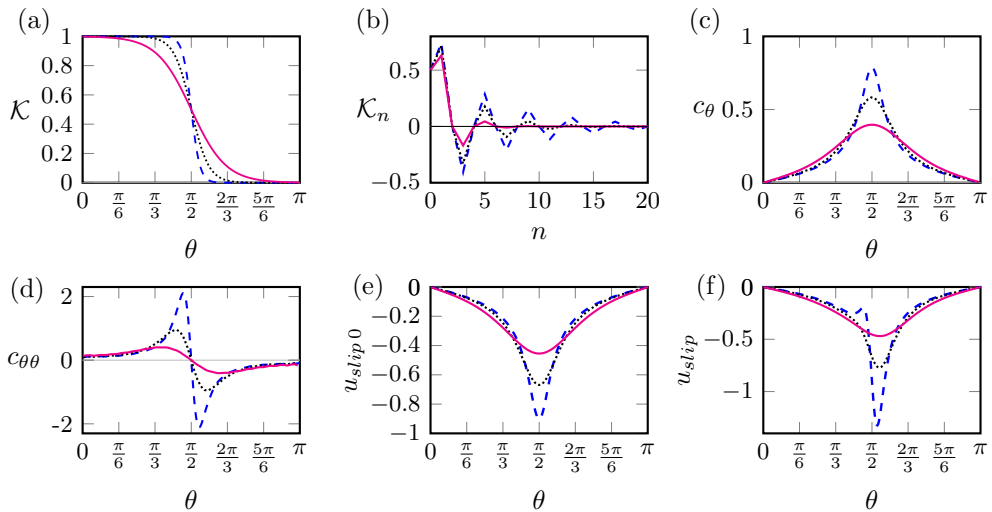
\begin{figure}
	\centering
	\begin{tikzpicture}
	
	\begin{axis}[clip=false,
	height=0.26\textwidth,width=.33\textwidth, extra y ticks= 0, xshift=0cm,yshift=0cm,	ylabel shift= -3 pt,
	ylabel style={rotate=-90},
	extra y tick labels = ,
	extra y tick style  = { grid = major }, 
	xtick={
		0, 0.5233, 1.0467, 1.5708, 2.093, 2.617, 3.14
	},
	xticklabels={
		0, $\frac{\pi}{6}$, $\frac{\pi}{3}$, $\frac{\pi}{2}$,	$\frac{2\pi}{3}$, $\frac{5\pi}{6}$, $\pi$
	},
	xlabel= $ \theta $, ylabel= { $ \mathcal{K} $ },%
	, xmin=0, xmax=3.14, ymin=0, ymax=1 , thick
	,	legend style={draw=none,at={(1,0.3)},anchor=south east},	legend style={nodes={scale=0.7, transform shape}}
	]
	\addplot[line width=0.7pt, blue,dashed] table[x=x, y=gam16] {stepVSsmooth.txt};
	\addplot[line width=0.7pt, black, densely dotted] table[x=x, y=gam8] {stepVSsmooth.txt};
	\addplot[line width=0.7pt, magenta] table[x=x, y=gam4] {stepVSsmooth.txt};
	\node[scale=1] at (axis cs: -0.7, 1.1) {(a)}; 
	\node[scale=1] at (axis cs: +4.2, 1.1) {(b)}; 
	\node[scale=1] at (axis cs: +9, 1.1) {(c)}; 
	\end{axis}
	\begin{axis}[
	height=0.26\textwidth,width=.33\textwidth, extra y ticks= 0, xshift=4.6cm,yshift=0cm,	ylabel shift= -15 pt,
	ylabel style={rotate=-90},
	extra y tick labels = ,
	extra y tick style  = { grid = major }, 
	xlabel= $ n $, ylabel= { $ \mathcal{K}_{n} $ },%
	, xmin=0, xmax=20, ymin=-0.5, ymax=0.8 , thick
	,	legend style={draw=none,at={(1,0.45)},anchor=south east},	legend style={nodes={scale=0.7, transform shape}}
	]
	\addplot[line width=0.8pt, blue,dashed] table[x=x, y=new] {Smooth_K_suppl.txt};
	\addplot[line width=0.8pt, black, densely dotted] table[x=x, y=gam8] {Smooth_K_suppl_gamma.txt};
	\addplot[line width=0.8pt, magenta] table[x=x, y=gam4] {Smooth_K_suppl_gamma.txt};
	\addplot[line width=0.1pt, black] coordinates { (0,0) (20,0) }; 
	\end{axis}
	\begin{axis}[
	height=0.26\textwidth,width=.33\textwidth, extra y ticks= 0, xshift=9cm,yshift=0cm,	ylabel shift= -5 pt,
	ylabel style={rotate=-90},
	extra y tick labels = ,
	extra y tick style  = { grid = major }, 
	ytick={
		0, 0.5,1
	},
	yticklabels={
		0, 0.5,1
	},
	xtick={
		0, 0.5233, 1.0467, 1.5708, 2.093, 2.617, 3.14
	},
	xticklabels={
		0, $\frac{\pi}{6}$, $\frac{\pi}{3}$, $\frac{\pi}{2}$,	$\frac{2\pi}{3}$, $\frac{5\pi}{6}$, $\pi$
	},
	xlabel= $ \theta $, ylabel= { $ c_{\theta} $ },%
	, xmin=0, xmax=3.14, ymin=0, ymax=1 , thick
	,	legend style={draw=none,at={(1,0.45)},anchor=south east},	legend style={nodes={scale=0.7, transform shape}}
	]
	\addplot[line width=0.8pt, blue,dashed] table[x=x, y=cx] {Smooth_conc_suppl.txt};
	\addplot[line width=0.8pt, black, densely dotted] table[x=x, y=c_th_gam44] {conc_gamma.txt};
	\addplot[line width=0.8pt, magenta] table[x=x, y=c_th_gam4] {conc_gamma.txt};
	\end{axis}
	\begin{axis}[clip=false,
	height=0.26\textwidth,width=.33\textwidth, extra y ticks= 0, xshift=0cm,yshift=-3.3cm,	ylabel shift= -3 pt,
	ylabel style={rotate=-90},
	extra y tick labels = ,
	extra y tick style  = { grid = major }, 
	ytick={
		-2,0,2
	},
	yticklabels={
		-2,0,2
	},
	xtick={
		0, 0.5233, 1.0467, 1.5708, 2.093, 2.617, 3.14
	},
	xticklabels={
		0, $\frac{\pi}{6}$, $\frac{\pi}{3}$, $\frac{\pi}{2}$,	$\frac{2\pi}{3}$, $\frac{5\pi}{6}$, $\pi$
	},
	xlabel= $ \theta $, ylabel= { $ c_{\theta \theta} $ },%
	, xmin=0, xmax=3.14, ymin=-2.3, ymax=2.3 , thick
	,	legend style={draw=none,at={(1,0.45)},anchor=south east},	legend style={nodes={scale=0.7, transform shape}}
	]
	\addplot[line width=0.8pt, blue,dashed] table[x=x, y=cxx] {Smooth_conc_suppl.txt};
	\addplot[line width=0.8pt, black, densely dotted] table[x=x, y=c_th_th_gam44] {conc_gamma.txt};
	\addplot[line width=0.8pt, magenta] table[x=x, y=c_th_th_gam4] {conc_gamma.txt};
	\node[scale=1] at (axis cs: -0.8, 2.2) {(d)}; 
	\end{axis}
	\begin{axis}[clip=false,
	height=.26\textwidth,width=.33\textwidth, extra y ticks= 0, xshift=4.6cm,yshift=-3.3cm,	
	ylabel shift= -5 pt,
	xtick={
		0, 0.5233, 1.0467, 1.5708, 2.093, 2.617, 3.14
	},
	xticklabels={
		0, $\frac{\pi}{6}$, $\frac{\pi}{3}$, $\frac{\pi}{2}$,	$\frac{2\pi}{3}$, $\frac{5\pi}{6}$, $\pi$
	},
	xlabel=$ \theta $ \normalsize, ylabel= { $ u_{slip\,0} $ \normalsize},%
	, xmin=0, xmax=3.14, ymin=-1, ymax=0 , thick
	,	legend style={draw=none,at={(0.42,0.03)},anchor=south east},	legend style={nodes={scale=0.6, transform shape}}
	]
	\node[scale=1] at (axis cs: -0.9,0) {(e)}; 
	\addplot[line width=0.9pt, blue, dashed] table[x=x, y=michelin] {Smooth_slip_suppl.txt};
	\addplot[line width=0.8pt, black, densely dotted] table[x=x, y=michelin] {Smooth_slip_suppl_gam_8.txt};
	\addplot[line width=0.8pt, magenta] table[x=x, y=michelin] {Smooth_slip_suppl_gam_4.txt};
	\end{axis}
	\begin{axis}[clip=false,
	height=.26\textwidth,width=.33\textwidth, extra y ticks= 0, xshift=9cm,yshift=-3.3cm,	
	ylabel shift= -5 pt,
	xtick={
		0, 0.5233, 1.0467, 1.5708, 2.093, 2.617, 3.14
	},
	xticklabels={
		0, $\frac{\pi}{6}$, $\frac{\pi}{3}$, $\frac{\pi}{2}$,	$\frac{2\pi}{3}$, $\frac{5\pi}{6}$, $\pi$
	},
	xlabel=$ \theta $ \normalsize, ylabel= { $ u_{slip} $ \normalsize},%
	, xmin=0, xmax=3.14, ymin=-1.4, ymax=0 , thick
	,	legend style={draw=none,at={(0.42,0.03)},anchor=south east},	legend style={nodes={scale=0.6, transform shape}}
	]
	\node[scale=1] at (axis cs: -0.75,0) {(f)}; 
	\addplot[line width=0.9pt, blue, dashed] table[x=x, y=aku02] {Smooth_slip_suppl.txt};
	\addplot[line width=0.8pt, black, densely dotted] table[x=x, y=aku02] {Smooth_slip_suppl_gam_8.txt};
	\addplot[line width=0.8pt, magenta] table[x=x, y=aku02] {Smooth_slip_suppl_gam_4.txt};
	\end{axis}
	\end{tikzpicture}
	\caption{Analysis of activity, spectral modes, concentration gradients and slip velocity for three different transition parameters: $ \zeta=4 $ (\ref{gam4}), $ \zeta=8 $ (\ref{gam8}), $ \zeta=16 $ (\ref{gam16}). (a) Variation of activity with polar angle $ \theta $. (b) Convergence of spectral modes. (c) Tangential concentration gradient. (d) Second tangential derivative of concentration field. (e) Slip velocity of a Newtonian fluid. (f) Slip velocity of the second-order fluid. Parameters: $ De=0.1 $, $ \delta=-0.5 $, $ \Phi_{0}=-1 $.} 
	\label{fig:smooth_gamma}
\end{figure}

Fig. \ref{fig:smooth_gamma} demonstrates the effect of the transition parameter ($ \zeta $) on the spectral coefficients, concentration gradients and slip velocity. Fig. \ref{fig:smooth_gamma}(a) shows  that as $ \zeta $ decreases, the transition from active to passive surface becomes smoother. Fig. \ref{fig:smooth_gamma}(b) depicts that the convergence is faster for lower $ \zeta $.
As the transition becomes smoother the magnitude of tangential gradients of concentration field reduces (fig.\ref{fig:smooth_gamma}c,d). As a result, the slip velocity for Newtonian and second-order fluid reduces (shown in fig.\ref{fig:smooth_gamma}e-f, respectively).

\begin{figure}
	\centering
	{{\includegraphics[scale=0.58]{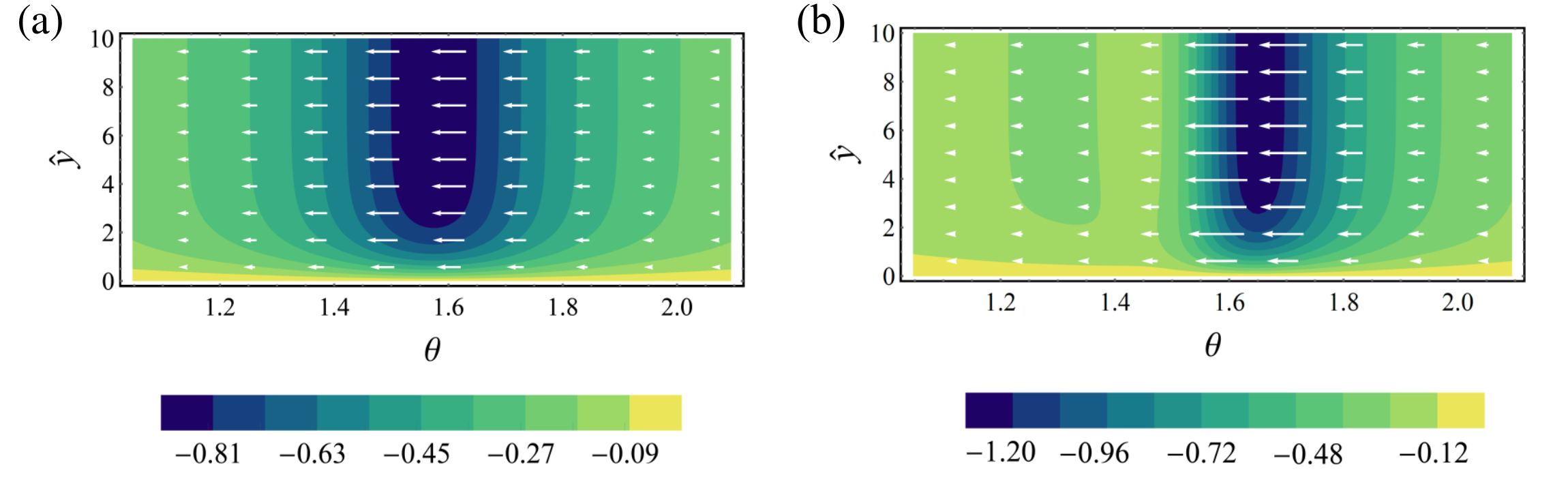} }}%
	\caption{(a) Newtonian tangential velocity contours inside the interaction layer for $ \theta_{c}=\pi/2 $. (b) Tangential velocity contours corresponding to the second-order fluid for  $ \hat{\psi}(\xi)=- e^{-\xi} $, $ \zeta=16 $, $ \delta=-0.5 $, and $ De=0.2 $. Arrows depict the velocity vectors ($ \hat{\IB{u}}_{1}^{(0)} $).}
	\label{fig:inner_vel_field}
\end{figure}

The effect of viscoelasticity on the tangential velocity field inside the interaction layer is shown in fig. \ref{fig:inner_vel_field} (for $ \theta_{c}=\pi/2 $). 
Fig. \ref{fig:inner_vel_field}(a) depicts the symmetry of velocity field across $ \theta_{c} $ for Newtonian fluid.
Addition of viscoelasticity breaks this symmetry: the contours of the SOF in \ref{fig:inner_vel_field}(b) depict that the magnitude of velocity field is reduced before the transition of activity ($ \theta < \theta_{c} $) and increased after it ($ \theta > \theta_{c} $).
These results suggest that the elastic effects (characterized by $ De $) tend to locally reverse the slip velocity. At higher Deborah numbers, a reversal in slip velocity can generate a local extensional flow near the transition of activity (as shown by \cite{SOF_natale2017}), resembling a puller or pusher-type flow field.
Interestingly, there have been similar observations in the context of electrophoresis.
Recently, \cite{li2020electrophoresis} reported that, in weakly viscoelastic fluids, polymer elasticity in the inner region changes the electrophoretic particle into a puller-type squirmer: the strong shear flow in the double layer causes the polymers to be stretched tangentially, resulting in a pulling-flow at the front \& rear end, and a pushing-flow from the sides.
Since electrophoretic and diffusiophoretic mechanisms share qualitative similarities (such as strong shear inside the thin layer), it is plausible that the effects of elasticity, observed in electrophoresis, emerge in diffusiophoresis as well.

\subsection{Swimming velocity}\label{VEswim}

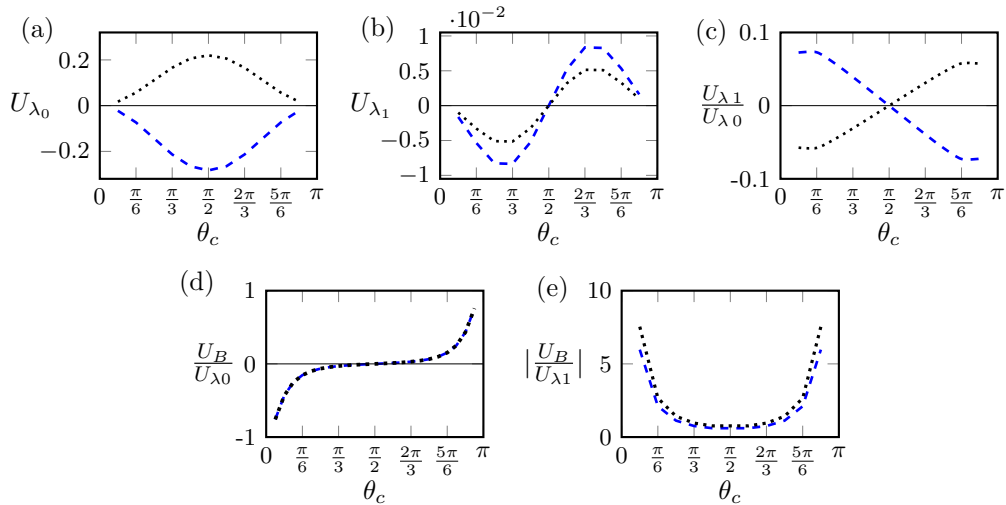
\begin{figure}
	\centering
	\begin{tikzpicture}
	\begin{axis}[clip=false,
	height=0.26\textwidth,width=.33\textwidth, xshift=0cm,yshift=0cm,	ylabel shift= -12 pt,
	xtick={
		0, 0.5233, 1.0467, 1.5708, 2.093, 2.617, 3.14
	},
	xticklabels={
		0, $\frac{\pi}{6}$, $\frac{\pi}{3}$, $\frac{\pi}{2}$,	$\frac{2\pi}{3}$, $\frac{5\pi}{6}$, $\pi$
	},
	xlabel shift= -5pt,
	ylabel style={rotate=-90},
	extra y tick labels = ,
	extra y tick style  = { grid = major }, 
	xlabel=$ \theta_{c}$, ylabel= {${U_{\lambda_0}}$},%
	, xmin=0, xmax=3.14, ymin=-0.32, ymax=0.32 , thick
	,	legend style={draw=none,at={(1,0.2)},anchor=south east},	legend style={nodes={scale=0.8, transform shape}}
	]
	\addplot[line width=1pt, blue,dashed] table[x=x, y=y1] {LRT_SOF.txt};
	\addplot[line width=1pt, black, dotted] table[x=x, y=y1] {LRT_SOF_Repulsion.txt};
	\node[scale=1] at (axis cs: -0.9, 0.33) {(a)}; 
	\addplot[line width=0.1pt, black]  coordinates { (0,0) (3.14,0) }; 
	\end{axis}
	\begin{axis}[clip=false,
	height=0.26\textwidth,width=.33\textwidth, xshift=4.5cm,yshift=0cm,	ylabel shift= -12 pt,
	xtick={
		0, 0.5233, 1.0467, 1.5708, 2.093, 2.617, 3.14
	},
	xticklabels={
		0, $\frac{\pi}{6}$, $\frac{\pi}{3}$, $\frac{\pi}{2}$,	$\frac{2\pi}{3}$, $\frac{5\pi}{6}$, $\pi$
	},
	xlabel shift= -5pt,
	ylabel style={rotate=-90},
	extra y tick labels = ,
	extra y tick style  = { grid = major }, 
	xlabel=$ \theta_{c}$, ylabel= { ${U_{\lambda_1}}$},%
	, xmin=0, xmax=3.14, ymin=-0.0105, ymax=0.0105 , thick
	,	legend style={draw=none,at={(1,0.2)},anchor=south east},	legend style={nodes={scale=0.8, transform shape}}
	]
	\addplot[line width=1pt, blue,dashed] table[x=x, y=y2] {LRT_SOF.txt};
	\addplot[line width=1pt, black, dotted] table[x=x, y=y2] {LRT_SOF_Repulsion.txt};
	\addplot[line width=0.1pt, black]  coordinates { (0,0) (3.14,0) }; 
	\node[scale=1] at (axis cs: -0.9, 0.011) {(b)}; 
	\end{axis}
	
	\begin{axis}[clip=false,
	height=0.26\textwidth,width=.33\textwidth, xshift=9cm,yshift=0cm,	ylabel shift= -13pt,
	xtick={
		0, 0.5233, 1.0467, 1.5708, 2.093, 2.617, 3.14
	},
	xticklabels={
		0, $\frac{\pi}{6}$, $\frac{\pi}{3}$, $\frac{\pi}{2}$,	$\frac{2\pi}{3}$, $\frac{5\pi}{6}$, $\pi$
	},
	ytick={
		-0.1, 0, 0.1
	},
	yticklabels={
		-0.1, 0, 0.1
	},
	xlabel shift= -5pt,
	ylabel style={rotate=-90},
	extra y tick labels = ,
	extra y tick style  = { grid = major }, 
	xlabel=$ \theta_{c}$, ylabel= {\large $\frac{U_{\lambda\,1}}{U_{\lambda\,0}}$ \normalsize},%
	, xmin=0, xmax=3.14, ymin=-0.1, ymax=0.1 , thick
	,	legend style={draw=none,at={(1,0.2)},anchor=south east},	legend style={nodes={scale=0.8, transform shape}}
	]
	\addplot[line width=1pt, blue,dashed,smooth] table[x=x, y=y3] {LRT_SOF.txt};
	\addplot[line width=1pt, black, dotted,smooth] table[x=x, y=y3] {LRT_SOF_Repulsion.txt};
	\addplot[line width=0.1pt, black]  coordinates { (0,0) (3.14,0) }; 
	\node[scale=1] at (axis cs: -1, 0.10) {(c)}; 
	\end{axis}
	\begin{axis}[clip=false,
	height=0.26\textwidth,width=.33\textwidth, xshift=2.2cm,yshift=-3.4cm,	ylabel shift= -7pt,
	xtick={
		0, 0.5233, 1.0467, 1.5708, 2.093, 2.617, 3.14
	},
	xticklabels={
		0, $\frac{\pi}{6}$, $\frac{\pi}{3}$, $\frac{\pi}{2}$,	$\frac{2\pi}{3}$, $\frac{5\pi}{6}$, $\pi$
	},
	ytick={
		-1, 0, 1
	},
	yticklabels={
		-1, 0, 1
	},
	xlabel shift= -5pt,
	ylabel style={rotate=-90},
	extra y tick labels = ,
	extra y tick style  = { grid = major }, 
	xlabel=$ \theta_{c}$, ylabel= {{\large $\frac{U_{B}}{U_{\lambda {0}}}$}},%
	, xmin=0, xmax=3.14, ymin=-1, ymax=+1 , thick
	,	legend style={draw=none,at={(1,0.2)},anchor=south east},	legend style={nodes={scale=0.8, transform shape}}
	]
	\addplot[line width=1pt, blue,dashed] table[x=x, y=y] {UB_SOF.txt};
	\addplot[line width=1.4pt, black,dotted] table[x=x, y=y] {UB_SOF.txt};
	\addplot[line width=0.1pt, black]  coordinates { (0,0) (3.14,0) }; 
	\node[scale=1] at (axis cs: -1, 1.1) {(d)}; 
	\end{axis}
	\begin{axis}[clip=false,
	height=0.26\textwidth,width=.33\textwidth, xshift=6.9cm,yshift=-3.4cm,	ylabel shift= -6pt,
	xtick={
		0, 0.5233, 1.0467, 1.5708, 2.093, 2.617, 3.14
	},
	xticklabels={
		0, $\frac{\pi}{6}$, $\frac{\pi}{3}$, $\frac{\pi}{2}$,	$\frac{2\pi}{3}$, $\frac{5\pi}{6}$, $\pi$
	},
	ytick={
		0, 5,10, 15
	},
	yticklabels={
		0, 5,10, 15
	},
	xlabel shift= -5pt,
	ylabel style={rotate=-90},
	extra y tick labels = ,
	extra y tick style  = { grid = major }, 
	xlabel=$ \theta_{c}$, ylabel= {{\large $| \frac{U_{B}}{U_{\lambda {1}}}| $}},%
	, xmin=0, xmax=3.14, ymin=0, ymax=+10 , thick
	,	legend style={draw=none,at={(1,0.2)},anchor=south east},	legend style={nodes={scale=0.8, transform shape}}
	]
	\addplot[line width=1pt, blue,dashed] table[x=x, y=y] {UB_UL1.txt};\label{att}
	\addplot[line width=1.2pt, black,dotted] table[x=x, y=y] {UB_UL1_REP.txt};\label{rep}
	\addplot[line width=0.1pt, black]  coordinates { (0,0) (3.14,0) }; 
	\node[scale=1] at (axis cs: -1, 10.1) {(e)}; 
	\end{axis}
	
	\end{tikzpicture}
	\caption{(a) Variation of swimming velocity due to Newtonian surface slip ($ U_{\lambda_{0}} $) for various surface coverages. (b) Perturbation to swimming velocity due to modification in the slip ($ U_{\lambda_{1}} $). (c) Ratio of the swimming velocity components i.e. (b) and (a). (d) Variation of modification to the swimming velocity due to bulk polymeric stresses ($ U_{B} $). (e) Ratio of (d) and (c): comparison of two components arising due to viscoelasticity. Parameters: $ De=0.1 $, $ \delta=-0.5 $, $\zeta=16$,  $ \Phi_{0}=-1 $ (attractive;\ref{att}), and $ \Phi_{0}=+1 $ (repulsive;\ref{rep}). } 
	\label{fig:LRT_SOF}
\end{figure}

The motion of the Janus sphere is found by using the reciprocal theorem. Following \cite{stone1996propulsion} and \cite{ho1976migration}, we obtain the swimming velocity as
\begin{equation}
U=\frac{-1}{4\pi} \int_{S_{p}} \left.\left( \IB{u}_{0} + De \,\IB{u}_{1}  \right)\right\vert_{r=1} \bcdot \IB{e}_{z} \, {\rm d}S \; - \; \frac{1}{6\pi} De \int_{V_{f}}  \bten{S}_{0}\IB{:} \bnabla{\IB{u}^{t}} {\rm d}V. \label{LRT}
\end{equation}
Here, $ \IB{u}_{0} $ is the Newtonian slip velocity ($  M_{0}  c_{\theta}(1,\theta) \, \IB{e}_{\theta} $); $ De \, \IB{u}_{1} $ is the modification to slip due to viscoelasticity ($ De \, M_{1} \delta \, c_{\theta}(1,\theta) \, c_{\theta 
\theta}(1,\theta) \, \IB{e}_{\theta} $)
 ; $ \IB{u}^{t} $ is the test flow field which governs the motion of a rigid sphere in $ z $-direction with unit velocity in a quiescent Newtonian medium \citep{michelin2014phoretic,datt2015squirmingST}, and  $ \bten{S}_{0} $ is the polymeric stress.
The first integral ($ U_{\lambda} $) denotes the contribution to swimming velocity arising from the slip, which contains a Newtonian ($ U_{\lambda_0} $) and a non-Newtonian ($ U_{\lambda_1} $) component:
\begin{equation}\label{ULambda}
	U_{\lambda_{0}} = \frac{M_{0}}{2} \int_{0}^{\pi} c_{\theta}(1,\theta) \sin^{2}{\theta} \, {\rm{d}}\theta;
\quad 	U_{\lambda_{1}} = \frac{De\, \delta M_{1}}{2} \int_{0}^{\pi} c_{\theta}(1,\theta) c_{\theta \theta}(1,\theta) \sin^{2}{\theta} \, {\rm{d}}\theta.
\end{equation}
We evaluate the above components and compare them in fig. \ref{fig:LRT_SOF} for different surface coverages. 
For attractive interactions between the solute molecules and particle, the Newtonian swimming velocity is in the negative z-direction (see fig. \ref{fig:LRT_SOF}(a)).
Figure(s). \ref{fig:LRT_SOF}(b,c) show that the modification to the swimming velocity is relatively small and changes sign with the surface coverage: the swimming velocity increases for $ \theta_{c}<\pi/2 $ and reduces for $ \theta_{c}>\pi/2 $. 

For a repulsive interaction, $ U_{\lambda_{0}} $ changes sign (depicting swimming in the opposite direction) as also reported by earlier studies \citep{sharifi2013,michelin2014phoretic,SOF_natale2017}.
However, $ U_{\lambda_{1}} $ for a repulsive interaction is in the same direction as that for attractive interaction. 
In this case, the contribution of $ U_{\lambda_{1}} $ is such that the swimming velocity is reduced for $ \theta_{c} < \pi/2 $ and enhanced for $ \theta_{c} > \pi/2 $.
This outcome can be understood by comparing the flow field inside the interaction layer for attractive and repulsive cases i.e. fig.\ref{fig:innerVel} and fig.\ref{fig:innerVel2}(a), respectively. 
In comparison to the attractive interaction, Newtonian velocity field was reversed for repulsive interactions, whereas the modification (arising from viscoelasticity) was found to be in the same direction.
It is also interesting to note that, contrary to $ U_{\lambda_{0}} $, $ U_{\lambda_{1}} $ is an odd function (antisymmetric about $ \pi/2 $)  which is due to the proportionality to double tangential derivative of surface concentration, whereas the former is solely proportional to the single tangential derivative.

The second integral ($ U_{B} $) accounts for the contribution from polymeric stresses in the bulk. 
In their evaluation of swimming velocity, \cite{datt2017activeComplex} accounted for the bulk viscoelastic effects ($ U_{B} $), and provided the following analytical expression (re-expressed here using the scales in eq.\ref{definitions}):
\begin{equation}\label{UB}
\frac{U_{B}}{U_{\lambda_{0}}} =  - De (1+\delta) \sum_{n=1}^{\infty} \frac{6n}{(n+1)^{2} (n+2)} \frac{\alpha_{n} \alpha_{n+1}}{\alpha_{1}^{2}},
\end{equation} 
where $ \alpha_{n} = {n \mathcal{K}_{n}}/{(2n+1)} $ and $ \mathcal{K}_{n} $ is determined using (\ref{sigmoid}) and (\ref{sigmoid_modes}).
The modification due to the bulk stresses ($ U_{B}/U_{\lambda_{0}} $) is plotted in fig.\ref{fig:LRT_SOF}(d) for smooth activity.
The figure shows that $ U_{B} $ can significantly enhance (or impede) the active swimming, depending upon the surface coverage ($ \theta_{c} $) being more (or less) than $ \pi/2 $. This was also reported by \cite{datt2017activeComplex} for step activity.
A comparison of the two effects of viscoelasticity, $ U_{B} $ and $ U_{\lambda_{1}} $ (shown in fig.\ref{fig:LRT_SOF}(e)), reveals that the contribution to the swimming velocity from the modification in slip ($ U_{\lambda_{1}} $) is comparable to that from $ U_{B} $ for $ 5\pi/6 < \theta_{c}<\pi/6 $. 

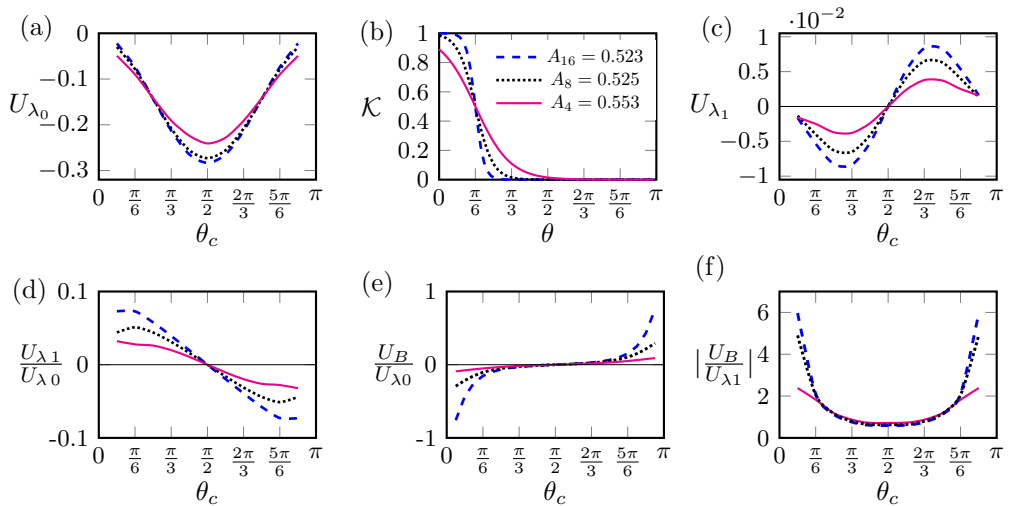
\begin{figure}
	\centering
	\begin{tikzpicture}
	\begin{axis}[clip=false,
	height=0.26\textwidth,width=.33\textwidth, xshift=0cm,yshift=0cm,	ylabel shift= -12 pt,
	xtick={
		0, 0.5233, 1.0467, 1.5708, 2.093, 2.617, 3.14
	},
	xticklabels={
		0, $\frac{\pi}{6}$, $\frac{\pi}{3}$, $\frac{\pi}{2}$,	$\frac{2\pi}{3}$, $\frac{5\pi}{6}$, $\pi$
	},
	xlabel shift= -5pt,
	ylabel style={rotate=-90},
	extra y tick labels = ,
	extra y tick style  = { grid = major }, 
	xlabel=$ \theta_{c}$, ylabel= {${U_{\lambda_0}}$},%
	, xmin=0, xmax=3.14, ymin=-0.32, ymax=0 , thick
	,	legend style={draw=none,at={(1,0.2)},anchor=south east},	legend style={nodes={scale=0.8, transform shape}}
	]
	\addplot[line width=1pt, blue,dashed,smooth] table[x=x, y=NewtGam16] {U_Lambda_Gamma.txt};
	\addplot[line width=1pt, black, densely dotted,smooth] table[x=x, y=NewtGam8] {U_Lambda_Gamma.txt};
	\addplot[line width=0.8pt, magenta,smooth] table[x=x, y=NewtGam4] {U_Lambda_Gamma.txt};
	\node[scale=1] at (axis cs: -0.9, 0.01) {(a)}; 
	\end{axis}
	\begin{axis}[clip=false,
	height=0.26\textwidth,width=.33\textwidth, xshift=4.5cm,yshift=0cm,	ylabel shift= -1 pt,
	xtick={
		0, 0.5233, 1.0467, 1.5708, 2.093, 2.617, 3.14
	},
	xticklabels={
		0, $\frac{\pi}{6}$, $\frac{\pi}{3}$, $\frac{\pi}{2}$,	$\frac{2\pi}{3}$, $\frac{5\pi}{6}$, $\pi$
	},
	xlabel shift= -5pt,
	ylabel style={rotate=-90},
	extra y tick labels = ,
	extra y tick style  = { grid = major }, 
	xlabel=$ \theta$, ylabel= { $ \mathcal{K} $},%
	, xmin=0, xmax=3.14, ymin=0, ymax=1 , thick
	,	legend style={draw=none,at={(1,0.4)},anchor=south east},	legend style={nodes={scale=0.7, transform shape}}
	]
	\addplot[line width=1pt, blue,dashed] table[x=x, y=k16] {K.txt};
	\addplot[line width=1pt, black, densely dotted] table[x=x, y=k8] {K.txt};
	\addplot[line width=0.8pt, magenta] table[x=x, y=k4] {K.txt};
    \legend{$ A_{16}=0.523 $,$ A_{8}=0.525 $, $ A_{4}=0.553 $} 
	\node[scale=1] at (axis cs: -0.9,1) {(b)}; 
	\end{axis}
	\begin{axis}[clip=false,
	height=0.26\textwidth,width=.33\textwidth, xshift=9cm,yshift=0cm,	ylabel shift= -12 pt,
	xtick={
		0, 0.5233, 1.0467, 1.5708, 2.093, 2.617, 3.14
	},
	xticklabels={
		0, $\frac{\pi}{6}$, $\frac{\pi}{3}$, $\frac{\pi}{2}$,	$\frac{2\pi}{3}$, $\frac{5\pi}{6}$, $\pi$
	},
	xlabel shift= -5pt,
	ylabel style={rotate=-90},
	extra y tick labels = ,
	extra y tick style  = { grid = major }, 
	xlabel=$ \theta_{c}$, ylabel= { ${U_{\lambda_1}}$},%
	, xmin=0, xmax=3.14, ymin=-0.0105, ymax=0.0105 , thick
	,	legend style={draw=none,at={(1,0.2)},anchor=south east},	legend style={nodes={scale=0.8, transform shape}}
	]
	\addplot[line width=1pt, blue,dashed,smooth] table[x=x, y=VEGam16] {U_Lambda_Gamma.txt};
	\addplot[line width=1pt, black, densely dotted,smooth] table[x=x, y=VEGam8] {U_Lambda_Gamma.txt};
	\addplot[line width=0.8pt, magenta,smooth] table[x=x, y=VEGam4] {U_Lambda_Gamma.txt};
	\addplot[line width=0.1pt, black]  coordinates { (0,0) (3.14,0) }; 
	\node[scale=1] at (axis cs: -0.9, 0.011) {(c)}; 
	\end{axis}
	
	\begin{axis}[clip=false,
	height=0.26\textwidth,width=.33\textwidth, xshift=0cm,yshift=-3.4cm,	ylabel shift= -13pt,
	xtick={
		0, 0.5233, 1.0467, 1.5708, 2.093, 2.617, 3.14
	},
	xticklabels={
		0, $\frac{\pi}{6}$, $\frac{\pi}{3}$, $\frac{\pi}{2}$,	$\frac{2\pi}{3}$, $\frac{5\pi}{6}$, $\pi$
	},
	ytick={
		-0.1, 0, 0.1
	},
	yticklabels={
		-0.1, 0, 0.1
	},
	xlabel shift= -5pt,
	ylabel style={rotate=-90},
	extra y tick labels = ,
	extra y tick style  = { grid = major }, 
	xlabel=$ \theta_{c}$, ylabel= {\large $\frac{U_{\lambda\,1}}{U_{\lambda\,0}}$ \normalsize},%
	, xmin=0, xmax=3.14, ymin=-0.1, ymax=0.1 , thick
	,	legend style={draw=none,at={(1,0.2)},anchor=south east},	legend style={nodes={scale=0.8, transform shape}}
	]
	\addplot[line width=1pt, blue,dashed,smooth] table[x=x, y=RatioGam16] {U_Lambda_Gamma.txt};
	\addplot[line width=1pt, black, densely dotted,smooth] table[x=x, y=RatioGam8] {U_Lambda_Gamma.txt};
	\addplot[line width=0.8pt, magenta,smooth] table[x=x, y=RatioGam4] {U_Lambda_Gamma.txt};
	\addplot[line width=0.1pt, black]  coordinates { (0,0) (3.14,0) }; 
	\node[scale=1] at (axis cs: -1, 0.10) {(d)}; 
	\end{axis}
	\begin{axis}[clip=false,
	height=0.26\textwidth,width=.33\textwidth, xshift=4.6cm,yshift=-3.4cm,	ylabel shift= -7pt,
	xtick={
		0, 0.5233, 1.0467, 1.5708, 2.093, 2.617, 3.14
	},
	xticklabels={
		0, $\frac{\pi}{6}$, $\frac{\pi}{3}$, $\frac{\pi}{2}$,	$\frac{2\pi}{3}$, $\frac{5\pi}{6}$, $\pi$
	},
	ytick={
		-1,0,1
	},
	yticklabels={
		-1,0,1
	},
	xlabel shift= -5pt,
	ylabel style={rotate=-90},
	extra y tick labels = ,
	extra y tick style  = { grid = major }, 
	xlabel=$ \theta_{c}$, ylabel= {{\large $\frac{U_{B}}{U_{\lambda {0}}}$}},%
	, xmin=0, xmax=3.14, ymin=-1, ymax=+1 , thick
	,	legend style={draw=none,at={(1,0.2)},anchor=south east},	legend style={nodes={scale=0.8, transform shape}}
	]
	\addplot[line width=1pt, blue,dashed] table[x=x, y=y] {UB_SOF.txt};
	\addplot[line width=1.1pt, black,densely dotted] table[x=x, y=gam8] {UB_SOF_gamma.txt};
	\addplot[line width=0.8pt, magenta] table[x=x, y=gam4] {UB_SOF_gamma.txt};
	\addplot[line width=0.1pt, black]  coordinates { (0,0) (3.14,0) }; 
	\node[scale=1] at (axis cs: -1, 1.1) {(e)}; 
	\end{axis}
	\begin{axis}[clip=false,
	height=0.26\textwidth,width=.33\textwidth, xshift=9cm,yshift=-3.4cm,	ylabel shift= -6pt,
	xtick={
		0, 0.5233, 1.0467, 1.5708, 2.093, 2.617, 3.14
	},
	xticklabels={
		0, $\frac{\pi}{6}$, $\frac{\pi}{3}$, $\frac{\pi}{2}$,	$\frac{2\pi}{3}$, $\frac{5\pi}{6}$, $\pi$
	},
	ytick={
		0, 2,4,6
	},
	yticklabels={
		0, 2,4,6
	},
	xlabel shift= -5pt,
	ylabel style={rotate=-90},
	extra y tick labels = ,
	extra y tick style  = { grid = major }, 
	xlabel=$ \theta_{c}$, ylabel= {{\large $| \frac{U_{B}}{U_{\lambda {1}}}| $}},%
	, xmin=0, xmax=3.14, ymin=0, ymax=+7, thick
	,	legend style={draw=none,at={(1,0.2)},anchor=south east},	legend style={nodes={scale=0.8, transform shape}}
	]
		\addplot[line width=0.8pt, magenta,smooth] table[x=x, y=gam4] {UB_UL1_gamma.txt};\label{gam4}
	\node[scale=1] at (axis cs: -1, 8.1) {(f)}; 
	\addplot[line width=1.1pt, blue,dashed,smooth] table[x=x, y=gam16] {UB_UL1_gamma.txt};;\label{gam16}
	\addplot[line width=1.1pt, black,densely dotted,smooth] table[x=x, y=gam8] {UB_UL1_gamma.txt};\label{gam8}
	\end{axis}
	
	\end{tikzpicture}
	\caption{Analysis of different components of swimming velocity for three different transition parameters: $ \zeta=4 $ (\ref{gam4}), $ \zeta=8 $ (\ref{gam8}), $ \zeta=16 $ (\ref{gam16}). (a) Variation of $ U_{\lambda_{0}} $ with different surface coverages. (b) Variation of activity $ \mathcal{K} $ with polar angle $ \theta $. (c) Variation of $ U_{\lambda_{1}} $ for different surface coverages. (d) Contribution from the slip modification to the swimming velocity for different surface coverages. (e) Bulk stress contribution to the swimming velocity for different surface coverages. (f) Comparison of (e) with (d) (i.e. $ U_{B} $ with $ U_{\lambda_{1}} $) for different surface coverages. Parameters: $ De=0.1 $, $ \delta=-0.5 $, $ \Phi_{0}=-1 $.} 
	\label{fig:LRT_SOF_gamma}
\end{figure}

Fig. \ref{fig:LRT_SOF_gamma} shows the effect of transition parameter ($ \zeta $) on the components of swimming velocity for various surface coverages.
As $ \zeta $ decreases the Newtonian swimming velocity reduces (fig.\ref{fig:LRT_SOF_gamma}a) because the slip velocity decreases (shown in fig.\ref{fig:smooth_gamma}). 
However, for $ \theta_{c}<\pi/4 \; \&\; \theta_{c}>3\pi/4 $, $ U_{\lambda_{0}} $ is greatest for $ \zeta=4 $. 
This is because, for such coverages, the total area of catalytic activity is more for $ \zeta=4 $ than steeper activity transitions ($ \zeta=8,16 $). 
Fig.\ref{fig:LRT_SOF_gamma}(b) shows the area under the $ \mathcal{K}-\theta $ curve for $ \zeta=4,8,16 $, representing the total area of activity for $ \theta_{c}=\pi/6 $. 
For $ \pi/4 < \theta_{c} < 3\pi/4 $, the area of catalytic activity is independent of $ \zeta $ and thus a direct comparison can be made.
Fig.\ref{fig:LRT_SOF_gamma}(c),(d) show that the contribution from $ U_{\lambda_{1}} $ reduces as $ \zeta $ decreases due to significant reduction in slip velocity. Similarly, the contribution from bulk polymeric stresses reduces as $ \zeta $ decreases, as shown in fig.\ref{fig:LRT_SOF_gamma}(e),(f).

\section{Active particle in a shear-thinning fluid}


The equations governing a weak shear-thinning flow are described by
\begin{equation}
\bnabla^{*} \bcdot \IB{u}^{*}=0, \quad 	-\bnabla^{*} p^{*} + \bnabla^{*} \bcdot \left( \mu^{*} \mathsfbi{A}^{*} \right) -C^{*} \bnabla^{*} \mathcal{P}^{*} =\IB{0}.
\end{equation}
The viscosity follows a general non-linear relation with respect to Newtonian shear-rate ($ \gamma_{0} $) \citep{khair2012coupling}:
$\mu^{*}(\gamma_{0})=\mu^{*}_{0} \left({1+\chi \mu_{1}(\gamma_{0})}\right)$. Here, $ \chi $ is a small parameter which represents the viscosity ratio: $ (\mu_{0}^{*} - \mu_{\infty}^{*})/{\mu_{\infty}^{*}} $; $ \mu_{0}^{*} $ is the zero shear rate viscosity and $ \mu_{\infty}^{*} $ is the infinite shear rate viscosity.
The variables are non-dimensionalized as in \S 2.
Since the concentration field is decoupled from the hydrodynamics (provided $ \epsilon^{2} Pe \ll 1 $), the solution to concentration field is unchanged and is given by (\ref{innerConcSol}).

\subsection{Evaluation of the diffusio-osmotic slip}
Similar to \S2.1, we use matched asymptotic expansion (assuming $ Pe \ll \epsilon^{-2} $ and $ \epsilon \ll1 $) and perturbation expansion (in $ \epsilon $) to obtain the leading order slip for a generalized weakly shear-thinning fluid.
\subsubsection{Outer region}

At the leading order, the equations in the outer region are:
\begin{subequations}\label{ST_outerGE}
	\begin{gather}
	\bnabla  \bcdot \IB{u}^{(0)} = 0,  \quad	 -  \bnabla p^{(0)} + {\mu(\gamma_{0})} \nabla^{2}   \IB{u}^{(0)}  +  \bnabla {\mu(\gamma_{0})} \bcdot \bten{A}  = \IB{0}, 
	\end{gather}
\end{subequations}
subject to the following boundary conditions
\begin{equation}
\IB{u}^{(0)} \rightarrow \IB{0} , \; p^{(0)} \rightarrow 0  \quad \mbox{ as\ } y \rightarrow \infty.  \label{ST_outerBC}
\end{equation}
Here $ \mu(\gamma_{0}) = \mu^{*} / \mu^{*}_{0} = 1+\chi \mu_{1}(\gamma_{0}) $.

\subsubsection{Inner region}

We rescale the variables (similar to \S 2.1.2) and obtain the governing equations in the inner region (at the leading order) as:
\begin{subequations} \label{ST_innerGE}
	\begin{gather}
	\frac{\p \hat{u}^{(0)}}{\p \hat{x}} + \frac{\p \hat{v}^{(0)}}{\p \hat{y}} =0 , \label{ST_innerGE:CE}\\
	- \frac{\p \hat{p}^{(0)}}{\p \hat{x}} +  {\mu(\gamma_{0})} \frac{\p^2 \hat{u}^{(0)}}{\p \hat{y}^2} +  \frac{\p \mu(\gamma_{0})}{\p \hat{y}} \frac{\p \hat{u}^{(0)}}{\p \hat{y}}    =  (\hat{c}^{(0)} + C_{\infty})\frac{\p \hat{\psi}^{(0)}}{\p \hat{x}}  ,  \label{ST_innerGE:NSx}\\
	- \frac{\p \hat{p}^{(0)}}{\p \hat{y}}  =  (\hat{c}^{(0)}+C_{\infty}) \frac{\p \hat{\psi}^{(0)}}{\p \hat{y}}    ,  \label{ST_innerGE:NSy}
	\end{gather}
\end{subequations}
subject to the pressure decay and no-slip boundary condition
\begin{equation}\label{ST_innerBC}
\hat{p} \rightarrow 0 \; {\rm{as}} \; \hat{y} \rightarrow 0; \quad  \left.	\hat{u}^{(0)} \right\vert_{\hat{y}=0} = 0.
\end{equation}

Assuming that the difference between zero shear and infinite shear viscosity is small (i.e. $ \chi \ll 1 $), we perform a regular perturbation expansion in $ \chi $ and represent the velocity and pressure fields as
\begin{equation}\label{ST_pert}
f^{(i)}(x,y)=f^{(i)}_{0}(x,y)+\chi \, f^{(i)}_{1}(x,y)+\cdots  .
\end{equation}
Substituting the perturbed field variables in (\ref{ST_innerGE}-\ref{ST_innerBC}) gives the equations governing the system at $ O(1) $ and $ O(\chi) $. 

At $ O(1) $, the equations and their solutions are identical to (\ref{innersol}). The hydrodynamics at $ O(\chi) $ is governed by
\begin{subequations}\label{STinnerGE}
	\begin{gather}
	\frac{\partial \hat{u}_{1}^{(0)}}{\partial \hat{x}} + \frac{\partial \hat{v}_{1}^{(0)}}{\partial \hat{y}} =0 , \label{STinnerGE:CE}\\
	- \frac{\partial \hat{p}_{1}^{(0)}}{\partial \hat{x}} +  \frac{\partial^2 \hat{u}_{1}^{(0)}}{\partial \hat{y}^2}  +   \mu_{1}(\gamma_{0})  \frac{\partial^2 \hat{u}_{0}^{(0)}}{\partial \hat{y}^2} +  \frac{\partial \mu_{1}(\gamma_{0})}{\partial \hat{y}}  \frac{\p \hat{u}_{0}^{(0)}}{\p \hat{y}}  =  0  ,  \label{STinnerGE:NSx}\\
	- \frac{\partial \hat{p}_{1}^{(0)}}{\partial \hat{y}}  =  0. \label{STinnerGE:NSy}
	\end{gather}
\end{subequations}
Integrating (\ref{STinnerGE:NSy}) and using the pressure decay condition, we get $ \hat{p}_{1}^{(0)} = 0 $. 
(\ref{STinnerGE:NSx}) thus yields
\begin{equation}
	\frac{\partial^{2} \hat{u}_{1}^{(0)}}{\partial \hat{y}^{2}} = - \frac{\partial}{\partial \hat{y}} \left( \mu_{1}(\gamma_{0}) \frac{\partial \hat{u}_{0}^{(0)}}{\partial \hat{y}}   \right)
\end{equation}
Integrating the above equation twice gives
\begin{equation}\label{STinnersol}
	\hat{u}_{1}^{(0)} = \mathcal{I}'(x) \int_{0}^{\hat{y}} \mu_{1}(\gamma_{0}) \int_{t}^{\infty} \mathcal{F}(s) {\rm d}s \, {\rm d}t   - \mathcal{J}_{0}(\hat{x}) \int_{0}^{\hat{y}} \mu_{1} (\gamma_{0}) {\rm d}s \;  +  \mathcal{J}_{1}(\hat{x}) \hat{y},
\end{equation}
where $ \mathcal{F}(s) =  -1 + e^{-\hat{\psi}(s)}   $.

\subsubsection{Matching}
As in \S 2.1.3, upon matching the inner and outer solutions, we find that $ \mathcal{J}_{0}=\mathcal{J}_{1}=0 $ and obtain:
\begin{equation}
\left. u^{(0)} \right\vert_{y=0} = \left. c^{(0)}_{x}\right\vert_{y=0} \left( M_{0} + \chi  \, M_{1}  \right),
\label{FINALSTSol}
\end{equation}
where $ M_{0} $ is the mobility coefficient representing the Newtonian contribution and is expressed in (\ref{FINALDeSol}). The expression for $ M_{1} $ is
\begin{equation}
 M_{1} = {\displaystyle\int_{0}^{\infty} } \mu_{1}\left(\gamma_{0}^{(0)}\right) \int_{t}^{\infty}  \mathcal{F}(s)  {\rm d}s \, {\rm d}t  .
\label{ST_M1}
\end{equation}
Since the power-law model diverges for asymptotically small shear-rates \citep{bird1987dynamics}, we choose the Carreau model to capture viscosity variation:
$\mu_{1}=\left( 1+ \tau^{*\,2} |\gamma_{0}^{*}|^{2}  \right)^{\frac{n-1}{2}}   -   1 $ \citep{bird1987dynamics}.
Here $ \tau^{*} $ is the fluid relaxation time scale and $ n $ characterizes the degree of shear-thinning ($ n<1 $).  For Carreau fluid, the non-Newtonian mobility coefficient ($ M_{1} $) is
\begin{equation}
M_{1}  =  	{\displaystyle\int_{0}^{\infty} } \left\lbrace \left[ 1    +  \left( C\!u_{\lambda} \, \left. c_{x}^{(0)}\right\vert_{y=0}   \int_{t}^{\infty} \mathcal{F}(s) {\rm{d}}s \right)^{2} \right]^{\frac{n-1}{2}} -1  \right\rbrace  \int_{t}^{\infty} \mathcal{F}(s) {\rm d}s \, {\rm d}t  ,
\label{FINALSTSolFINAL}
\end{equation}
where $ C\!u_{\lambda} $ (based on shear in the interaction layer) is the ratio of timescales associated with relaxation ($ \tau^{*} $) to that of shear in the flow ($ \lambda_{I}^{*}/U_{ch} $): $ C\!u_{\lambda} = \tau^{*}/(\lambda_{I}^{*}/U_{ch}) $. 
This is different than the Carreau number corresponding to shear in the bulk fluid (i.e. outer region): $ C\!u_{B} = \tau^{*}/(a^{*}/U_{ch}) $.
The ratio of the two Carreau numbers ($ C\!u_{\lambda}/C\!u_{B} = a^{*}/\lambda_{I}^{*} $) provides the comparison of shear rate inside to that outside the interaction layer i.e. $ \sim O(\epsilon^{-1}) $.

\begin{figure}
	\centering
	{{\includegraphics[scale=0.60]{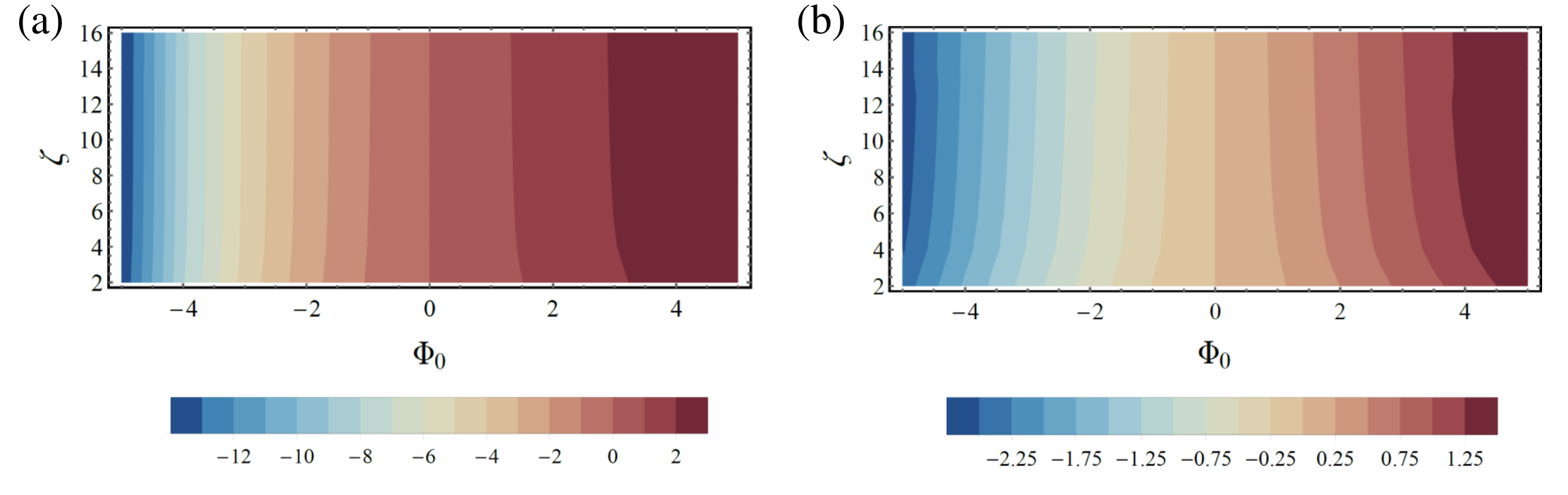} }}%
	\caption{Non-Newtonian mobility ($ M_{1} $) at $ \theta=\pi/2 $ for two interaction potentials (a) exponential-type and (b) van der Waals-type. Other parameters: $ n=0.25 $, $ C\!u_{\lambda}=10^{2} $, $ \theta_{c}=\pi/2 $.}
	\label{fig:ST_mobility}
\end{figure}

\subsection{Diffusio-osmotic slip on an active particle}
As in \S2.3, we now apply the above results  to an axi-symmetric Janus sphere.
Since the shear in the thin interaction layer is generally high (i.e. $ U_{ch}/\lambda_{I}^{*} \sim 10^{2} s^{-1} $) and relaxation time scales of the biological fluids are $ \sim 1s $ \citep{zare2019analysis}, large Carreau numbers are possible ($ C\!u_{\lambda} \sim 10^{2} $).
Contrary to the second-order fluid, $ M_{1} $ for Carreau fluid depends both on the magnitude of interaction ($ \Phi_{0} $) as well as the concentration gradient (determined by the activity transition parameter $ \zeta $). 
Since the concentration field varies in tangential direction, $ M_{1} $ varies tangentially. 
Thus, we analyze the effect of $ \Phi_{0}$ and $ \zeta $ on $ M_{1} $ for $ \theta=\theta_{c}=\pi/2 $ in fig.\ref{fig:ST_mobility}. 
For both short-range exponential and long-range van der Waals interactions, $ M_{1} $ is a weak function of $ \zeta $. $ M_{1} $ for Carreau fluid is of different signs for attractive ($ \Phi_{0}<0 $) and repulsive interactions ($ \Phi_{0}>0 $), as opposed to the case of second-order fluid (fig.\ref{fig:Mobility}b).

\begin{figure}
	\centering
	{{\includegraphics[scale=0.23]{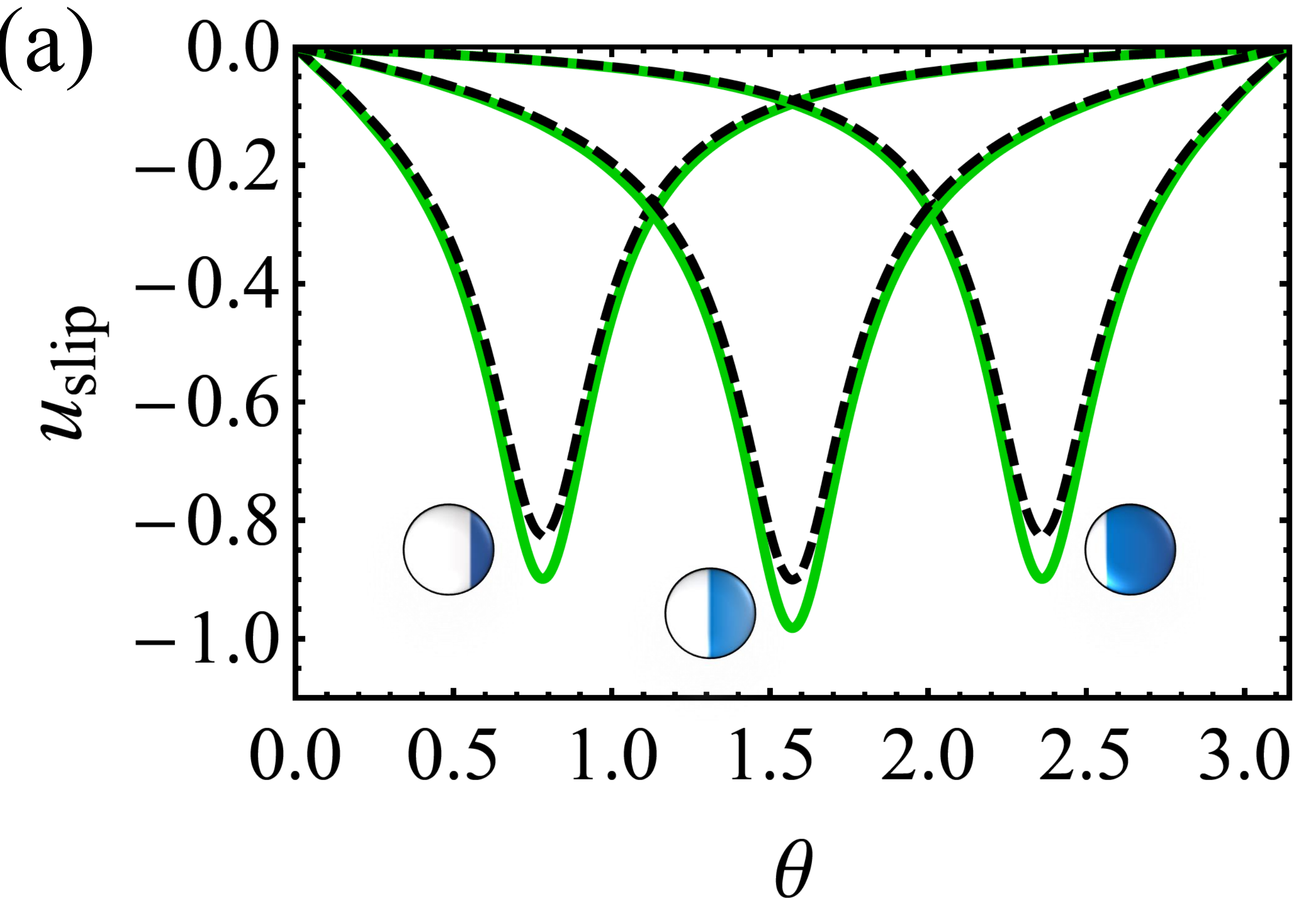} }}%
	$ \; \; \; $
	\begin{tikzpicture}
	\pgfplotsset{
		set layers,
		domain=-1:1,
		tick label style={
			/pgf/number format/.cd,
			fixed, fixed zerofill,
			precision=1,
			/tikz/.cd,
		},
	}
	\begin{axis}[clip=false,
	height=0.35\textwidth,width=.45\textwidth,
	xlabel=$ {\small C\!u_{\lambda}}$, ylabel= {\large $\frac{U_{\lambda\,1}}{U_{\lambda\,0}}$ \normalsize} , xshift=0cm,yshift=0cm,	ylabel shift= -5 pt, 	ylabel style={rotate=-90}, xmin=0, xmax=100, ymin=0, ymax=0.1,
	ytick={
		0, 0.02, 0.04, 0.06, 0.08, 0.10
	},
	yticklabels={
		0, 0.02, 0.04, 0.06, 0.08, 0.10
	},
	xtick={
		0, 20, 40, 60, 80, 100
	},
	xticklabels={
		0, 20, 40, 60, 80, 100
	}
	,	legend style={draw=none,at={(0.9,0.07)},anchor=south east},	legend style={nodes={scale=0.75, transform shape}}
	]
	\addplot[line width=1pt, black] table[x=x, y=y2] {ST_LRT.txt};
	\addplot[line width=1pt, black, dashed] table[x=x, y=y1] {ST_LRT.txt};
	\node[scale=1] at (axis cs: -27, 0.1) {(b)}; 
	%
	\end{axis}
	\begin{semilogxaxis}[
	height=0.20\textwidth,width=.28\textwidth, xshift=2cm,yshift=0.7cm,	ylabel shift= -12 pt,
	ytick={
		0.9,1
	},
	yticklabels={
		0.9,1
	},
	xtick={
		0.1, 10
	},
	xticklabels={
		$ 0.1 $, $ 10 $
	},
	xlabel shift= -8pt,
	ylabel style={rotate=-90},
	extra y tick labels = ,
	extra y tick style  = { grid = major }, 
	xlabel=\small $ { C\!u_{\lambda}} $, ylabel= \small {${ \mu}$},%
	, xmin=0.01, xmax=100, ymin=0.89, ymax=1.02 , thick
	,	legend style={draw=none,at={(1,0.2)},anchor=south east},	legend style={nodes={scale=0.8, transform shape}}
	]
	\addplot[line width=0.9pt, blue] table[x=x, y=y] {CY.txt};
	\end{semilogxaxis}
	\end{tikzpicture}
	
	\begin{tikzpicture}
	\begin{semilogxaxis}[clip=false,
	height=0.32\textwidth,width=.45\textwidth,
	xlabel=$ {\small C\!u_{B}}$, ylabel= {\large $\frac{U_{B}}{U_{\lambda\,0}}$ \normalsize} , xshift=0cm,yshift=0.8cm,	ylabel shift= -4 pt, 	ylabel style={rotate=-90}, xmin=0.02, xmax=100, ymin=-0.020, ymax=0,
	legend style={draw=none,at={(0.9,0.07)},anchor=south east},	legend style={nodes={scale=0.75, transform shape}}
	]
	\addplot[line width=0.8pt, black,smooth] table[x=x, y=PB2Gam16] {ST_LRT_UB_Gamma.txt};\label{pb2}
	
	\addplot[line width=1pt, black, dashed,smooth] table[x=x, y=PB4Gam16] {ST_LRT_UB_Gamma.txt};\label{pb4}

	\node[scale=1] at (-6, 200) {(c)}; 
	\node[scale=1] at (+6.48, 200) {(d)}; 
	\end{semilogxaxis}
	\begin{semilogxaxis}[
	height=0.32\textwidth,width=.45\textwidth,
	xlabel=$ \epsilon $, ylabel= {\large $\frac{U_{B}}{U_{\lambda_{1}}}$ \normalsize} , xshift=6.6cm,yshift=0.8cm,	ylabel shift= -4 pt, 	ylabel style={rotate=-90}, xmin=0.0002, xmax=0.01, ymin=-0.24, ymax=0,
	,	legend style={draw=none,at={(0.9,0.07)},anchor=south east},	legend style={nodes={scale=0.75, transform shape}}
	]
	\addplot[line width=0.8pt, black,smooth] table[x=x, y=INV_PB2ratio] {ST_UB_UL1_comparison.txt};
	
	\addplot[line width=1pt, black, dashed,smooth] table[x=x, y=INV_PB4ratio] {ST_UB_UL1_comparison.txt};
	\end{semilogxaxis}
	\end{tikzpicture}
	\caption{(a) The dashed  black line represents the Newtonian slip velocity. The solid green line represents the variation of total slip velocity for a shear thinning fluid along the polar angle for $ \hat{\psi}(\xi)=- e^{-\xi} $ and $ C\!u_{\lambda}=100 $. 
		(b) Variation of non-Newtonian contribution to the swimming velocity (arising from the surface slip) with $ C\!u_{\lambda} $, for three different surface coverage ($ \theta_{c} $) and $ \hat{\psi}(\xi) = -e^{-\xi} $. Solid (\ref{pb2}) and dashed lines (\ref{pb4}) correspond to $ \theta_{c}=\pi/2 $ and $ \theta_{c}=\pi/4 $ or $ 3\pi/4 $, respectively. Inset showing the variation in non-dimensional viscosity $ \mu $ with $ C\!u_{\lambda} $ at $ y=0 $ and $ \theta=\theta_{c}=\pi/2 $. (c) Variation of non-Newtonian contribution to the swimming velocity (arising from the bulk stresses) with $ C\!u_{B} $. The convergence was obtained by accounting for first 15 modes. (d) Comparison between the two non-Newtonian components of swimming velocity for $ C\!u_{\lambda}=10^{2} $.
		Other parameters: $ \chi=0.1 $, $ n=0.25 $, $ \zeta=16 $.}
	\label{fig:slipST}
\end{figure}
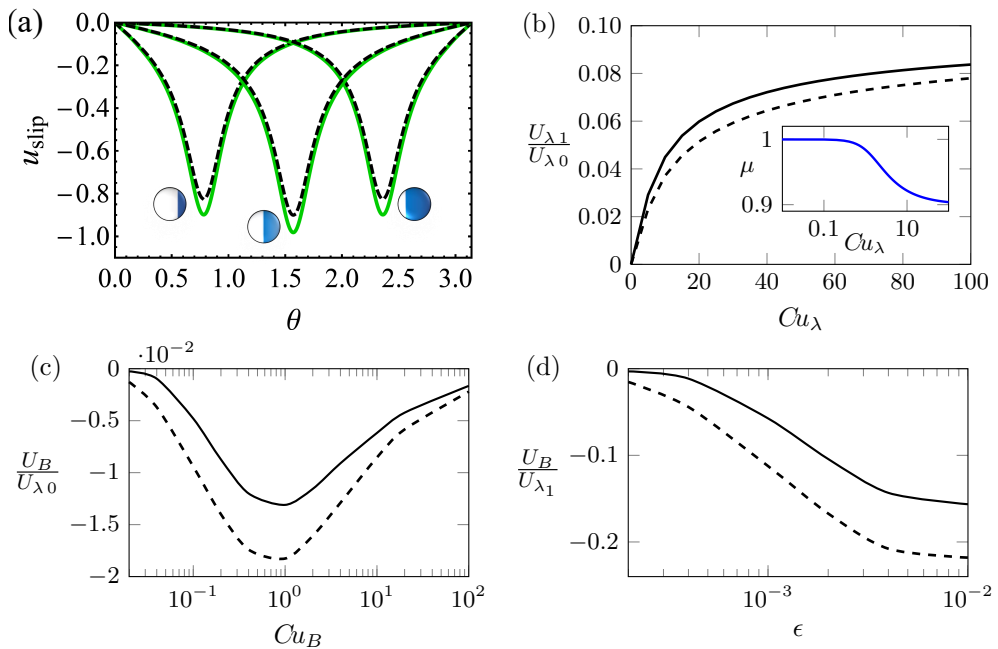

Using (\ref{concGold}), (\ref{sigmoid}), (\ref{FINALSTSol}), and (\ref{FINALSTSolFINAL}), we numerically evaluate the slip for shear-thinning fluid ($ n=0.25 $) and depict its variation along the surface of Janus particle in fig. \ref{fig:slipST} (a).
For large Carreau numbers ($ C\!u_{\lambda} \sim 10^{2} $) and attractive interactions, we find a marginal increase in the total slip velocity. For repulsive interactions, the enhancement in slip velocity is qualitatively similar (see Appendix A).
The modification in slip (in both attractive and repulsive cases) is maximum at the transition of activity, depicting a local symmetry about $ \theta_{c} $.

\subsection{Swimming velocity}
We now evaluate the swimming velocity using the reciprocal theorem \citep{stone1996propulsion,khair2012coupling}:
\begin{equation}
U=\frac{-1}{4\pi} \int_{S_{p}} \left.\left( \IB{u}_{0} + \chi \,\IB{u}_{1}  \right)\right\vert_{r=1} \bcdot \IB{e}_{z} \,  {\rm d}S \; - \; \frac{1}{6\pi} \chi \int_{V_{f}}  \mu_{1} \left( \gamma_{0} \right)  \bten{A}_{0} \IB{:} \bnabla{\IB{u}^{t}} {\rm d}V. \label{LRT_ST}
\end{equation}
As in \S \ref{VEswim}, the first integral contains the Newtonian component $ U_{\lambda_0} $ (given by eq.\ref{ULambda}) and non-Newtonian component $ U_{\lambda_1} $:
\begin{equation}\label{ST_ULambda1}
	U_{\lambda_{1}} = \frac{\chi}{2} \int_{0}^{\pi} M_{1}(\theta) c_{\theta}(1,\theta) \sin^{2} \theta {\rm{d}}\theta. 
\end{equation}
Substituting (\ref{FINALSTSolFINAL}) in (\ref{LRT_ST}) and integrating numerically, we find that the contribution from $ U_{\lambda_1} $ modestly affects the swimming of a Janus particle. This is shown in figure \ref{fig:slipST}(b) for three different surface coverages.
The positive sign of the ratio shows that $ U_{\lambda_1} $ adds to the contribution arising from the Newtonian slip ($ U_{\lambda_{0}} $). 
As $ C\!u_{\lambda} $ increases, the viscosity in the interaction layer decreases, which results in a faster diffusio-osmotic flow. However, this enhancement reaches a plateau as the viscosity reduction stagnates at high $ C\!u_{\lambda} $ (see inset in fig.\ref{fig:slipST}b).
For repulsive interactions, this enhancement in swimming velocity ($ U_{\lambda_{1}} $) is qualitatively similar (see Appendix A); shear thinning enhances the swimming velocity, irrespective of the interaction between solute molecules and the particle.

The second integral ($ U_{B} $) accounts for the bulk stresses arising from viscosity variations. 
The modification due to bulk non-Newtonian stresses is quantified in fig.\ref{fig:slipST}(c) for three different surface coverages\footnote{We follow the approach of \cite{blake1971spherical, datt2017activeComplex} to find $ U_{B} $ for smooth activity. The details of implementation and reproduced results (for step activity) are provided in the supplementary material.}.
The negative sign of $ U_{B}/U_{\lambda_{0}} $ denotes that the bulk stresses always reduce the swimming velocity, for all surface coverages (also reported by \cite{datt2017activeComplex} for step activity). 
The non-monotonic behavior is consistent with the fact that a viscosity, following Carreau model, reduces to a Newtonian fluid of a lower viscosity ($ \mu_{\infty}^{*} $) as $ C\!u_{B} \rightarrow \infty $. As a result, the bulk non-Newtonian stresses are maximum at intermediate $ C\!u_{B} $ and vanish at high $ C\!u_{B} $ \citep{khair2012coupling,datt2015squirmingST,datt2017activeComplex}.
Fig.\ref{fig:slipST}(d) compares the contribution from the two non-Newtonian components to swimming velocity i.e. $ U_{B} $ and $ U_{\lambda_{1}} $. 
We take $ C\!u_{\lambda}=10^{2} $, as shear rate in the interaction layer is generally high. This yields $ C\!u_{B} = (\epsilon) 10^{2} $.
Thus, in fig.\ref{fig:slipST}(d), we compare the two components, keeping $ \epsilon $ as an independent variable. Since the current analysis is valid for asymptotically thin interaction layers, we vary $ \epsilon $ from $ 10^{-4} $ to $ 10^{-2} $.
Physically, $ C\!u_{\lambda}/C\!u_{B} = \epsilon^{-1} $ corresponds to the disparity in shear rates in the inner and outer regions.
 For $ \epsilon \sim 10^{-4} $, the shear rate inside the interaction layer is very high and thus the contribution from $ U_{\lambda_{1}} $ dominates that from $ U_{B} $.
 As $ \epsilon $ increases (i.e. thickness of interaction layer), the shear in the interaction layer decreases, which increases the magnitude of $ U_{B}/U_{\lambda_{1}} $. This continues until $ U_{B} $ reaches a plateau.\footnote{The above results are for $ \zeta=16 $. Appendix A shows the effect $ \zeta $ on the swimming velocity.}

The contribution from both non-Newtonian components is modest in magnitude. In context of electrophoresis, \cite{khair2012coupling} also reported a similar enhancement and trends in a shear-thinning medium. Analogous to their study, our result (\ref{FINALSTSol}) also reveals that the shear-thinning modification of the slip does not add a size dependency to the motion induced due to self-diffusiophoresis. On the other hand, the modification due to bulk non-Newtonian stresses adds a size dependency (it enters through the bulk Carreau number).


\section{Conclusions}
The results of the current study reveal the effects of fluid rheology on the diffusio-osmotic slip over an active surface and its consequence on the mobility of a Janus particle.
Using matched asymptotic expansions in conjunction with perturbation theory, we derived the modification to slip velocity for a second-order fluid. 
Our result (eq.\ref{FINALDeSol_alt}), for an axisymmetrically active Janus particle, shows that the polymeric stress significantly alters the slip velocity, and is valid for low to moderate advective effects. 
The proportionality to second tangential gradients results in a sharp reversal of the surface slip, triggered by large gradient in polymeric stress across the discontinuity of surface activity. This explains the generation of extensional flow across a step change in activity observed by \cite{SOF_natale2017}. 
An examination of characteristic scales reveals that the localized reversal of slip velocity is a result of mathematical inconsistency: a consequence of employing the activity as a step function, which can be overcome by choosing a smooth (sigmoidal) function. The results reveal that the polymer elasticity tends to generate local elongational flows across the transition of surface activity.
We also explored the effects of attractive and repulsive interaction between the solute molecules and Janus particle.
Using the reciprocal theorem, we found that the modification in the slip has an effect on the swimming velocity which is comparable to that arising from bulk viscoelastic stresses \citep{datt2017activeComplex}.

We further applied the framework to a generalized weakly shear-thinning fluid and obtained the modification to slip velocity (eq.\ref{FINALSTSol}). 
Employing a Carreau-fluid model, we showed that the shear-thinning effects marginally increase the slip velocity, provided the time scale associated with the shear in the interaction layer is asymptotically smaller than the fluid relaxation time. Using the reciprocal theorem, we showed an enhancement in the swimming velocity due to modification in slip. 
For an asymptotically thin interaction layer, this enhancement dominates the retardation caused by the bulk non-Newtonian stress.

Another key implication of our results is that the modification in slip provides an estimate of the hydrodynamic disturbance around an active particle for weak non-Newtonian effects. 
This can be a significant contribution towards accurate modeling of the interaction of two or more active particles in polymeric fluids \citep{rallabandi2019motion,stark2018artificial}. 
Recently, \cite{Pump_Geo_michelin2015SoftMatter,Pump_Patch_michelin2019nature} have designed `phoretic pumps' which can transport fluids without the need of applying pressure difference across the channel. This can be helpful in transportation of biological fluids through narrow channels. The flow in such systems occurs due to local concentration gradients, arising either from geometric variation \citep{Pump_Geo_lisicki2016phoretic,Pump_Geo_michelin2015SoftMatter} or variation in surface activity \citep{Pump_Patch_michelin2019nature}. Since our results are applicable to a general diffusio-osmotic slip, they should be useful to model the flow of complex fluids through such pumps.

The current work and previous investigations corresponding to the active motion in complex fluids \citep{datt2017activeComplex,SOF_natale2017} have assumed the diffusivity of solute molecules to be constant. This might be reasonable in the limit of weak non-Newtonian effects, but may lead to imprecise conclusions for fluid mediums exhibiting strong non-Newtonian behavior, as the Stokes-Einstein equation fails in describing diffusion in complex media. 
Recently, \cite{makuch2020diffusion} devised a relationship between translational and rotational diffusion coefficients which depends on the size of solute. Such theoretical formulation can provide a database for precise description of diffusion in various complex fluids. Such data can be employed in a theoretical framework similar to that devised by \cite{vrentas2003steady} and \cite{tiefenbruck1980note} who studied the effect of diffusion in non-Newtonian flows. Further research in this direction can help in a better understanding of self-propulsion through the mediums which exhibit significant deviations from the Newtonian behavior.
\\ \\ 

 The financial support from Indian Ministry of Human Resource Development is gratefully acknowledged.
The authors would like to thank anonymous referees of this work for their valuable comments and suggesting us to directly compare the two non-Newtonain components of swimming velocity. A.C thanks Sebastian Michelin for pointing out the importance of transition length for sigmoidal activity.
The authors also thank Marco De Corato and Dipin Pillai for their useful suggestions.\\ \\ \\
	\textbf{Declaration of Interests}\\
	 The authors report no conflict of interest.\\ \\ \\

	\vspace{3cm}

  \appendix

  \section{Additional results for the shear-thinning fluid}\label{appB} 
\begin{figure}
 	\centering
 	{{\includegraphics[scale=0.55]{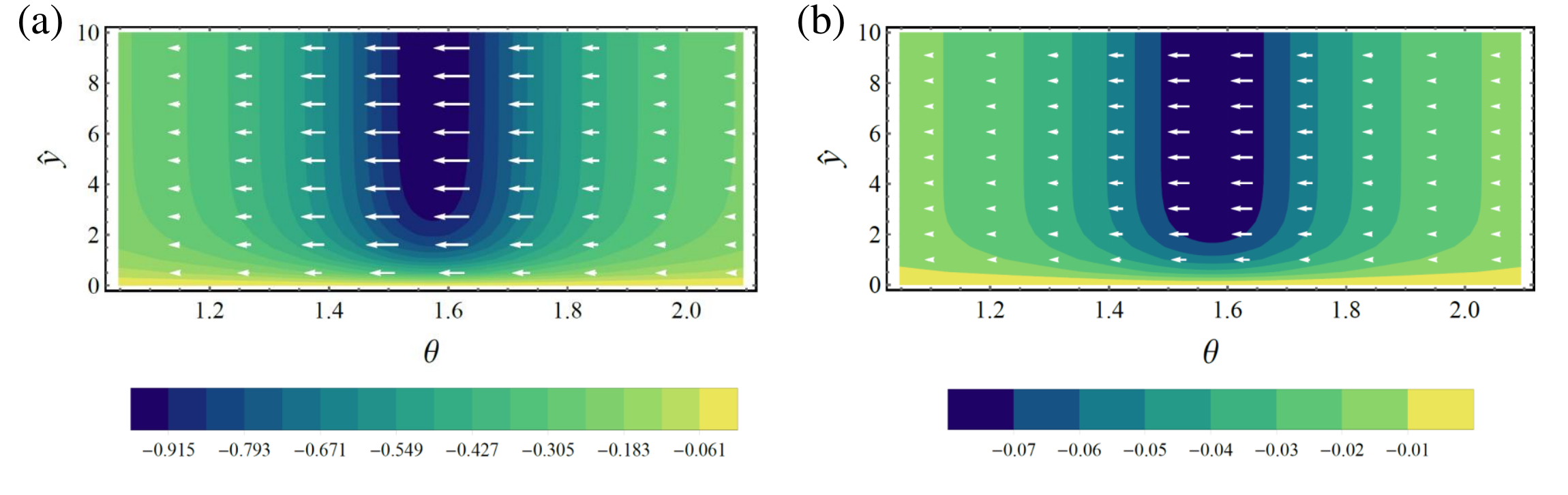} }}%
 	\caption{Tangential velocity contours inside the interaction layer for: (a)Newtonian fluid; (b) a weakly shear-thinning fluid. The arrows depict the velocity field vector $ \hat{\IB{u}}_{1}^{(0)} $.
 		Other parameters: $ \chi=0.1 $, $ n=1/4 $, $\zeta=16$, $ \hat{\psi}(\xi) = -e^{-\xi} $.}
 	\label{fig:ST_inner_vel}
 \end{figure}

Fig.\ref{fig:ST_inner_vel}(a) shows the velocity field inside the thin interaction layer for Carreau fluid and a sigmoidal activity transition (\ref{sigmoid}). The contours bear close resemblance to that for Newtonian fluids \ref{fig:inner_vel_field}(a). Thus, we plot the difference between the two in fig.\ref{fig:ST_inner_vel}(b). 
The local viscosity reduction due to shear-thinning increases the velocity inside the inner region and consequently the slip (as observed in fig.\ref{fig:slipST}(a)). This enhancement is symmetric about $ \theta_{c} $.

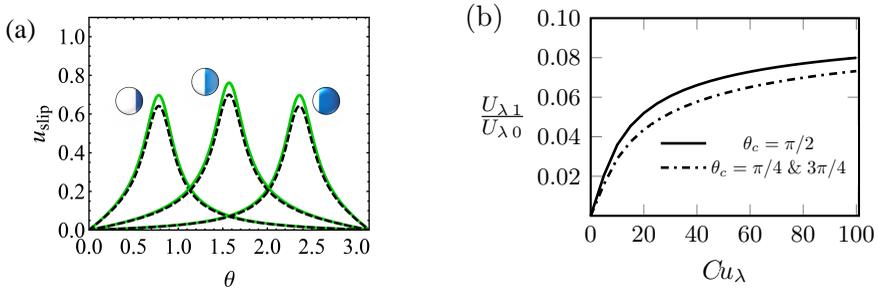
\begin{figure}
	\centering
	{{\includegraphics[scale=0.19]{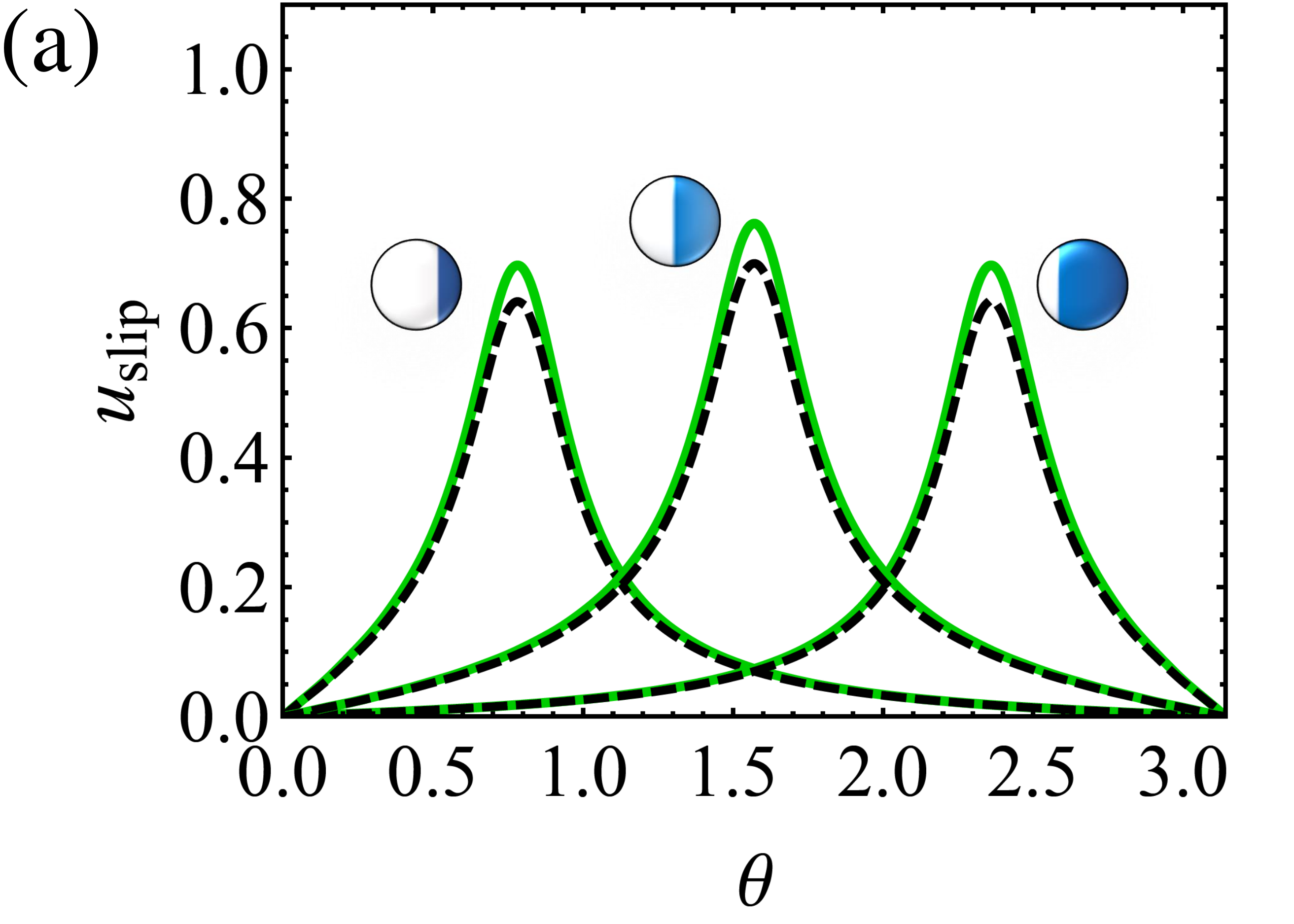} }}%
	$ \; \; \; \quad $
	\begin{tikzpicture}
	\begin{axis}[clip=false,
	height=0.31\textwidth,width=.38\textwidth, xshift=-0.5cm,yshift=-0cm,	ylabel shift= -3 pt,
	ylabel style={rotate=-90},
	ytick={0.02,0.04,0.06,0.08,0.1}, 
	y tick label style={
		/pgf/number format/.cd,
		fixed,
		fixed zerofill,
		/tikz/.cd
	},
	extra y tick labels = ,
	extra y tick style  = { grid = major }, 
	xlabel=$ {\small C\!u_{\lambda}}$, ylabel= {\large $\frac{U_{\lambda\,1}}{U_{\lambda\,0}}$ \normalsize},%
	, xmin=0, xmax=101, ymin=0, ymax=0.1 , thick
	,	legend style={draw=none,at={(1,0.15)},anchor=south east},	legend style={nodes={scale=0.7, transform shape}}
	]
	\addplot[line width=1pt, black] table[x=x, y=piby2] {ST_LRT_Repulsive.txt};
	\addplot[line width=1pt, black, dash dot] table[x=x, y=piby4] {ST_LRT_Repulsive.txt};
	\node[scale=1.1] at (axis cs: -40, 0.1) {(b)}; 
	\legend{$ \theta_{c}=\pi/2 $, $ \theta_{c}=\pi/4 \; \& \; 3\pi/4 $}
	\end{axis}
	\end{tikzpicture}
	\caption{(a) The dashed line represents the Newtonian slip velocity. The solid line represents the variation of total slip velocity along the polar angle for $ \hat{\psi}(\xi)=+ e^{-\xi} $ and $ C\!u_{\lambda}=100 $. 
		(b) Variation of swimming velocity (arising from the surface slip) with $ C\!u_{\lambda} $, for three different surface coverage ($ \theta_{c} $). Other parameters: $ \chi=0.1 $, $ n=1/4 $, $ \zeta=16 $.}
	\label{fig:slipST_app}
\end{figure}

Fig.\ref{fig:slipST_app} shows the results for repulsive exponential interactions between the solute molecules and active surface.
 Fig.\ref{fig:slipST_app}(a) depicts the slip velocity, which is qualitatively reversed in comparison to that depicted for the attractive interactions (see fig. \ref{fig:slipST}(a)).
 The magnitude is lower in this case because the adsorption coefficient is reduced in repulsive interactions \citep{anderson1982motion}.
 Figure \ref{fig:slipST_app}(b) shows that the enhancement caused by $ U_{\lambda_{I}} $ is qualitatively similar to that observed in the case of attractive interactions. As the ratio $ U_{B}/U_{\lambda_{0}} $ is independent of mobility, its profile is identical to fig.\ref{fig:slipST}(c).
 
 
\begin{figure}
	\centering
	\begin{tikzpicture}
	\begin{axis}[clip=false,
	height=0.26\textwidth,width=.31\textwidth, xshift=0cm,yshift=0cm,	ylabel shift= 0 pt,
	xlabel shift= 1pt,
	ylabel style={rotate=-90},
	extra y tick labels = ,
	extra y tick style  = { grid = major }, 
	xlabel=$ C\!u_{\lambda}$, ylabel= \large{ $\frac{U_{\lambda_1}}{U_{\lambda_0}}$} \normalsize,%
	, xmin=0, xmax=100, ymin=0, ymax=0.09 , thick
	,	legend style={draw=none,at={(1,0.2)},anchor=south east},	legend style={nodes={scale=0.8, transform shape}}
	]
	\addplot[line width=1pt, blue,dashed] table[x=x, y=gam16] {ST_LRT_UL1_Gamma.txt};
	\addplot[line width=1pt, black, densely dotted] table[x=x, y=gam8] {ST_LRT_UL1_Gamma.txt};
	\addplot[line width=1pt, magenta] table[x=x, y=gam4] {ST_LRT_UL1_Gamma.txt};
	\node[scale=1] at (axis cs: -20, 0.09) {(a)}; 
	\end{axis}
	\end{tikzpicture}
	\begin{tikzpicture}
	\begin{semilogxaxis}[clip=false,
	height=0.26\textwidth,width=.31\textwidth,
	xlabel=$ {\small C\!u_{B}}$, ylabel= {\large $\frac{U_{B}}{U_{\lambda\,0}}$ \normalsize} , xshift=0cm,yshift=0cm,	ylabel shift= -6 pt, xlabel shift = -3 pt,	ylabel style={rotate=-90}, xmin=0.01, xmax=100, ymin=-0.015, ymax=0,
	ytick={
		-0.015,-0.01, -0.005,0
	},
	yticklabels={
		-0.015,-0.01,-0.005, 0
	}
	,	legend style={draw=none,at={(0.9,0.07)},anchor=south east},	legend style={nodes={scale=0.75, transform shape}}
	]
	\addplot[line width=0.8pt, blue,dashed,smooth] table[x=x, y=PB2Gam16] {ST_LRT_UB_Gamma_alt.txt};
	\addplot[line width=1pt, black,densely dotted,smooth] table[x=x, y=PB2Gam8] {ST_LRT_UB_Gamma_alt.txt};
	\addplot[line width=0.8pt, magenta,smooth] table[x=x, y=PB2Gam4] {ST_LRT_UB_Gamma_alt.txt};

	\node[scale=1] at (-7, 155) {(b)}; 
	\node[scale=1] at (+6.9, 155) {(c)}; 
	\end{semilogxaxis}
	\begin{semilogxaxis}[
	height=0.26\textwidth,width=.31\textwidth,
	xlabel=$ {\epsilon}$, ylabel= {\large $\frac{U_{B}}{U_{\lambda\,1}}$ \normalsize} , xshift=4.1cm,yshift=0cm,	ylabel shift= -6 pt, xlabel shift = -3 pt,	ylabel style={rotate=-90}, xmin=0.0002, xmax=0.01, ymin=-0.2, ymax=0,
	ytick={
		-0.2,-0.1, 0
	},
	yticklabels={
		-0.2,-0.1, 0
	}
	,	legend style={draw=none,at={(0.9,0.07)},anchor=south east},	legend style={nodes={scale=0.75, transform shape}}
	]
	\addplot[line width=1.1pt, blue,dashed,smooth] table[x=x, y=INV_RatioGam16] {ST_UB_UL1_comparison.txt};
	\addplot[line width=1pt, black,densely dotted,smooth] table[x=x, y=INV_RatioGam8] {ST_UB_UL1_comparison.txt};
	\addplot[line width=0.9pt, magenta,smooth] table[x=x, y=INV_RatioGam4] {ST_UB_UL1_comparison.txt};
	%
	\end{semilogxaxis}
	\end{tikzpicture}
	\caption{Analysis of different components of swimming velocity for three different transition parameters: $ \zeta=4 $ (\ref{gam4}), $ \zeta=8 $ (\ref{gam8}), $ \zeta=16 $ (\ref{gam16}). 
		(a) Contribution from the slip modification to the swimming velocity for different surface coverages and $ \hat{\psi}(\xi) = -e^{-\xi} $. (b) Bulk stress contribution to the swimming velocity for different surface coverages. (c) Comparison of two non-Newtonian contributions to the swimming for $ C\!u_{\lambda}=10^{2} $.
		Other parameters: $ \chi=0.1 $, $ n=0.25 $, $ \theta_{c}=\pi/2 $.} 
	\label{fig:ST_gamma}
\end{figure}
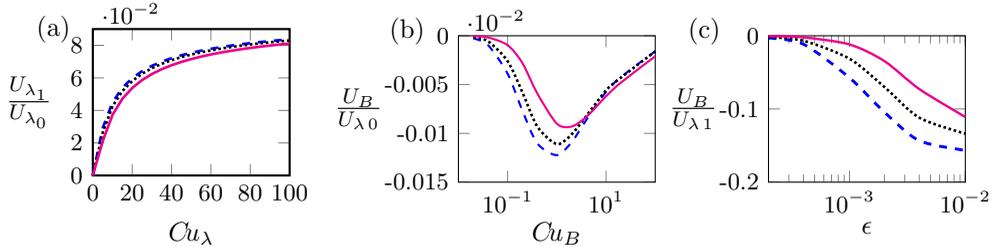

As in \S 2.4, we now analyze the effect of transition parameter ($ \zeta $) on the non-Newtonian components of the swimming velocity.
$ \zeta $ determines the magnitude of shear inside the interaction layer and slip velocity.
Fig.\ref{fig:ST_gamma}(a) shows that the transition parameter has a small effect on the contribution due to slip modification. This is because, $ M_{1} $ has a weak dependency on $ \zeta $ (see fig.\ref{fig:ST_mobility}), which endows $ U_{\lambda_{1}} $ to have a $ \zeta $ dependency similar to that of $ U_{\lambda_{0}} $ (c.f eq.\ref{ST_ULambda1} \& eq.\ref{ULambda}). Thus, the ratio ($ U_{\lambda_{1}}/U_{\lambda_{0}} $) does not change substantially with $ \zeta $.
Fig.\ref{fig:ST_gamma}(b) depicts that, at low to moderate $ C\!u_{B} $, the contribution due to bulk stresses decreases significantly with decrease in $ \zeta $ (as the shear rate $ |\gamma_{0}| $ reduces). 
As $ C\!u_{B} $ increases, the bulk non-Newtonian stresses reduce because the flow approaches a Newtonian state i.e. $ \mu^{*} \rightarrow \mu^{*}_{\infty} $. 
For a lower $ \zeta $, the shear ($ |\gamma_{0}| $) is reduced and thus, the reduction in non-Newtonian stresses occurs at a higher $ C\!u_{B} $. As a result, the trend reverses for $ C\!u_{B} > 10 $ in fig.\ref{fig:ST_gamma}(b): $ \zeta=4 $ is greater than $ \zeta=8,16 $.
Fig.\ref{fig:ST_gamma}(c) compares the two components over the range of interaction layer thickness.

%
\bibliographystyle{jfm}

\newpage

%

\vspace{1cm}

\begin{center}
	\textbf{\large Supplementary Material}\\[.2cm]
	Akash Choudhary,$^{1}$, T. Renganathan$ ^{1} $, and S. Pushpavanam$^{1,\dagger}$\\[.1cm]
	{\itshape  \textsuperscript{1}Department of Chemical Engineering, Indian Institute of Technology Madras, Chennai, 600036 TN, India}\\ \vspace{1cm}
\end{center}

\setcounter{equation}{0}
\setcounter{figure}{0}
\setcounter{page}{1}
\renewcommand{\theequation}{S\arabic{equation}}
\renewcommand{\thefigure}{S\arabic{figure}}



In this Supplementary Material we provide details of the derivation of slip velocity in a second-order fluid.

\vspace{2mm}
\textbf{S.1 Evaluation of $ O(De) $ solution in the inner region}
\vspace{1mm}

The flow field at $ \mathcal{O}(De) $ is governed by
\begin{subequations}\label{innerDeGE_SUPP}
	\begin{gather}
	\frac{\partial \hat{u}_{1}}{\partial \hat{x}} + \frac{\partial \hat{v}_{1}}{\partial \hat{y}} =0 , \label{innerDeGE:CE_SUPP}\\
	- \frac{\partial \hat{p}_{1}}{\partial \hat{x}} +  \frac{\partial^2 \hat{u}_{1}}{\partial \hat{y}^2} + \frac{\partial \hat{S}_{xx \, 0}}{\partial \hat{x}}  +  \frac{\partial \hat{S}_{xy \, 0}}{\partial \hat{y}}   =  0  ,  \label{innerDeGE:NSx_SUPP}\\
	- \frac{\partial \hat{p}_{1}}{\partial \hat{y}} +  \frac{\partial \hat{S}_{yy \, 0}}{\partial \hat{y}} =  0. \label{innerDeGE:NSy_SUPP}
	\end{gather}
\end{subequations}
Here the subscripts on $ S $ denote the components of polymeric stress tensor.
The components of the polymeric stress tensor ($ \mathsfbi{S} $) are: 
\begin{subequations}
	\begin{gather}\label{polyStress_SUPP}
	\hat{S}_{xx} = \left( \frac{\partial \hat{u}}{\partial \hat{y}} \right)^{2},\quad  	\hat{S}_{yy} = \left( 1 +2 \delta \right)  \left( \frac{\partial \hat{u}}{\partial \hat{y}} \right)^{2}, \\
	\hat{S}_{xy} = \hat{S}_{yx} = 2 \delta \left( \frac{\partial \hat{u}}{\partial \hat{y}} \frac{\partial \hat{u}}{\partial \hat{x}} + \frac{\hat{v} \frac{\partial^2 \hat{u}}{\partial \hat{y}^2}  +  \hat{u} \frac{\partial^2 \hat{u}}{\partial \hat{y} \partial \hat{x}} }{2}  \right).
	\end{gather}
\end{subequations}

Using the pressure decay condition: $ \hat{p} \rightarrow 0  $ as $ \hat{y} \rightarrow \infty $ \cite[pg.579]{michelin2014phoretic}, the solution to (\ref{innerDeGE:NSy_SUPP}) yields $ \hat{p}_{1}=\hat{S}_{yy \, 0} - \mathcal{J}_{0}(\hat{x})^{2}(1+2\delta) $, where $ \mathcal{J}_{0} $ is later shown to be zero through matching.
Substituting $ \hat{p}_1 $ in (\ref{innerDeGE:NSx_SUPP}), we obtain
\begin{subequations} \label{du1_simplify_1_SUPP}
	\begin{gather}
	\frac{\partial^{2} \hat{u}_{1}}{\partial \hat{y}^{2}}  + 2\mathcal{J}_{0}(\hat{x}) \mathcal{J}_{0}'(\hat{x})(1+2\delta)  = \frac{\partial \hat{S}_{yy0}}{\partial \hat{x}}  -  \left[  \frac{\partial \hat{S}_{xx0}}{\partial \hat{x}} + \frac{\partial \hat{S}_{xy0}}{\partial \hat{y}}   \right], \label{du1_simplify_1a_SUPP} \\
	= 2 (\textcolor{blue}{1}+2\delta) \textcolor{blue}{u_{y}u_{yx}} - \left[  \textcolor{blue}{2u_{y}u_{yx}} + 2 \delta \left( u_{yy}u_{x} + u_{y} u_{yx}  + \frac{v_{y} u_{yy} + vu_{yyy}+ u_{y}u_{yx} + u u_{yyx}}{2}  \right)  \right]. \label{du1_simplify_1b_SUPP}
	\end{gather}
\end{subequations}
Here we have substituted (\ref{polyStress_SUPP}) in (\ref{du1_simplify_1a_SUPP})  to obtain (\ref{du1_simplify_1b_SUPP}). The subscripts in above equation (and the next) denote the derivative of $ u $ and $ v $ with respect to $ x $ or $ y $. We also have temporarily dropped the hat symbol and subscript $ 0 $ in the right hand side. The coloured terms (blue) get canceled out and using the continuity equation ($ v_{y}=-u_{x} $), we simplify the equation as:
\small
\begin{subequations} \label{du1_simplify_2_SUPP}
	\begin{gather}
	\frac{\partial^{2} \hat{u}_{1}}{\partial \hat{y}^{2}}  +  2\mathcal{J}_{0}(\hat{x}) \mathcal{J}_{0}'(\hat{x})(1+2\delta) = 2 \delta \left\lbrace  2 u_{y}u_{yx} - \left[   \textcolor{green}{u_{yy}u_{x}} + \textcolor{red}{u_{y} u_{yx}} +  \frac{\textcolor{green}{-u_{x} u_{yy}} + vu_{yyy}+ \textcolor{red}{u_{y}u_{yx}} + u u_{yyx}}{2}   \right]   \right\rbrace  , \label{du1_simplify_2a_SUPP}\\
	=  2 \delta \left\lbrace  \textcolor{blue}{2 u_{y}u_{yx}} - \left[   \frac{u_{yy}u_{x}}{2}+ \textcolor{blue}{\frac{3 \, u_{y} u_{yx}}{2}} +  \frac{ vu_{yyy}+ u u_{yyx}}{2}   \right]   \right\rbrace  , \label{du1_simplify_2b_SUPP}\\ 
	=  2 \delta \left\lbrace  \frac{u_{y}u_{yx}}{2} - \left[   \frac{u_{yy}u_{x}}{2} +  \frac{ vu_{yyy}+ u u_{yyx}}{2}   \right]   \right\rbrace  , \label{du1_simplify_2c_SUPP}\\
	=  - \delta \left\lbrace  - u_{y}u_{yx} +   u_{yy}u_{x} +  vu_{yyy}+ u u_{yyx}  \right\rbrace  \label{du1_simplify_2d_SUPP} .
	\end{gather}
\end{subequations}
\normalsize
Dropping the subscript notation (for derivative), we obtain the differential equation for $ \hat{u}_{1} $ as:
\begin{equation}
\frac{\partial^{2} \hat{u}_{1}}{\partial \hat{y}^{2}} = - \delta  \left\lbrace   -\frac{\partial \hat{u}_{0}}{\partial \hat{y}} \frac{\partial^{2} \hat{u}_{0}}{\partial \hat{y} \partial \hat{x}} +  \frac{\partial \hat{u}_{0}}{\partial \hat{x}} \frac{\partial^{2} \hat{u}_{0}}{\partial \hat{y}^{2}}  +  \hat{v}_{0} \frac{\partial^{3} \hat{u}_{0}}{\partial \hat{y}^{3}}  +  \hat{u}_{0} \frac{\partial^{3} \hat{u}_{0}}{\partial \hat{x}  \partial \hat{y}^{2}}         \right\rbrace   - 2\mathcal{J}_{0}(\hat{x}) \mathcal{J}_{0}'(\hat{x})(1+2\delta)   .
\label{Deu1eq_SUPP}
\end{equation}
We will now substitute the $ O(De^{0}) $ solution in the above equation. $ \hat{u}_{0} $ and its derivatives, and $ \hat{v}_{0} $ (obtained from the continuity equation) are:
\begin{subequations} \label{zero_order_sol_SUPP}
	\begin{gather}
	\hat{u}_{0} = -\mathcal{I}'(\hat{x}) \int_{0}^{\hat{y}} \int_{t}^{\infty} \mathcal{F}(s) ds dt  \; + \mathcal{J}_{0}(x) \hat{y} , \label{zero_order_sol_u_SUPP}\\
	\frac{\partial \hat{u}_{0} }{\partial \hat{y}}= -\mathcal{I}'(\hat{x}) \int_{\hat{y}}^{\infty} \mathcal{F}(s) ds  \; + \mathcal{J}_{0}(x)  , \label{zero_order_sol_uy_SUPP}\\
	\frac{\partial^{2} \hat{u}_{0} }{\partial \hat{y}^{2}}= \mathcal{I}'(\hat{x}) \mathcal{F}(\hat{y}) , \label{zero_order_sol_uyy_SUPP}\\
	\hat{v}_{0}  = - \int_{0}^{\hat{y}} \frac{\partial \hat{u}_{0}}{\partial \hat{x}} dy + \mathcal{C}(\hat{x}) = + \mathcal{I}''(\hat{x}) \int_{0}^{\hat{y}} \int_{0}^{r} \int_{t}^{\infty} \mathcal{F}(s) \, ds dt dr -  \frac{\mathcal{J}_{0}'(\hat{x}) \hat{y}^{2}}{2} + \mathcal{C}(\hat{x}). \label{zero_order_sol_v_SUPP}
	\end{gather}
\end{subequations}
Using the no-penetration condition at the surface, we find that $ \mathcal{C}(\hat{x}) =0 $. 

Substituting (\ref{zero_order_sol_SUPP}) in the bracket terms of (\ref{Deu1eq_SUPP}), we obtain the four terms $ \left\lbrace  A+B+C+D  \right\rbrace  $ as:
\begin{subequations} \label{ABCD1}
	\begin{gather}
	A = - \left[  \left( -\mathcal{I}'(\hat{x}) \int_{\hat{y}}^{\infty} \mathcal{F}(s) ds  \; + \mathcal{J}_{0}(x)    \right)    \left( -\mathcal{I}''(\hat{x}) \int_{\hat{y}}^{\infty} \mathcal{F}(s) ds  \; + \mathcal{J}_{0}'(\hat{x}) \right) \right], \label{A1}\\
	B =  \left( -\mathcal{I}''(\hat{x}) \int_{0}^{\hat{y}} \int_{t}^{\infty} \mathcal{F}(s) ds dt  \; + \mathcal{J}_{0}'(x) \hat{y}  \right)    \left( \mathcal{I}'(\hat{x}) \mathcal{F}(\hat{y})  \right) , \label{B1}\\
	C =  \left( \mathcal{I}''(\hat{x}) \int_{0}^{\hat{y}} \int_{0}^{r} \int_{t}^{\infty} \mathcal{F}(s) \, ds dt dr -  \frac{\mathcal{J}_{0}'(\hat{x}) \hat{y}^{2}}{2}   \right)  \left(  \mathcal{I}'(\hat{x}) \mathcal{F}'(\hat{y})  \right) , \label{C1}\\
	D  = \left( -\mathcal{I}'(\hat{x}) \int_{0}^{\hat{y}} \int_{t}^{\infty} \mathcal{F}(s) ds dt  \; + \mathcal{J}_{0}(\hat{x}) \hat{y}  \right)  \left( \mathcal{I}''(\hat{x}) \mathcal{F}(\hat{y})  \right). \label{D1}
	\end{gather}
\end{subequations}
These terms can be further simplified as
\begin{subequations} \label{ABCD2}
	\begin{gather}
	A = - \left[ + \mathcal{I}'(\hat{x}) \mathcal{I}''(\hat{x}) \left( \int_{\hat{y}}^{\infty} \mathcal{F}(s) ds \right)^{2} \, - \, \mathcal{J}_{0}(\hat{x}) \mathcal{I}''(\hat{x}) \left( \int_{\hat{y}}^{\infty} \mathcal{F} (s) ds \right) \, \right.  \nonumber \\
    \left.   	- \, \mathcal{I}'(\hat{x}) \mathcal{J}_{0}'(\hat{x}) \left( \int_{\hat{y}}^{\infty} \mathcal{F}(s) ds \right) \, + \, \mathcal{J}_{0}(\hat{x}) \mathcal{J}_{0}'(\hat{x}) \right]
	\normalsize, \label{A2}\\
	B =  -\mathcal{I}'(\hat{x}) \mathcal{I}''(\hat{x}) \mathcal{F}(\hat{y}) \left( \int_{0}^{\hat{y}} \int_{t}^{\infty} \mathcal{F}(s) dsdt  \right) \, + \, \mathcal{I}'(\hat{x}) \mathcal{J}_{0}'(\hat{x}) \hat{y} \mathcal{F}(\hat{y})   , \label{B2}\\
	C = \mathcal{I}'(\hat{x}) \mathcal{I}''(\hat{x}) \mathcal{F}'(\hat{y}) \left( \int_{0}^{\hat{y}} \int_{0}^{r} \int_{t}^{\infty} \mathcal{F}(s) dsdtdr \right) \, - \, \mathcal{J}_{0}'(\hat{x})\mathcal{I}'(\hat{x}) \mathcal{F}'(\hat{y}) \frac{\hat{y}^{2}}{2} , \label{C2}\\
	D  = - \mathcal{I}'(\hat{x}) \mathcal{I}''(\hat{x}) \mathcal{F} (\hat{y}) \left( \int_{0}^{\hat{y}} \int_{t}^{\infty} \mathcal{F}(s) ds dt \right) \, + \, \mathcal{J}_{0}(\hat{x}) \mathcal{I}''(\hat{x}) \hat{y} \mathcal{F}(\hat{y}). \label{D2}
	\end{gather}
\end{subequations}
Substituting the above four terms in (\ref{Deu1eq_SUPP}) and simplifying we get:
\small
\begin{align}
\frac{\partial^{2} \hat{u}_{1}}{\partial y^{2}} = 
& -  \delta \mathcal{I}'(\hat{x})\mathcal{I}''(\hat{x})  \left\lbrace   
- \left(  \int_{\hat{y}}^{\infty} \mathcal{F}(s) \, {\rm{d}}s  \right)^{2} 
-  2\,\mathcal{F}(\hat{y}) \left( \int_{0}^{\hat{y}} \int_{t}^{\infty} \mathcal{F}(s) ds dt \right) 
\right.  \nonumber \\
&  \qquad \qquad \quad \qquad \left.  	+ \mathcal{F}'(\hat{y}) \left( \int_{0}^{r} \int_{0}^{\omega} \int_{t}^{\infty} \mathcal{F}(s) ds dt d\omega  \right) +
 \right.  \nonumber \\
&  \qquad \qquad \quad \qquad \left.   +  \left( \frac{\mathcal{J}_{0}(\hat{x})}{\mathcal{I}'(\hat{x})} + \frac{\mathcal{J}_{0}'(\hat{x})}{\mathcal{I}''(\hat{x})} \right)  \left( \hat{y} \mathcal{F}(\hat{y}) + \int_{\hat{y}}^{\infty} \mathcal{F}(s) ds  \right) 
\right.  \nonumber \\
&  \qquad \qquad \quad \qquad \left.   -   \frac{\mathcal{J}_{0}(\hat{x})}{\mathcal{I}''(\hat{x})} \mathcal{F}'(\hat{y}) \frac{\hat{y}^{2}}{2}  -  \frac{\mathcal{J}_{0}(\hat{x}) \mathcal{J}_{(0)}'(\hat{x})}{\mathcal{I}'(\hat{x}) \mathcal{I}''(\hat{x})}
 +  \frac{2\mathcal{J}_{0}(\hat{x}) \mathcal{J}_{0}'(\hat{x})(1+2\delta) }{\delta \mathcal{I}'(\hat{x}) \mathcal{I}''(\hat{x})}
\right\rbrace . 		
\label{d2u1_SUPP}
\end{align}
\normalsize
Integrating the above equation once, we obtain
\small
\begin{align}
\frac{\partial \hat{u}_{1}}{\partial y} = C_{1}
& -  \delta \mathcal{I}'(\hat{x})\mathcal{I}''(\hat{x}) \int_{0}^{\hat{y}}  \left\lbrace   
- \left(  \int_{r}^{\infty} \mathcal{F}(s) \, {\rm{d}}s  \right)^{2} 
-  2\,\mathcal{F}(r) \left( \int_{0}^{r} \int_{t}^{\infty} \mathcal{F}(s) ds dt \right) 
\right.  \nonumber \\
&  \qquad \qquad \qquad \qquad  \left.  + \mathcal{F}'(r) \left( \int_{0}^{r} \int_{0}^{\omega} \int_{t}^{\infty} \mathcal{F}(s) ds dt d\omega  \right)
\right.  \nonumber \\
&  \qquad \qquad \qquad \qquad  \left.	+ \left( \frac{\mathcal{J}_{0}(\hat{x})}{\mathcal{I}'(\hat{x})} +  \frac{\mathcal{J}_{0}'(\hat{x})}{\mathcal{I}''(\hat{x})} \right) \left( r \mathcal{F}(r) + \int_{r}^{\infty} \mathcal{F}(s) ds  \right) 
\right.  \nonumber \\
&  \qquad \qquad \qquad \qquad \left. -   \frac{\mathcal{J}_{0}(\hat{x})}{\mathcal{I}''(\hat{x})} \mathcal{F}'(r) \frac{r^{2}}{2}  -  \frac{\mathcal{J}_{0}(\hat{x}) \mathcal{J}_{(0)}'(\hat{x})}{\mathcal{I}'(\hat{x}) \mathcal{I}''(\hat{x})}  
+ \frac{2\mathcal{J}_{0}(\hat{x}) \mathcal{J}_{0}'(\hat{x})(1+2\delta) }{\delta \mathcal{I}'(\hat{x}) \mathcal{I}''(\hat{x})}
\right\rbrace dr . 		
\label{d1u1_SUPP}
\end{align}
\normalsize
As $ \hat{y} \rightarrow \infty $, $ \frac{\partial \hat{u}_{1}}{\partial y}  = \mathcal{J}_{1}(\hat{x}) $ which is to be determined through matching. Simplifying the above equation (replacing $ C_{1} = \mathcal{J}_{1} + \delta \mathcal{I}' \mathcal{I}'' \int_{0}^{\infty} \cdots dr $), we get: 
\small
\begin{align}
\frac{\partial \hat{u}_{1}}{\partial y} = \mathcal{J}_{1}(\hat{x})
& +  \delta \mathcal{I}'(\hat{x})\mathcal{I}''(\hat{x}) \int_{\hat{y}}^{\infty}  \left\lbrace   
- \left(  \int_{r}^{\infty} \mathcal{F}(s)  {\rm{d}}s  \right)^{2} 
-  2 \mathcal{F}(r) \left( \int_{0}^{r} \int_{t}^{\infty} \mathcal{F}(s) ds dt \right) 
\right.  \nonumber \\
&  \qquad \qquad \qquad \qquad  \left.  + \mathcal{F}'(r) \left( \int_{0}^{r} \int_{0}^{\omega} \int_{t}^{\infty} \mathcal{F}(s) ds dt d\omega  \right)
\right.  \nonumber \\
&  \qquad \qquad \qquad \qquad \quad  \left.  	+ \left( \frac{\mathcal{J}_{0}(\hat{x})}{\mathcal{I}'(\hat{x})} +  \frac{\mathcal{J}_{0}'(\hat{x})}{\mathcal{I}''(\hat{x})} \right) \left( r \mathcal{F}(r) + \int_{r}^{\infty} \mathcal{F}(s) ds  \right) -   \frac{\mathcal{J}_{0}(\hat{x})}{\mathcal{I}''(\hat{x})} \mathcal{F}'(r) \frac{r^{2}}{2}  
\right.  \nonumber \\
&  \qquad \qquad \quad \qquad \qquad  \left. 
-  \frac{\mathcal{J}_{0}(\hat{x}) \mathcal{J}_{(0)}'(\hat{x})}{\mathcal{I}'(\hat{x}) \mathcal{I}''(\hat{x})}
+ \frac{2\mathcal{J}_{0}(\hat{x}) \mathcal{J}_{0}'(\hat{x})(1+2\delta) }{\delta \mathcal{I}'(\hat{x}) \mathcal{I}''(\hat{x})} 
\right\rbrace dr . 		
\label{d1u1_BC_SUPP}
\end{align}
\normalsize
Integrating once more and using the no-slip condition (renders the constant $ C_{2} $ zero), we get:
\small
\begin{align} \label{u1_unreduced}
\hat{u}_{1}= \mathcal{J}_{1}(\hat{x}) \hat{y}
& -  \delta \mathcal{I}'(\hat{x})\mathcal{I}''(\hat{x}) \int_{0}^{\hat{y}}  dp \int_{p}^{\infty}  \left\lbrace   
\left(  \int_{r}^{\infty} \mathcal{F}(s) \, {\rm{d}}s  \right)^{2} 
+  2\,\mathcal{F}(r) \left( \int_{0}^{r} \int_{t}^{\infty} \mathcal{F}(s) ds dt \right) 
\right.  \nonumber \\
&  \qquad \qquad \quad \qquad \qquad \quad \qquad  \left.  - \mathcal{F}'(r) \left( \int_{0}^{r} \int_{0}^{\omega} \int_{t}^{\infty} \mathcal{F}(s) ds dt d\omega  \right)
\right.  \nonumber \\
&  \qquad \qquad \quad \qquad \qquad \quad \qquad  \left.  	- \left( \frac{\mathcal{J}_{0}(\hat{x})}{\mathcal{I}'(\hat{x})} +  \frac{\mathcal{J}_{0}'(\hat{x})}{\mathcal{I}''(\hat{x})} \right)  \left( r \mathcal{F}(r) + \int_{r}^{\infty} \mathcal{F}(s) ds  \right) 
\right.  \nonumber \\
&  \qquad \qquad \quad \qquad \qquad \qquad \quad \left. +   \frac{\mathcal{J}_{0}(\hat{x})}{\mathcal{I}''(\hat{x})} \mathcal{F}'(r) \frac{r^{2}}{2}  +  \frac{\mathcal{J}_{0}(\hat{x}) \mathcal{J}_{(0)}'(\hat{x})}{\mathcal{I}'(\hat{x}) \mathcal{I}''(\hat{x})}
- \frac{2\mathcal{J}_{0}(\hat{x}) \mathcal{J}_{0}'(\hat{x})(1+2\delta) }{\delta \mathcal{I}'(\hat{x}) \mathcal{I}''(\hat{x})} 
\right\rbrace dr . 		
\end{align}
\normalsize
Here $ \mathcal{F}(r) = -1 + e^{-\hat{\psi}(r)} $ and $ \mathcal{F}'(r) = - \hat{\psi}'(r) e^{-\hat{\psi}(r)} $.

We now reduce the integrals (inside the bracket) by changing the order of integration:
\begin{align}
& 2 \mathcal{F}(r) \left( \int_{0}^{r} \int_{t}^{\infty} \mathcal{F}(s) {\rm{d}}s \, {\rm{d}}t \right) {\rm{d}}r  = 
2 \mathcal{F}(r) \left(  \int_{0}^{r}s\mathcal{F}(s) {\rm{d}}s   + r  \int_{r}^{\infty} \mathcal{F}(s) {\rm{d}}s  \right) {\rm{d}}r \mbox{ and\ }\nonumber \\ 
& \int_{0}^{r} \int_{0}^{\omega} \int_{t}^{\infty}  \mathcal{F}(s)  {\rm{d}}s \, {\rm{d}}t \, {\rm{d}}\omega  =  
\int_{0}^{r} \left( r - \frac{s}{2} \right)s\mathcal{F}(s) {\rm{d}}s \,
+ \,
\frac{r^{2}}{2} \int_{r}^{\infty} \mathcal{F}(s) {\rm{d}}s. 
\end{align}
Substituting the above simplification in (\ref{u1_unreduced}), we obtain
\small
\begin{align} \label{u1_SUPP}
\hat{u}_{1}= \mathcal{J}_{1}(\hat{x}) \hat{y}
& -  \delta \mathcal{I}'(\hat{x})\mathcal{I}''(\hat{x}) \int_{0}^{\hat{y}}  dp \int_{p}^{\infty}  \left\lbrace   
\left(  \int_{r}^{\infty} \mathcal{F}(s) \, {\rm{d}}s  \right)^{2} 
+  2 \mathcal{F}(r) \left(  \int_{0}^{r}s\mathcal{F}(s) {\rm{d}}s   + r  \int_{r}^{\infty} \mathcal{F}(s) {\rm{d}}s  \right) {\rm{d}}r  
\right.  \nonumber \\
&  \qquad \qquad \qquad \qquad \qquad  \qquad \left.  + \, \hat{\psi}'(r) e^{-\hat{\psi}(r)} \left( \int_{0}^{r} \left( r - \frac{s}{2} \right)s\mathcal{F}(s) {\rm{d}}s \,
+ \, \frac{r^{2}}{2} \int_{r}^{\infty} \mathcal{F}(s) {\rm{d}}s  \right)	
\right. \nonumber \\
&  \qquad \qquad \qquad \qquad \qquad \qquad \left.  - \left( \frac{\mathcal{J}_{0}(\hat{x})}{\mathcal{I}'(\hat{x})} +  \frac{\mathcal{J}_{0}'(\hat{x})}{\mathcal{I}''(\hat{x})} \right) \left( r \mathcal{F}(r) + \int_{r}^{\infty} \mathcal{F}(s) ds  \right)+   \frac{\mathcal{J}_{0}(\hat{x})}{\mathcal{I}''(\hat{x})} \mathcal{F}'(r) \frac{r^{2}}{2}  
\right.  \nonumber \\
&\qquad \qquad \qquad \qquad \qquad \qquad \left.  +  \frac{\mathcal{J}_{0}(\hat{x}) \mathcal{J}_{(0)}'(\hat{x})}{\mathcal{I}'(\hat{x}) \mathcal{I}''(\hat{x})} 
- \frac{2\mathcal{J}_{0}(\hat{x}) \mathcal{J}_{0}'(\hat{x})(1+2\delta) }{\delta \mathcal{I}'(\hat{x}) \mathcal{I}''(\hat{x})} 
\right\rbrace dr . 		
\end{align}
\normalsize

\vspace{2mm}
\textbf{S.2 Matching inner and outer solutions}
\vspace{1mm}

For any field variable $ f $ (representing concentration and velocity), the matching condition at $\mathcal{O}$($ \epsilon^{0} $) is
\begin{equation}
\lim\limits_{y \rightarrow 0} (f^{(0)}_{0}+De f^{(0)}_{1}+\cdots)=\lim\limits_{\hat{y} \rightarrow \infty}(\hat{f}^{(0)}_{0}+De \hat{f}^{(0)}_{1}+\cdots).
\label{Matching_SUPP}
\end{equation}
The matching condition for the concentration field ($ \hat{c} $) yields:	$\mathcal{I} = \lim\limits_{y \rightarrow 0} c^{(0)}(x,y) + C_{\infty}$. 
At $ O(1) $, the matching condition for velocity yields:
\begin{equation}
\left.{u}^{(0)}_{0}\right\vert_{y=0}   =  \lim\limits_{\hat{y} \rightarrow \infty}  \hat{u}^{(0)}_{0} . \nonumber
\end{equation}
We substitute (\ref{zero_order_sol_u_SUPP}) in the above equation, which yields
\begin{align}
\left. u^{(0)}_{0}\right\vert_{y=0} 
& = -\left(  \left. \grad{c}{x}\right\vert_{y=0}  \right) \int_{0}^{\infty} \int_{t}^{\infty} \left(  e^{-\hat{\psi}(s)} -1  \right) {\rm{d}s} \, {\rm{d}t } + \lim\limits_{\hat{y} \rightarrow \infty}  \mathcal{J}_{0} \, \hat{y}.
\label{FINALu0SUPP}
\end{align}
For a bounded solution, $ \mathcal{J}_{0}=0 $. We thus obtain the solution reported previously in the literature \cite{derjaguin1947kinetic,anderson1982motion,michelin2014phoretic}. 
At $ O(De) $, we use (\ref{u1_SUPP}) and obtain 
\begin{align}
\left. u^{(0)}_{1}\right\vert_{y=0} 
& = - \delta   \left( \left.\grad{c}{x}\right\vert_{y=0}   \left.\lap{c}{x}\right\vert_{y=0}  \right){\displaystyle\int_{0}^{\infty} }
{\rm{d}}p  
\int_{p}^{\infty} \mathcal{G}(r) {\rm{d}}r + \lim\limits_{\hat{y} \rightarrow \infty}  \mathcal{J}_{1} \, \hat{y}.
\label{FINALu1aSUPP}
\end{align}
\begin{align}
{\rm{Here},}\;\mathcal{G}(r)= 
& \left\lbrace   
\left(  \int_{r}^{\infty} \mathcal{F}(s) \, {\rm{d}}s  \right)^{2} 
+  2\,\mathcal{F}(r) \left( 
\int_{0}^{r}  s \mathcal{F}(s) {\rm{d}} s   +   r  \int_{r}^{\infty}  \mathcal{F}(s) {\rm{d}} s
\right) \right.\nonumber \\
& \; \;\, \left.  	+ \, \hat{\psi}'(r) \, e^{-\hat{\psi}(r)}  \left(
\int_{0}^{r} \left( r - \frac{s}{2}\right) s\, \mathcal{F}(s) {\rm{d}}s + \frac{r^{2}}{2} \int_{r}^{\infty} \mathcal{F}(s) {\rm{d}}s
\right) 
\right\rbrace.
\end{align}
Similar to the $ O(1) $ solution, we obtain $ \mathcal{J}_{1}=0 $.

\begin{center}
	\textit{2.1 Intermediate matching}
\end{center}
The above results for velocity can also be obtained using intermediate matching \cite{hinch_1991}. In an arbitrary intermediate region ($ y \sim \epsilon^{\alpha} $, where $ 0 < \alpha < 1 $), the matching condition for both outer and inner region is
\begin{equation}
\lim\limits_{\epsilon \rightarrow 0} (u^{(0)}_{0}+De u^{(0)}_{1}+\cdots)=\lim\limits_{\epsilon \rightarrow 0}(\hat{u}^{(0)}_{0}+De \hat{u}^{(0)}_{1}+\cdots).
\label{intMatching}
\end{equation}
Using the Taylor series in the LHS of (\ref{intMatching}), at $ O(1) $ the matching condition yields
\begin{equation}
\lim\limits_{\epsilon \rightarrow 0}  \left.{u}^{(0)}_{0}\right\vert_{y=0} +y \left( \left.\grad{u^{(0)}_{0}}{y}\right\vert_{y=0}  \right)+ \cdots  = \lim\limits_{\epsilon \rightarrow 0}   \hat{u}^{(0)}_{0}   .
\end{equation}
We substitute (\ref{zero_order_sol_u_SUPP}) in the RHS of the above equation and obtain
\begin{equation}
\lim\limits_{\epsilon \rightarrow 0}  \left[  \left.{u}^{(0)}_{0}\right\vert_{y=0} +y \left( \left.\grad{u^{(0)}_{0}}{y}\right\vert_{y=0} \right) + \cdots  \right]
=  \lim\limits_{\epsilon \rightarrow 0}   \left[   -\left(  \left. \grad{c}{x}\right\vert_{y=0}  \right) \int_{0}^{\hat{y}} \int_{t}^{\infty} \mathcal{F}(s) ds dt  \; + \mathcal{J}_{0} \, \hat{y}   \right] . 
\end{equation}
Rescaling $ y $ and $\hat{y}$ in terms of the intermediate coordinate ($\bar{y}$): $ y = \bar{y} \, \epsilon^{\alpha}   \mbox{ and\ }   \hat{y} = \bar{y} \, \epsilon^{-\alpha}$. 
\begin{equation}\label{int_Rescale_match}
\lim\limits_{\epsilon \rightarrow 0}  \left[   \left.{u}^{(0)}_{0}\right\vert_{y=0} +\epsilon^{\alpha} \bar{y} \left( \left.\grad{u^{(0)}_{0}}{y}\right\vert_{y=0} \right) + \cdots  \right]  = 
\lim\limits_{\epsilon \rightarrow 0}   \left[ -\left(  \left. \grad{c}{x}\right\vert_{y=0}  \right) \int_{0}^{\bar{y} \epsilon^{-\alpha}} \int_{t}^{\infty} \mathcal{F}(s) ds dt  \; + \mathcal{J}_{0} \, \bar{y} \epsilon^{-\alpha} \right] .
\end{equation}
Comparing the coefficients of $ \bar{y} $, we obtain
\begin{equation}
\mathcal{J}_{0} = \epsilon^{2 \alpha} \left( \left.\grad{u^{(0)}_{0}}{y}\right\vert_{y=0} \right) .  
\end{equation}
Since the velocity gradient at the surface ($ y=0 $) is less than or equal to $ O(1) $, $ \mathcal{J}_{0} $ can be neglected at the leading order (as it is $ O(\epsilon^{2\alpha}) $, where $ 0<\alpha<1 $). In the limit $ \epsilon \rightarrow 0 $, equation (\ref{int_Rescale_match}) yields 
\begin{equation}
\lim\limits_{\epsilon \rightarrow 0}     \left.{u}^{(0)}_{0}\right\vert_{y=0}  = 
-\left(  \left. \grad{c}{x}\right\vert_{y=0}  \right) \int_{0}^{\infty} \int_{t}^{\infty} \mathcal{F}(s) ds dt .
\end{equation}


\vspace{2mm}
\textbf{S.3 Validation with literature}
\vspace{1mm}

\begin{figure}[H]
	\centering
	{{\includegraphics[scale=0.5]{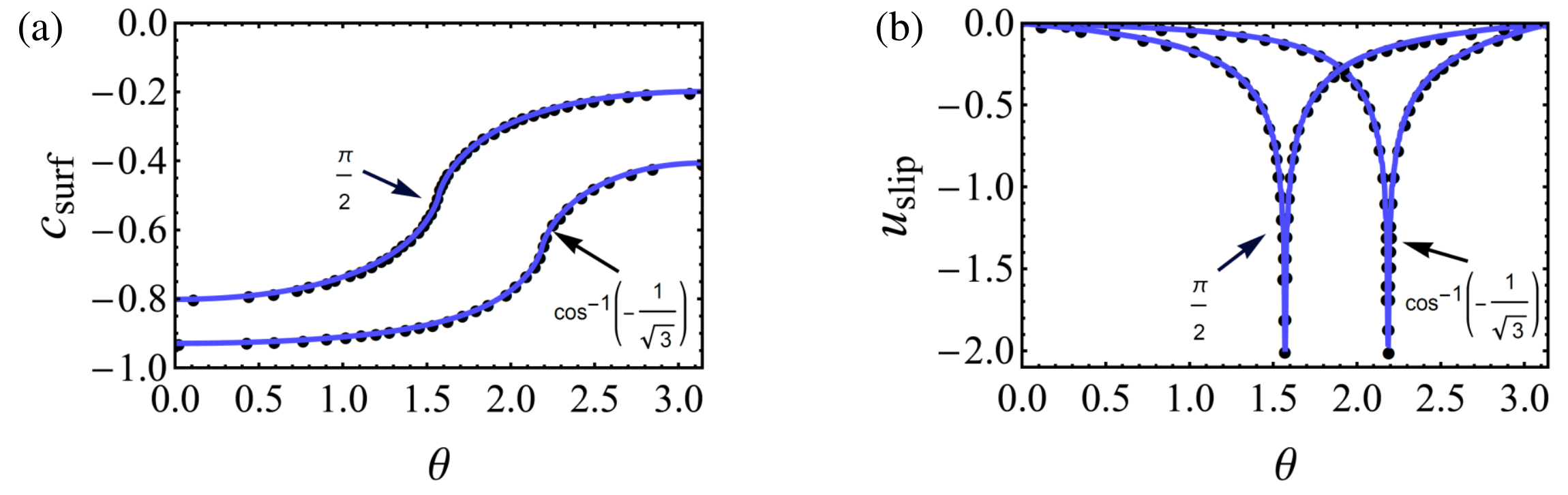} }}%
	\caption{(a) Surface concentration profile and (b) slip velocity profile for $ \cos \theta_{c}=0 \mbox{ and\ } -1/\sqrt{3} $. The filled circles represent the results of \cite{michelin2014phoretic}.}
	\label{fig:VelSUPP}
\end{figure}

\begin{center}
	\textit{3.1 Concentration field and Newtonian slip velocity}
\end{center}
The slip velocity and concentration field for different surface coverage is shown in fig. \ref{fig:VelSUPP}. 
For step change in activity, our results agree well with that of \cite{michelin2014phoretic} (i.e. Newtonian fluid). 
To obtain these results, the Newtonian mobility coefficient $ M_{0} $  is fixed to be -1 (as performed by \cite{michelin2014phoretic}). 
It should be noted that $ M_{0} $ in the main text is $ -1.1465 $ (corresponding to $ \Phi_{0}=-1 $).

\begin{figure}[H]
	\begin{tikzpicture}
	\begin{semilogxaxis}[
	height=0.29\textwidth,width=.37\textwidth,
	xlabel=$ {\small C\!u_{B}}$, ylabel= {\large $\frac{U_{B}}{U_{\lambda\,0}}$ \normalsize} , xshift=0cm,yshift=0.8cm,	ylabel shift= -4 pt, 	ylabel style={rotate=-90}, xmin=0.01, xmax=100, ymin=-0.020, ymax=0,
	legend style={draw=none,at={(0.9,0.07)},anchor=south east},	legend style={nodes={scale=0.75, transform shape}}
	]
	\addplot[line width=1pt, black,smooth] table[x=x, y=y] {ST_UB_aku.txt};\label{aku}
	
	\addplot[mark=o, black,mark size=1.5pt] table[x=x, y=y] {ST_UB_elf.txt};\label{elf}
	%
	
	\end{semilogxaxis}
	\end{tikzpicture}
	\caption{Comparison of numerical calculation of $ U_{B} $ (\ref{aku})  with \cite{datt2017activeComplex} (\ref{elf}) for $ \chi=0.1 $, $ \theta_{c}=\pi/2 $, $ n =0.25, m=15 $.}
	\label{fig:ST_UB_reproduced}
\end{figure}
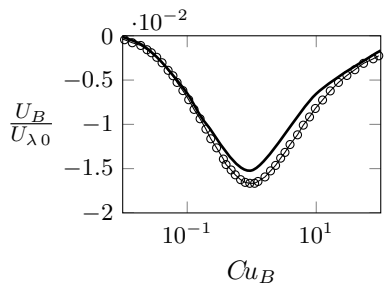

\begin{center}
	\textit{3.2 $ \, U_{B} $ for shear-thinning fluid}
\end{center}
The velocity field around an axisymmetric squirmer was provided by \cite{blake1971spherical}, which was later used by \cite{datt2017activeComplex} to find the bulk non-Newtonian effects on the swimming of an axisymmetric Janus sphere. The radial component of the outer region disturbance field is
\begin{equation}
u_{r} = \alpha_{1} \frac{P_{1}}{r^{3}} + \sum_{m=2}^{\infty} \left( \frac{1}{r^{m+2}} -\frac{1}{r^{m}} \right) \left( m+\frac{1}{2} \right) \alpha_{n} P_{m},
\end{equation}
and the tangential component is
\begin{equation}
u_{\theta} = \alpha_{1}  \frac{V_{1}  }{2 r^{3}}  + \sum_{m=2}^{\infty} \left[ \frac{m}{2 r^{m+2}}  - \left( \frac{n}{2} - 1 \right) \frac{1}{r^{n}} \right] \left( m+\frac{1}{2} \right)  \alpha_{m} V_{m}.
\end{equation}
Here, $ V_{m}= \left[-2 \sin \theta/(m(m+1))\right] P_{m}^{1}(\cos \theta)  $, $ P_{m}^{1} $ is an associated Legendre polynomial of the first kind, $ \alpha_{m} = {m \mathcal{K}_{m}}/{(2m+1)} $, and $ \mathcal{K}_{m} $ is the m\textsuperscript{th} spectral mode for the step activity which is given by (2.26) in the main text.
We convert  the above field in Cartesian coordinates and substitute it in the expression for $ U_{B} $:
\begin{equation}
U_{B}= -  \frac{1}{6\pi} \chi \int_{V_{f}}  \mu_{1} \left( \gamma_{0} \right)  \bten{A}_{0} \IB{:} \nabla{\IB{u}^{t}} {\rm d}V. 
\label{ST_UB}
\end{equation}
Here $ \mu_{1} = \left( 1+ C\!u_{B}^{2} |\gamma_{0}|^{2}  \right)^{\frac{n-1}{2}} - 1 $ and $ |\gamma_{0}| = \left(\bten{A}_{0} \IB{:} \bten{A}_{0}/2 \right)^{1/2} $. We use inbuilt Gauss-Kronrod rule in Mathematica 12 to numerically evaluate $ U_{B} $. 
Fig.\ref{fig:ST_UB_reproduced} shows an agreement with the results reported by \cite{datt2017activeComplex} for $ \theta_{c}=\pi/2 $.


\bibliography{Akash}

\begin{thebibliography}{50}
\expandafter\ifx\csname natexlab\endcsname\relax\def\natexlab#1{#1}\fi
\def\au#1{#1} \def\ed#1{#1} \def\yr#1{#1}\def\at#1{#1}\def\jt#1{\textit{#1}}
  \def\bt#1{#1}\def\bvol#1{\textbf{#1}} \def\vol#1{#1} \def\pg#1{#1}
  \def\publ#1{#1}\def\arxiv#1{#1}\def\org#1{#1}\def\st#1{\textit{#1}}

\bibitem[Anderson {\em et~al.\/}(1982)Anderson, Lowell \&
  Prieve]{anderson1982motion}
{\sc \au{Anderson, JL}, \au{Lowell, ME} \& \au{Prieve, DC}} \yr{1982}
  \at{Motion of a particle generated by chemical gradients part 1.
  non-electrolytes}.  \jt{Journal of Fluid Mechanics}  \bvol{117},
  \pg{107--121}.

\bibitem[Anderson(1989)]{anderson1989}
{\sc \au{Anderson, John~L}} \yr{1989}  \at{Colloid transport by interfacial
  forces}.  \jt{Annual review of fluid mechanics}  \bvol{21}~(1),  \pg{61--99}.

\bibitem[Aragones {\em et~al.\/}(2018)Aragones, Yazdi \&
  Alexander-Katz]{aragones2018diffusion}
{\sc \au{Aragones, Juan~L}, \au{Yazdi, Shahrzad} \& \au{Alexander-Katz,
  Alfredo}} \yr{2018}  \at{Diffusion of self-propelled particles in complex
  media}.  \jt{Physical Review Fluids}  \bvol{3}~(8),  \pg{083301}.

\bibitem[Baraban {\em et~al.\/}(2012)Baraban, Tasinkevych, Popescu, Sanchez,
  Dietrich \& Schmidt]{baraban2012transport}
{\sc \au{Baraban, L}, \au{Tasinkevych, M}, \au{Popescu, MN}, \au{Sanchez, S},
  \au{Dietrich, S} \& \au{Schmidt, OG}} \yr{2012}  \at{Transport of cargo by
  catalytic janus micro-motors}.  \jt{Soft Matter}  \bvol{8}~(1),  \pg{48--52}.

\bibitem[Bird {\em et~al.\/}(1987)Bird, Armstrong \&
  Hassager]{bird1987dynamics}
{\sc \au{Bird, Robert~Byron}, \au{Armstrong, Robert~Calvin} \& \au{Hassager,
  Ole}} \yr{1987}  \at{Dynamics of polymeric liquids. vol. 1: Fluid mechanics}
  .

\bibitem[Blake(1971)]{blake1971spherical}
{\sc \au{Blake, John~R}} \yr{1971}  \at{A spherical envelope approach to
  ciliary propulsion}.  \jt{Journal of Fluid Mechanics}  \bvol{46}~(1),
  \pg{199--208}.

\bibitem[Brady(2011)]{brady2011particle}
{\sc \au{Brady, John~F}} \yr{2011}  \at{Particle motion driven by solute
  gradients with application to autonomous motion: continuum and colloidal
  perspectives}.  \jt{Journal of Fluid Mechanics}  \bvol{667},  \pg{216--259}.

\bibitem[C{\'o}rdova-Figueroa \& Brady(2008)]{cordova2008osmotic}
{\sc \au{C{\'o}rdova-Figueroa, Ubaldo~M} \& \au{Brady, John~F}} \yr{2008}
  \at{Osmotic propulsion: the osmotic motor}.  \jt{Physical review letters}
  \bvol{100}~(15),  \pg{158303}.

\bibitem[Datt {\em et~al.\/}(2017)Datt, Natale, Hatzikiriakos \&
  Elfring]{datt2017activeComplex}
{\sc \au{Datt, Charu}, \au{Natale, Giovanniantonio}, \au{Hatzikiriakos,
  Savvas~G} \& \au{Elfring, Gwynn~J}} \yr{2017}  \at{An active particle in a
  complex fluid}.  \jt{Journal of Fluid Mechanics}  \bvol{823},  \pg{675--688}.

\bibitem[Datt {\em et~al.\/}(2015)Datt, Zhu, Elfring \&
  Pak]{datt2015squirmingST}
{\sc \au{Datt, Charu}, \au{Zhu, Lailai}, \au{Elfring, Gwynn~J} \& \au{Pak,
  On~Shun}} \yr{2015}  \at{Squirming through shear-thinning fluids}.
  \jt{Journal of Fluid Mechanics}  \bvol{784}.

\bibitem[De~Corato {\em et~al.\/}(2015)De~Corato, Greco \&
  Maffettone]{corato2015locomotion}
{\sc \au{De~Corato, M}, \au{Greco, F} \& \au{Maffettone, PL}} \yr{2015}
  \at{Locomotion of a microorganism in weakly viscoelastic liquids}.
  \jt{Physical Review E}  \bvol{92}~(5),  \pg{053008}.

\bibitem[Derjaguin {\em et~al.\/}(1947)Derjaguin, Sidorenkov, Zubashchenkov \&
  Kiseleva]{derjaguin1947kinetic}
{\sc \au{Derjaguin, BV}, \au{Sidorenkov, GP}, \au{Zubashchenkov, EA} \&
  \au{Kiseleva, EV}} \yr{1947}  \at{Kinetic phenomena in boundary films of
  liquids}.  \jt{Kolloidn. zh}  \bvol{9},  \pg{335--347}.

\bibitem[Ebbens \& Howse(2011)]{ebbens2011direct}
{\sc \au{Ebbens, Stephen~J} \& \au{Howse, Jonathan~R}} \yr{2011}  \at{Direct
  observation of the direction of motion for spherical catalytic swimmers}.
  \jt{Langmuir}  \bvol{27}~(20),  \pg{12293--12296}.

\bibitem[Elfring \& Goyal(2016)]{elfring2016effect}
{\sc \au{Elfring, Gwynn~J} \& \au{Goyal, Gaurav}} \yr{2016}  \at{The effect of
  gait on swimming in viscoelastic fluids}.  \jt{Journal of Non-Newtonian Fluid
  Mechanics}  \bvol{234},  \pg{8--14}.

\bibitem[Fournier-Bidoz {\em et~al.\/}(2005)Fournier-Bidoz, Arsenault, Manners
  \& Ozin]{ozin2005synthetic}
{\sc \au{Fournier-Bidoz, S{\'e}bastien}, \au{Arsenault, Andr{\'e}~C},
  \au{Manners, Ian} \& \au{Ozin, Geoffrey~A}} \yr{2005}  \at{Synthetic
  self-propelled nanorotors}.  \jt{Chemical Communications} ~(4),
  \pg{441--443}.

\bibitem[Gao \& Wang(2014)]{gao2014synthetic}
{\sc \au{Gao, Wei} \& \au{Wang, Joseph}} \yr{2014}  \at{Synthetic
  micro/nanomotors in drug delivery}.  \jt{Nanoscale}  \bvol{6}~(18),
  \pg{10486--10494}.

\bibitem[Golestanian {\em et~al.\/}(2007)Golestanian, Liverpool \&
  Ajdari]{golestanian2007designing}
{\sc \au{Golestanian, R}, \au{Liverpool, TB} \& \au{Ajdari, A}} \yr{2007}
  \at{Designing phoretic micro-and nano-swimmers}.  \jt{New Journal of Physics}
   \bvol{9}~(5),  \pg{126}.

\bibitem[Golestanian {\em et~al.\/}(2005)Golestanian, Liverpool \&
  Ajdari]{golestanian2005propulsion}
{\sc \au{Golestanian, Ramin}, \au{Liverpool, Tanniemola~B} \& \au{Ajdari,
  Armand}} \yr{2005}  \at{Propulsion of a molecular machine by asymmetric
  distribution of reaction products}.  \jt{Physical review letters}
  \bvol{94}~(22),  \pg{220801}.

\bibitem[Gomez-Solano {\em et~al.\/}(2016)Gomez-Solano, Blokhuis \&
  Bechinger]{gomez2016dynamics}
{\sc \au{Gomez-Solano, Juan~Ruben}, \au{Blokhuis, Alex} \& \au{Bechinger,
  Clemens}} \yr{2016}  \at{Dynamics of self-propelled janus particles in
  viscoelastic fluids}.  \jt{Physical review letters}  \bvol{116}~(13),
  \pg{138301}.

\bibitem[Hinch(1991)]{hinch_1991}
{\sc \au{Hinch, E.~J.}} \yr{1991} {\em Perturbation Methods\/}.
  \publ{Cambridge University Press}.

\bibitem[Ho \& Leal(1976)]{ho1976migration}
{\sc \au{Ho, BP} \& \au{Leal, LG}} \yr{1976}  \at{Migration of rigid spheres in
  a two-dimensional unidirectional shear flow of a second-order fluid}.
  \jt{Journal of Fluid Mechanics}  \bvol{76}~(4),  \pg{783--799}.

\bibitem[Howse {\em et~al.\/}(2007)Howse, Jones, Ryan, Gough, Vafabakhsh \&
  Golestanian]{howse2007self}
{\sc \au{Howse, Jonathan~R}, \au{Jones, Richard~AL}, \au{Ryan, Anthony~J},
  \au{Gough, Tim}, \au{Vafabakhsh, Reza} \& \au{Golestanian, Ramin}} \yr{2007}
  \at{Self-motile colloidal particles: from directed propulsion to random
  walk}.  \jt{Physical review letters}  \bvol{99}~(4),  \pg{048102}.

\bibitem[J{\"u}licher \& Prost(2009)]{julicher2009generic}
{\sc \au{J{\"u}licher, Frank} \& \au{Prost, Jacques}} \yr{2009}  \at{Generic
  theory of colloidal transport}.  \jt{The European Physical Journal E}
  \bvol{29}~(1),  \pg{27--36}.

\bibitem[Ke {\em et~al.\/}(2010)Ke, Ye, Carroll \& Showalter]{ke2010motion}
{\sc \au{Ke, Hua}, \au{Ye, Shengrong}, \au{Carroll, R~Lloyd} \& \au{Showalter,
  Kenneth}} \yr{2010}  \at{Motion analysis of self-propelled pt- silica
  particles in hydrogen peroxide solutions}.  \jt{The Journal of Physical
  Chemistry A}  \bvol{114}~(17),  \pg{5462--5467}.

\bibitem[Khair {\em et~al.\/}(2012)Khair, Posluszny \&
  Walker]{khair2012coupling}
{\sc \au{Khair, Aditya~S}, \au{Posluszny, Denise~E} \& \au{Walker, Lynn~M}}
  \yr{2012}  \at{Coupling electrokinetics and rheology: electrophoresis in
  non-newtonian fluids}.  \jt{Physical Review E}  \bvol{85}~(1),  \pg{016320}.

\bibitem[Li \& Koch(2020)]{li2020electrophoresis}
{\sc \au{Li, Gaojin} \& \au{Koch, Donald~L}} \yr{2020}  \at{Electrophoresis in
  dilute polymer solutions}.  \jt{Journal of Fluid Mechanics}  \bvol{884}.

\bibitem[Lisicki {\em et~al.\/}(2016)Lisicki, Michelin \&
  Lauga]{Pump_Geo_lisicki2016phoretic}
{\sc \au{Lisicki, Maciej}, \au{Michelin, S{\'e}bastien} \& \au{Lauga, Eric}}
  \yr{2016}  \at{Phoretic flow induced by asymmetric confinement}.  \jt{Journal
  of Fluid Mechanics}  \bvol{799}.

\bibitem[Makuch {\em et~al.\/}(2020)Makuch, Ho{\l}yst, Kalwarczyk, Garstecki \&
  Brady]{makuch2020diffusion}
{\sc \au{Makuch, Karol}, \au{Ho{\l}yst, Robert}, \au{Kalwarczyk, Tomasz},
  \au{Garstecki, Piotr} \& \au{Brady, John~F}} \yr{2020}  \at{Diffusion and
  flow in complex liquids}.  \jt{Soft matter} .

\bibitem[Maldonado-Camargo \& Rinaldi(2016)]{maldonado2016breakdown}
{\sc \au{Maldonado-Camargo, Lorena} \& \au{Rinaldi, Carlos}} \yr{2016}
  \at{Breakdown of the stokes--einstein relation for the rotational diffusivity
  of polymer grafted nanoparticles in polymer melts}.  \jt{Nano letters}
  \bvol{16}~(11),  \pg{6767--6773}.

\bibitem[Michelin \& Lauga(2014)]{michelin2014phoretic}
{\sc \au{Michelin, S{\'e}bastien} \& \au{Lauga, Eric}} \yr{2014}  \at{Phoretic
  self-propulsion at finite p{\'e}clet numbers}.  \jt{Journal of Fluid
  Mechanics}  \bvol{747},  \pg{572--604}.

\bibitem[Michelin \& Lauga(2019)]{Pump_Patch_michelin2019nature}
{\sc \au{Michelin, S{\'e}bastien} \& \au{Lauga, Eric}} \yr{2019}  \at{Universal
  optimal geometry of minimal phoretic pumps}.  \jt{Scientific reports}
  \bvol{9}~(1),  \pg{10788}.

\bibitem[Michelin {\em et~al.\/}(2015)Michelin, Montenegro-Johnson, De~Canio,
  Lobato-Dauzier \& Lauga]{Pump_Geo_michelin2015SoftMatter}
{\sc \au{Michelin, S{\'e}bastien}, \au{Montenegro-Johnson, Thomas~D},
  \au{De~Canio, Gabriele}, \au{Lobato-Dauzier, Nicolas} \& \au{Lauga, Eric}}
  \yr{2015}  \at{Geometric pumping in autophoretic channels}.  \jt{Soft matter}
   \bvol{11}~(29),  \pg{5804--5811}.

\bibitem[Natale {\em et~al.\/}(2017)Natale, Datt, Hatzikiriakos \&
  Elfring]{SOF_natale2017}
{\sc \au{Natale, Giovanniantonio}, \au{Datt, Charu}, \au{Hatzikiriakos,
  Savvas~G} \& \au{Elfring, Gwynn~J}} \yr{2017}  \at{Autophoretic locomotion in
  weakly viscoelastic fluids at finite p{\'e}clet number}.  \jt{Physics of
  Fluids}  \bvol{29}~(12),  \pg{123102}.

\bibitem[O'Brien(1983)]{o1983solution}
{\sc \au{O'Brien, RW}} \yr{1983}  \at{The solution of the electrokinetic
  equations for colloidal particles with thin double layers}.  \jt{Journal of
  Colloid and Interface Science}  \bvol{92}~(1),  \pg{204--216}.

\bibitem[Patteson {\em et~al.\/}(2016)Patteson, Gopinath \&
  Arratia]{patteson2016active}
{\sc \au{Patteson, Alison~E}, \au{Gopinath, Arvind} \& \au{Arratia, Paulo~E}}
  \yr{2016}  \at{Active colloids in complex fluids}.  \jt{Current Opinion in
  Colloid \& Interface Science}  \bvol{100}~(21),  \pg{86--96}.

\bibitem[Paxton {\em et~al.\/}(2004)Paxton, Kistler, Olmeda, Sen, St.~Angelo,
  Cao, Mallouk, Lammert \& Crespi]{paxton2004catalytic}
{\sc \au{Paxton, Walter~F}, \au{Kistler, Kevin~C}, \au{Olmeda, Christine~C},
  \au{Sen, Ayusman}, \au{St.~Angelo, Sarah~K}, \au{Cao, Yanyan}, \au{Mallouk,
  Thomas~E}, \au{Lammert, Paul~E} \& \au{Crespi, Vincent~H}} \yr{2004}
  \at{Catalytic nanomotors: autonomous movement of striped nanorods}.
  \jt{Journal of the American Chemical Society}  \bvol{126}~(41),
  \pg{13424--13431}.

\bibitem[Paxton {\em et~al.\/}(2006)Paxton, Sundararajan, Mallouk \&
  Sen]{paxton2006chemical}
{\sc \au{Paxton, Walter~F}, \au{Sundararajan, Shakuntala}, \au{Mallouk,
  Thomas~E} \& \au{Sen, Ayusman}} \yr{2006}  \at{Chemical locomotion}.
  \jt{Angewandte Chemie International Edition}  \bvol{45}~(33),
  \pg{5420--5429}.

\bibitem[Pietrzyk {\em et~al.\/}(2019)Pietrzyk, Nganguia, Datt, Zhu, Elfring \&
  Pak]{pietrzyk2019flow}
{\sc \au{Pietrzyk, Kyle}, \au{Nganguia, Herve}, \au{Datt, Charu}, \au{Zhu,
  Lailai}, \au{Elfring, Gwynn~J} \& \au{Pak, On~Shun}} \yr{2019}  \at{Flow
  around a squirmer in a shear-thinning fluid}.  \jt{Journal of Non-Newtonian
  Fluid Mechanics}  \bvol{268},  \pg{101--110}.

\bibitem[Rallabandi {\em et~al.\/}(2019)Rallabandi, Yang \&
  Stone]{rallabandi2019motion}
{\sc \au{Rallabandi, Bhargav}, \au{Yang, Fan} \& \au{Stone, Howard~A}}
  \yr{2019}  \at{Motion of hydrodynamically interacting active particles}.
  \jt{arXiv preprint arXiv:1901.04311} .

\bibitem[Saad \& Natale(2019)]{saad2019diffusiophoresis}
{\sc \au{Saad, Shabab} \& \au{Natale, Giovanniantonio}} \yr{2019}
  \at{Diffusiophoresis of active colloids in viscoelastic media}.  \jt{Soft
  matter}  \bvol{15}~(48),  \pg{9909--9919}.

\bibitem[Sabass \& Seifert(2012)]{sabass2012dynamics}
{\sc \au{Sabass, Benedikt} \& \au{Seifert, Udo}} \yr{2012}  \at{Dynamics and
  efficiency of a self-propelled, diffusiophoretic swimmer}.  \jt{The Journal
  of chemical physics}  \bvol{136}~(6),  \pg{064508}.

\bibitem[Sharifi-Mood {\em et~al.\/}(2013)Sharifi-Mood, Koplik \&
  Maldarelli]{sharifi2013}
{\sc \au{Sharifi-Mood, Nima}, \au{Koplik, Joel} \& \au{Maldarelli, Charles}}
  \yr{2013}  \at{Diffusiophoretic self-propulsion of colloids driven by a
  surface reaction: the sub-micron particle regime for exponential and van der
  waals interactions}.  \jt{Physics of Fluids}  \bvol{25}~(1),  \pg{012001}.

\bibitem[Stark(2018)]{stark2018artificial}
{\sc \au{Stark, Holger}} \yr{2018}  \at{Artificial chemotaxis of self-phoretic
  active colloids: Collective behavior}.  \jt{Accounts of chemical research}
  \bvol{51}~(11),  \pg{2681--2688}.

\bibitem[Stone \& Samuel(1996)]{stone1996propulsion}
{\sc \au{Stone, Howard~A} \& \au{Samuel, Aravinthan~DT}} \yr{1996}
  \at{Propulsion of microorganisms by surface distortions}.  \jt{Physical
  review letters}  \bvol{77}~(19),  \pg{4102}.

\bibitem[Su {\em et~al.\/}(2019)Su, Price, Jing, Tian, Liu \&
  Qian]{su2019janus}
{\sc \au{Su, Haiyang}, \au{Price, Cameron-Alexander~Hurd}, \au{Jing, Lingyan},
  \au{Tian, Qiang}, \au{Liu, Jian} \& \au{Qian, Kun}} \yr{2019}  \at{Janus
  particles: Design, preparation, and biomedical applications}.  \jt{Materials
  Today Bio}  \pg{p. 100033}.

\bibitem[Tiefenbruck \& Leal(1980)]{tiefenbruck1980note}
{\sc \au{Tiefenbruck, GF} \& \au{Leal, LG}} \yr{1980}  \at{A note on the slow
  motion of a bubble in a viscoelastic liquid}.  \jt{Journal of Non-Newtonian
  Fluid Mechanics}  \bvol{7}~(2-3),  \pg{257--264}.

\bibitem[Vrentas \& Vrentas(2003)]{vrentas2003steady}
{\sc \au{Vrentas, JS} \& \au{Vrentas, CM}} \yr{2003}  \at{Steady viscoelastic
  diffusion}.  \jt{Journal of applied polymer science}  \bvol{88}~(14),
  \pg{3256--3263}.

\bibitem[Zare {\em et~al.\/}(2019)Zare, Park \& Rhee]{zare2019analysis}
{\sc \au{Zare, Yasser}, \au{Park, Sang~Phil} \& \au{Rhee, Kyong~Yop}} \yr{2019}
   \at{Analysis of complex viscosity and shear thinning behavior in poly
  (lactic acid)/poly (ethylene oxide)/carbon nanotubes biosensor based on
  carreau--yasuda model}.  \jt{Results in Physics}  \bvol{13},  \pg{102245}.

\bibitem[Zhao \& Yang(2013)]{zhao2013electrokinetics}
{\sc \au{Zhao, Cunlu} \& \au{Yang, Chun}} \yr{2013}  \at{Electrokinetics of
  non-newtonian fluids: a review}.  \jt{Advances in colloid and interface
  science}  \bvol{201},  \pg{94--108}.

\bibitem[Zhu {\em et~al.\/}(2012)Zhu, Lauga \& Brandt]{zhu2012self}
{\sc \au{Zhu, Lailai}, \au{Lauga, Eric} \& \au{Brandt, Luca}} \yr{2012}
  \at{Self-propulsion in viscoelastic fluids: Pushers vs. pullers}.
  \jt{Physics of fluids}  \bvol{24}~(5),  \pg{051902}.

\end{thebibliography}


\begin{thebibliography}{6}
\expandafter\ifx\csname natexlab\endcsname\relax\def\natexlab#1{#1}\fi
\expandafter\ifx\csname bibnamefont\endcsname\relax
  \def\bibnamefont#1{#1}\fi
\expandafter\ifx\csname bibfnamefont\endcsname\relax
  \def\bibfnamefont#1{#1}\fi
\expandafter\ifx\csname citenamefont\endcsname\relax
  \def\citenamefont#1{#1}\fi
\expandafter\ifx\csname url\endcsname\relax
  \def\url#1{\texttt{#1}}\fi
\expandafter\ifx\csname urlprefix\endcsname\relax\def\urlprefix{URL }\fi
\providecommand{\bibinfo}[2]{#2}
\providecommand{\eprint}[2][]{\url{#2}}

\bibitem[{\citenamefont{Michelin and Lauga}(2014)}]{michelin2014phoretic}
\bibinfo{author}{\bibfnamefont{S.}~\bibnamefont{Michelin}} \bibnamefont{and}
  \bibinfo{author}{\bibfnamefont{E.}~\bibnamefont{Lauga}},
  \bibinfo{journal}{Journal of Fluid Mechanics} \textbf{\bibinfo{volume}{747}},
  \bibinfo{pages}{572} (\bibinfo{year}{2014}).

\bibitem[{\citenamefont{Derjaguin et~al.}(1947)\citenamefont{Derjaguin,
  Sidorenkov, Zubashchenkov, and Kiseleva}}]{derjaguin1947kinetic}
\bibinfo{author}{\bibfnamefont{B.}~\bibnamefont{Derjaguin}},
  \bibinfo{author}{\bibfnamefont{G.}~\bibnamefont{Sidorenkov}},
  \bibinfo{author}{\bibfnamefont{E.}~\bibnamefont{Zubashchenkov}},
  \bibnamefont{and} \bibinfo{author}{\bibfnamefont{E.}~\bibnamefont{Kiseleva}},
  \bibinfo{journal}{Kolloidn. zh} \textbf{\bibinfo{volume}{9}},
  \bibinfo{pages}{335} (\bibinfo{year}{1947}).

\bibitem[{\citenamefont{Anderson et~al.}(1982)\citenamefont{Anderson, Lowell,
  and Prieve}}]{anderson1982motion}
\bibinfo{author}{\bibfnamefont{J.}~\bibnamefont{Anderson}},
  \bibinfo{author}{\bibfnamefont{M.}~\bibnamefont{Lowell}}, \bibnamefont{and}
  \bibinfo{author}{\bibfnamefont{D.}~\bibnamefont{Prieve}},
  \bibinfo{journal}{Journal of Fluid Mechanics} \textbf{\bibinfo{volume}{117}},
  \bibinfo{pages}{107} (\bibinfo{year}{1982}).

\bibitem[{\citenamefont{Hinch}(1991)}]{hinch_1991}
\bibinfo{author}{\bibfnamefont{E.~J.} \bibnamefont{Hinch}},
  \emph{\bibinfo{title}{Perturbation Methods}}, Cambridge Texts in Applied
  Mathematics (\bibinfo{publisher}{Cambridge University Press},
  \bibinfo{year}{1991}).

\bibitem[{\citenamefont{Blake}(1971)}]{blake1971spherical}
\bibinfo{author}{\bibfnamefont{J.~R.} \bibnamefont{Blake}},
  \bibinfo{journal}{Journal of Fluid Mechanics} \textbf{\bibinfo{volume}{46}},
  \bibinfo{pages}{199} (\bibinfo{year}{1971}).

\bibitem[{\citenamefont{Datt et~al.}(2017)\citenamefont{Datt, Natale,
  Hatzikiriakos, and Elfring}}]{datt2017activeComplex}
\bibinfo{author}{\bibfnamefont{C.}~\bibnamefont{Datt}},
  \bibinfo{author}{\bibfnamefont{G.}~\bibnamefont{Natale}},
  \bibinfo{author}{\bibfnamefont{S.~G.} \bibnamefont{Hatzikiriakos}},
  \bibnamefont{and} \bibinfo{author}{\bibfnamefont{G.~J.}
  \bibnamefont{Elfring}}, \bibinfo{journal}{Journal of Fluid Mechanics}
  \textbf{\bibinfo{volume}{823}}, \bibinfo{pages}{675} (\bibinfo{year}{2017}).

\end{thebibliography}


\begin{thebibliography}{14}
\expandafter\ifx\csname natexlab\endcsname\relax\def\natexlab#1{#1}\fi

\bibitem[Batchelor(1971)]{Batchelor59}
{\sc Batchelor, G.~K.} 1971 Small-scale variation of convected quantities like
  temperature in turbulent fluid. part 1. general discussion and the case of
  small conductivity. {\em J.~Fluid Mech.\/} {\bf 5}, 113--133.

\bibitem[Brownell \& Su(2004)]{Brownell04}
{\sc Brownell, C.~J. \& Su, L.~K.} 2004 Planar measurements of differential
  diffusion in turbulent jets. {\em AIAA Paper 2004-2335\/}.

\bibitem[Brownell \& Su(2007)]{Brownell07}
{\sc Brownell, C.~J. \& Su, L.~K.} 2007 Scale relations and spatial spectra in
  a differentially diffusing jet. {\em AIAA Paper 2007-1314\/}.

\bibitem[Dennis(1985)]{Dennis85}
{\sc Dennis, S. C.~R.} 1985 {Compact explicit finite difference approximations
  to the Navier--Stokes equation}. In {\em Ninth Intl Conf. on Numerical
  Methods in Fluid Dynamics\/} (ed. Soubbaramayer \& J.~P. Boujot), {\em
  Lecture Notes in Physics\/}, vol. 218, pp. 23--51. Springer.

\bibitem[Hwang \& Tuck(1970)]{Hwang70}
{\sc Hwang, L.-S. \& Tuck, E.~O.} 1970 On the oscillations of harbours of
  arbitrary shape. {\em J.~Fluid Mech.\/} {\bf 42}, 447--464.

\bibitem[Koch(1983)]{Koch83}
{\sc Koch, W.} 1983 Resonant acoustic frequencies of flat plate cascades. {\em
  J.~Sound Vib.\/} {\bf 88}, 233--242.

\bibitem[Lee(1971)]{Lee71}
{\sc Lee, J.-J.} 1971 Wave-induced oscillations in harbours of arbitrary
  geometry. {\em J.~Fluid Mech.\/} {\bf 45}, 375--394.

\bibitem[Linton \& Evans(1992)]{Linton92}
{\sc Linton, C.~M. \& Evans, D.~V.} 1992 The radiation and scattering of
  surface waves by a vertical circular cylinder in a channel. {\em Phil.\
  Trans.\ R. Soc.\ Lond.\/} {\bf 338}, 325--357.

\bibitem[Martin(1980)]{Martin80}
{\sc Martin, P.~A.} 1980 On the null-field equations for the exterior problems
  of acoustics. {\em Q.~J. Mech.\ Appl.\ Maths\/} {\bf 33}, 385--396.

\bibitem[Miller(1991)]{Miller91}
{\sc Miller, P.~L.} 1991 Mixing in high schmidt number turbulent jets. PhD
  thesis, California Institute of Technology.

\bibitem[Rogallo(1981)]{Rogallo81}
{\sc Rogallo, R.~S.} 1981 Numerical experiments in homogeneous turbulence. {\em
  Tech. Rep.\/} 81835. NASA Tech.\ Mem.

\bibitem[Ursell(1950)]{Ursell50}
{\sc Ursell, F.} 1950 Surface waves on deep water in the presence of a
  submerged cylinder i. {\em Proc.\ Camb.\ Phil.\ Soc.\/} {\bf 46}, 141--152.

\bibitem[{van Wijngaarden}(1968)]{Wijngaarden68}
{\sc {van Wijngaarden}, L.} 1968 On the oscillations near and at resonance in
  open pipes. {\em J.~Engng Maths\/} {\bf 2}, 225--240.

\bibitem[Worster(1992)]{Worster92}
{\sc Worster, M.~G.} 1992 {The dynamics of mushy layers}. In {\em In
  Interactive dynamics of convection and solidification\/} (ed. S.~H. Davis,
  H.~E. Huppert, W.~Muller \& M.~G. Worster), pp. 113--138. Kluwer.

\end{thebibliography}

\end{document}